\documentclass[useAMS,usenatbib]{mn2e}
\usepackage{amsmath,amssymb}
\usepackage{fancyhdr}
\usepackage{graphicx}
\usepackage{pdfpages}
\usepackage{wasysym}
\usepackage[utf8x]{inputenc}
\usepackage{dcolumn}
\usepackage{url}

\newcommand\mnras{Mon. Not. R. Astron. Soc.}%
\newcommand\nat{Nature}
\newcommand\aap{Astron. Astrophys.}%
\newcommand\aj{Astron.~J.}%
\newcommand\apj{Astrophys.~J.}%
\newcommand\apjl{Astrophys. J. Lett.}%
\newcommand\apjs{Astrophys. J. Suppl. Ser.}
\newcommand\araa{ARA\&A}%Annual Review of Astronomy and Astrophysics
\newcommand\apss{Astrophys. Space Sci.}
\newcommand\prc{Phys. Rev.~C}

\newcommand\gca{Geochim.~Cosmochim.~Acta}%
\title[$s$-process production in rotating massive stars]{$s$-process production in rotating massive stars at solar and low metallicities}
\author[U. Frischknecht et. al]
{Urs Frischknecht$^{1,2}$, Raphael
Hirschi$^{1,3,8}$\thanks{E-mail: r.hirschi@keele.ac.uk}, Marco Pignatari$^{4}$, Andr\'e Maeder$^{5}$,
\newauthor George Meynet$^{5}$, Cristina Chiappini$^{6}$, Friedrich-Karl Thielemann$^{2}$, 
\newauthor  Thomas Rauscher$^{2,7,8}$, Cyril Georgy$^{1}$, Sylvia Ekstr\"om$^{5}$  
\\ \ \\
$^{1}$Astrophysics group, Lennard-Jones Laboratories, Keele University, ST5 5BG, Staffordshire, UK\\
$^{2}$Dept. of Physics, University of Basel, Klingelbergstr. 82, 4056, Basel, Switzerland\\
$^{3}$Kavli Institute for the Physics and Mathematics of the Universe (WPI),
University of Tokyo, 5-1-5 Kashiwanoha, Kashiwa, 277-8583, Japan\\
$^{4}$Konkoly Observatory, Hungarian Academy of Sciences, Konkoly Thege Miklos ut 15-17, H-1121 Budapest, Hungary\\
$^{5}$Geneva Observatory, Geneva University, 1290 Sauverny, Switzerland \\
$^{6}$Leibniz-Institut für Astrophysik Potsdam (AIP), An der Sternwarte 16, 14482 Potsdam, Germany\\
$^{7}$Centre for Astrophysics Research, School of Physics, Astronomy and Mathematics, University of Hertfordshire, Hatfield AL10 9AB, UK \\
$^{8}$UK Network for Bridging Disciplines of Galactic Chemical Evolution (BRIDGCE), {http://www.bridgce.net/}, UK
}
\begin{document}

\date{Accepted 1988 December 15. Received 1988 December 14; in original form 1988 October 11}

\pagerange{\pageref{firstpage}--\pageref{lastpage}} \pubyear{2002}

\maketitle

\label{firstpage}

\begin{abstract}
Rotation was shown to have a strong impact on the structure and light element nucleosynthesis in massive stars. In particular, models including rotation can reproduce the primary nitrogen observed in halo extremely metal-poor (EMP) stars. Additional exploratory models showed that rotation may enhance $s$-process production at low metallicity.

Here we present a large grid of massive star models including rotation and a full $s$-process network to study the impact of rotation on the weak $s$-process. We explore the possibility of producing significant amounts of elements beyond the strontium peak, which is where the weak $s$-process usually stops.

We used the Geneva stellar evolution code coupled to an enlarged reaction network with 737 nuclear species up to bismuth to calculate  $15-40\,\text{M}_\odot$ models at four metallicities ($Z = 0.014,10^{-3}$, $10^{-5}$, and $10^{-7}$) from the main sequence up to the end of oxygen burning.

We confirm that rotation-induced mixing between the convective H-shell and He-core enables an important production of primary $^{14}$N and $^{22}$Ne and $s$-process at low metallicity. 
At low metallicity, even though the production is still
limited by the initial number of iron seeds, rotation enhances the $s$-process
production, even for isotopes heavier than strontium, by increasing the neutron
to seed ratio. The increase in this ratio is a direct consequence of the primary production of $^{22}$Ne.
Despite nuclear uncertainties affecting the $s$-process production and stellar uncertainties affecting the rotation-induced mixing, our results show a robust production of $s$ process at low metallicity when rotation is taken into account.
Considering models with a distribution of initial rotation rates enables to reproduce the observed large range of the [Sr/Ba] ratios in (carbon-enhanced and normal) EMP stars.
\end{abstract}


\section{Introduction}

The classic view of the s-process nucleosynthesis in massive stars is that it occurs in He- and C-burning regions of the stars, producing 
only the low mass range of the s-process elements, typically the elements with an atomic mass number below about 90-100  \citep[e.g.][and references therein]{2011RvMP...83..157K}. 
It has also been shown, that in the regions where the s-process occurs,
the fact that, when the  metallicity decreases, 
1) the neutron source, mainly the $^{22}$Ne($\alpha,n)$ reaction, decreases; 
2) the neutron seeds (Fe) also decreases; 
3) the neutron poisons as for instance $^{16}$O remain independent of the metallicity;
implies that the s-process element production decreases with the metallicity and that there exists
some limiting metallicity below which the s-process becomes negligible. This limit was found to be around $Z$/Z$_\odot$=10$^{-2}$  \citep[][]{1990A&A...234..211P}.

First attempts to investigate the possible role of rotational mixing on the s-process production in massive stars have shown that
this classic picture could be significantly revised. 
The impact of rotation on the $s$-process nucleosynthesis in low-$Z$ massive rotating stars was studied by \citet{2008ApJ...687L..95P}.
In that study the $s$-process production was investigated by assuming different concentrations of primary $^{22}$Ne in the convective He-burning core, guided by the early results of \citet{2007A&A...461..571H}. 
\citet{2012A&A...538L...2F} presented $25\,\text{M}_\odot$ stellar models at various metallicities and with different initial rotation rates using an $s$-process network of 612  isotopes up to the end of core He-burning and 737 isotopes during the later stages. 
The main results of these works were that the s-process production could be boosted in models with strong rotational mixing, that isotopes with an atomic mass heavier than 100 can be synthesised and that 
very different ratios of first to second peak s-process element ratios can be obtained depending on the rotation rate.

The main reason for these changes comes from the following process: rotational mixing allows the production of large amounts of $^{14}$N in the H-burning shell, $^{14}$N, which, once engulfed into the He-burning core, is transformed into $^{22}$Ne via two $\alpha-$captures. Increasing the quantity of $^{22}$Ne favours s-process production since the main neutron source is the $^{22}$Ne($\alpha,n)$ reaction. Nevertheless, the limiting factors 
mentioned just above at low metallicity, namely the decrease of the seeds while the amount of important neutron poisons does not change,   remain whatever the star is rotating or not. Thus rotation act mainly on one of the aspect
of the s-process nucleosynthesis, the neutron source via the amount of $^{22}$Ne, leaving the other more or less the same as in the non-rotating models. Rotation can also have an impact on the s-process through its influence on the size of the
H- and He-burning cores, but these effects remain modest compared to the impact linked to the $^{22}$Ne. 

While the above-mentioned studies provide already the general trends of how rotation will impact the s-process production, they focus on only one initial mass.
In the present work we extend the mass range explored. In that respect, this is the first extended grid of this kind that is published and we hope that this will trigger new theoretical predictions in the future  exploring other physics, such as
the impact of an internal magnetic fields or of the presence of a close binary companion.

Before entering into the main body of this paper, we would like to emphasize an additional point, the fact 
that rotation has a particularly strong impact at low metallicity, and therefore on the evolution 
and nucleosynthesis of the first stellar generations in the Universe.

Due to their low metal content, they are more compact and rotate faster than their equivalents found in the Milky Way. This view is supported by observations of an increasing Be/B-type star ratio with decreasing metallicity \citep{2007A&A...472..577M} and by faster rotating massive stars in the SMC compared to the Milky Way \citep{2008A&A...479..541H}.

Fast rotating stellar models at low $Z$ have been calculated by \citet{2006A&A...447..623M} and \citet{2007A&A...461..571H}. In these 
models, nitrogen yields are much larger than in non-rotating models. When yields from these rotating models are used as input in chemical evolution models, a nice fit of the N/O in very metal poor halo stars \citep[see e.g.][]{2005A&A...430..655S} can be obtained \citep{2006A&A...449L..27C}. 
The nitrogen production in rotating low-$Z$ stellar models is accompanied by large production of other isotopes like $^{13}$C, and especially $^{22}$Ne, which is, as reminded above, the neutron source for $s$~process in massive stars \citep[e.g.][and references therein]{2011RvMP...83..157K}. 

The observation of large $s$-process enhancements in one of the oldest globular clusters in the bulge of our galaxy supports the view that massive stars could indeed be also important sources for these elements \citep{2011Natur.472..454C}, highlighting the need for comprehensive calculations of $s$~process in low-$Z$ massive rotating stars. 
This motivated us to produce a large grid of low-$Z$ massive rotating star models including a full $s$-process network. 
The observations by \cite{barbuy:09} and \cite{2011Natur.472..454C} were later updated by \cite{barbuy:14} and \cite{ness:14}. In particular, \cite{barbuy:14} confirmed that at least part of the stars in the globular cluster NGC 6522 is compatible with the $s$-process production in fast-rotating massive stars at low metallicity.
Galactic chemical evolution (GCE) models using the larger grid of models were presented in \citet{2013A&A...553A..51C,2015arXiv150302954C} (with some modifications explained in these papers) and showed that rotation-induced mixing is able to explain the large scatter for [Sr/Ba] observed in 
extremely metal poor stars.
In this paper, we present the large grid of low-$Z$ massive rotating star models including a full $s$-process network used in the GCE models listed above.

We describe our models in \textsection\ref{sec:models}. The mixing induced by rotation and the production of primary $^{22}$Ne are discussed in \textsection\ref{sec:mixing}. We revisit the 
s process in non-rotating stars and its dependence on initial metallicity in \textsection\ref{sec:mixing}. The impact of rotation on the $s$~process in massive stars at different metallicities is discussed in \textsection\ref{sec:boosteds}. We compare our models to the literature and observations in \textsection\ref{sec:comparison}.
Finally, we give our conclusions in \textsection\ref{sec:conclusions}.

\begin{table}%[ht!]
  \caption{Model parameters: initial mass (column 1), model label (2), initial ratio of surface velocity to critical velocity (3), time-averaged surface velocity during the MS phase (4), metallicity (5), [Fe/H] (6) and total lifetime, $\tau$, from the ZAMS until the advanced phases (7).}
  \centering
  \begin{tabular}{c|c|cr|ccc}
     \hline\hline
     Mass        &  Model  & $\frac{\upsilon_{\rm ini}}{\upsilon_{\rm crit}}$ & $\langle\upsilon\rangle_{\rm MS}$ & $Z$ & [Fe/H] & $\tau$ \\
     $[\text{M}_\odot]$ &      &      &  $[\text{km}\,\text{s}^{-1}]$  &   &  & [Myr] \\
     \hline     
     15            & A15s0 & 0.0      & 0    &   0.014     & 0.0    & 12.7 \\
                   & A15s4 & 0.4      & 200  &   0.014     & 0.0    & 15.0 \\
                   & B15s0 & 0.0      & 0    &   $10^{-3}$ & $-1.8$ & 13.1 \\
                   & B15s4 & 0.4      & 234  &   $10^{-3}$ & $-1.8$ & 15.4 \\
                   & C15s0 & 0.0      & 0    &   $10^{-5}$ & $-3.8$ & 12.9 \\
                   & C15s4 & 0.4      & 277  &   $10^{-5}$ & $-3.8$ & 15.0 \\
     \hline
     20            & A20s0 & 0.0      & 0    &   0.014     & 0.0    &  8.87 \\
                   & A20s4 & 0.4      & 216  &   0.014     & 0.0    & 10.5 \\
                   & B20s0 & 0.0      & 0    &   $10^{-3}$ & $-1.8$ &  9.37 \\
                   & B20s4 & 0.4      & 260  &   $10^{-3}$ & $-1.8$ & 11.1 \\
                   & C20s0 & 0.0      & 0    &   $10^{-5}$ & $-3.8$ &  9.28 \\
                   & C20s4 & 0.4      & 305  &   $10^{-5}$ & $-3.8$ & 10.8 \\
     \hline
     25            & A25s0 & 0.0      & 0    &   0.014     & 0.0    &  7.19 \\
                   & A25s4 & 0.4      & 214  &   0.014     & 0.0    &  8.43 \\
                   & B25s0 & 0.0      & 0    &   $10^{-3}$ & $-1.8$ &  7.62 \\
                   & B25s4 & 0.4      & 285  &   $10^{-3}$ & $-1.8$ &  8.85 \\
                   & C25s0 & 0.0      & 0    &   $10^{-5}$ & $-3.8$ &  7.53 \\
                   & C25s4 & 0.4      & 333  &   $10^{-5}$ & $-3.8$ &  8.68 \\
                   & C25s4b$^{\rm a}$ & 0.4      & 333  &   $10^{-5}$ & $-3.8$ &  8.68 \\
                   & C25s5 & 0.5      & 428  &   $10^{-5}$ & $-3.8$ &  8.85 \\
                   & C25s5b$^{\rm a}$ & 0.5      & 428  &   $10^{-5}$ & $-3.8$ &  8.85 \\
                   & D25s0 & 0.0      & 0    &   $10^{-7}$ & $-5.8$ &  7.18 \\
                   & D25s4 & 0.4      & 383  &   $10^{-7}$ & $-5.8$ &  8.26 \\
                   & D25s4b$^{\rm a}$ & 0.4      & 383  &   $10^{-7}$ & $-5.8$ &  8.26 \\
                   & D25s6 & 0.6      & 588  &   $10^{-7}$ & $-5.8$ &  8.70 \\
                   & D25s6b$^{\rm a}$ & 0.6      & 588  &   $10^{-7}$ & $-5.8$ &  8.70 \\
     \hline
     40            & A40s4 & 0.4      & 186  &   0.014     & 0.0    &  5.75 \\
                   & B40s4 & 0.4      & 334  &   $10^{-3}$ & $-1.8$ &  5.99 \\
                   & C40s4 & 0.4      & 409  &   $10^{-5}$ & $-3.8$ &  5.89 \\
     \hline
  \end{tabular}
      Note.\\
  $^{\rm a}$ Models calculated with a lower $^{17}$O$(\alpha,\gamma)$, see text for details.
\label{tab:sprocgrid}
\end{table}

\section{Models and yield calculations}
\label{sec:models}
\subsection{Model ingredients} 
We calculated the stellar evolution models with the Geneva stellar evolution code (GENEC), which is described in detail in \citet{2008Ap&SS.316...43E}. The main improvement brought to GENEC for these models is the integration of a large nuclear reaction network (613 isotopes up to the end of He-burning and 737 from thereon). The smaller network is almost identical to the $s$-process network used by \citet[][see their table 1]{2000ApJ...533..998T}. This version of GENEC with an enhanced nucleosynthesis network size and the nucleosynthesis network coupled to the structure is the same as in \citet{2012A&A...538L...2F} and \citet{2000ApJ...533..998T}. Since rotation induced mixing is of prime importance in this work, we briefly review here the input physics used. We used the horizontal diffusion coefficient of \citet{1992A&A...265..115Z} and the shear diffusion coefficient from \citet{1997A&A...317..749T}, which is a conservative choice since this prescription includes a strong reduction of mixing across mean molecular weight gradients. 

In the reaction library used for the network calculations, theoretical neutron capture and charged-particle rates from \citet{2000ADNDT..75....1R} were used unless experimental information was available as outlined below. The charged particle reaction rates from \citet{1999NuPhA.656....3A} were used except for the following reactions: $^{22}$Ne($\alpha$,n) and the 3$\alpha$-rate were taken from \citet{2001PhRvL..87t2501J} and from \citet{2005Natur.433..136F}, respectively. 
Neutron capture rates present in the KADoNiS compilation \citep[v0.1][]{2006AIPC..819..123D} were implemented. Beta-decay rates derived from experimental beta-decay half-lives were used except for the temperature-dependent rates given in \citet{1987ADNDT..36..375T}.
The {\sc Reaclib} parameters for $3\alpha$, $^{12}$C$(\alpha,\gamma)^{16}$O, $^{14}$N$(p,\gamma)^{15}$O, and the constant $\beta$-decays rates beyond Pd were obtained from the JINA-REACLIB website (\url{groups.nscl.msu.edu/jina/reaclib/db}).
Two of the most important nuclear reaction rates for s~process in massive stars are $^{22}$Ne$(\alpha,n)$ and $^{22}$Ne$(\alpha,\gamma)$. The rates used in this study, taken from \citet{2001PhRvL..87t2501J} and NACRE, respectively, result in an equal strength of both channels at $T \approx 2.8 \times 10^{8}$\,K ($T_8\approx 2.8$). Below this temperature the $(\alpha,\gamma)$-channel dominates, while above the $(\alpha,n)$-channel is stronger. In our models, an important fraction of $^{22}$Ne is burned when $^{22}$Ne$(\alpha,\gamma)$ dominates over the neutron source. More recent rate determinations of $^{22}$Ne$(\alpha,\gamma)$ from \citet{2006ApJ...643..471K} or \citet{2010NuPhA.841...31I}, \cite{longland:12} and \cite{bisterzo:15} are not used in this work, but are all lower than the NACRE rate. This means that the yields from He-core burning could be higher,
depending also on the ratio between the $(\alpha,n)$ and $(\alpha,\gamma)$ channels.
Previous impact studies of the $^{22}$Ne$(\alpha,\gamma)$ and $^{22}$Ne$(\alpha,n)$
rates on the $s$~process in massive stars are e.g., \cite{kaeppeler:94}, \cite{rauscher:02}, \cite{2010ApJ...710.1557P} and \cite{2014AIPC.1594..146N}.

 In the stellar models presented in this work, for $^{17}$O$(\alpha,\gamma)$ and $^{17}$O$(\alpha,n)$ reaction rates we used the rates of \citet{1988ADNDT..40..283C} (hereafter CF88) and \citet{1999NuPhA.656....3A}, respectively. Their ratio determines the strength of $^{16}$O as a neutron poison \citep[e.g.,][]{1992A&A...258..357B,2008IAUS..255..297H}. 
\citet{1993PhRvC..48.2746D} predicted that the $^{17}$O$(\alpha,\gamma)$ should be a factor of $1000$ smaller than the CF88 rate. 
More recently, two independent groups measured the $^{17}$O$(\alpha,\gamma)$ rate 
\citep{2011nuco.confT,2011PhRvC..83e2802B,best:13}, obtaining a rate lower than CF88 at relevant temperatures, but not as low as \citet{1993PhRvC..48.2746D}. \cite{best:13} also provided a new rate for the $^{17}$O$(\alpha,n)$.
In order to assess the impact of a lower $^{17}$O$(\alpha,\gamma)$
rate, we calculated the rotating $25\,\text{M}_\odot$ models at $Z=10^{-5}$ (C25s4, C25s5) and $10^{-7}$ (D25s4, D25s6) with the CF88 rate divided by a factor 10, which 
is consistent with the new measurements within the uncertainties.
These models are in the following text labeled by and additional ``b'' at the end of their name.  Although the $25\,\text{M}_\odot$ models have already been discussed in \citet{2012A&A...538L...2F}, we provide more details about these models in this paper and it is important to present models of all masses in a single paper.

The mass range from 15 and $40\,\text{M}_\odot$ was investigated, with models of 15, 20, 25 and $40\,\text{M}_\odot$ and for each mass a model without rotation and at least one with rotation was calculated.
The stellar models were calculated from zero-age main-sequence (ZAMS) up to O-burning for the grid of models, which is shown in Table~\ref{tab:sprocgrid}. 
Models with masses below $15\,\text{M}_\odot$ were not followed, because the temperature is not high enough to efficiently activate the neutron source. The observed s~process nuclei are usually also not considered to originate from stars beyond $40\,\text{M}_\odot$, because more massive stars are thought to collapse directly to black holes at the end of their life without an explosion, while stars between 25 and $40\,\text{M}_\odot$ lead to black hole formation by matter falling back on the remnant  neutron star \citep[e.g.][]{2002RvMP...74.1015W, 2003ApJ...591..288H}. In the latter case an explosion still happens, ejecting fractions of the synthesised elements.    
Let us note, however, that
the above mass limits between the different scenarios for the ultimate
explosion are very uncertain and depend on many factors such
as the metallicity and the input physics used in the stellar modelling
\citep[see the recent review by][and references therein]{2012ARNPS..62..407J}.
All masses were calculated at initial metallicities, $Z=0.014$ (solar metallicity models, starting with letter A), $10^{-3}$ (B), and $10^{-5}$ (C), to investigate the metallicity dependence of the s~process in massive rotating stars. Additionally $25\,\text{M}_\odot$ stars at $Z=10^{-7}$ were modelled. The [Fe/H]-values corresponding to these four metallicites are $0$, $-1.8$, $-3.8$ and $-5.8$. For $Z=0.014$ we have adopted the elemental composition of \citet{2005ASPC..336...25A}, with the modified Ne abundance of \citet{2006ApJ...647L.143C}, and the isotopic ratios from \citet{2003ApJ...591.1220L}. At all  three sub-solar metallicities, we assumed an $\alpha$-enhanced composition with the $\alpha$-elements ($^{12}$C, $^{16}$O, $^{20}$Ne, $^{24}$Mg, $^{28}$Si, $^{32}$S, $^{36}$Ar, $^{40}$Ca, and $^{48}$Ti) enhanced with respect to iron, i.e. [X/Fe]$=-A$[Fe/H] for $-1\ge$[Fe/H]$>0$ and [X/Fe]$=A=$constant for [Fe/H]$\le-1$ where $A=+0.562$, $+0.886$, $+0.500$, $+0.411$, $+0.307$, $+0.435$, $+0.300$, $+0.222$, and $+0.251$ for the different $\alpha$-enhanced isotopes. This $\alpha$-enhanced composition was derived by fitting the abundance trends [X/Fe] vs [Fe/H] derived from halo and thick disk F- and G-dwarfs \citep{2006MNRAS.367.1329R} between [Fe/H]$=0$ and $-1$. The linear fits were fixed to the solar value, i.e. [X/Fe]$=0$ at [Fe/H]$=0$, and below [Fe/H]$=-1$ a plateau was assumed. The values for the noble gases were adopted from the galactic chemical evolution models of \citet{2006ApJ...653.1145K}. This $\alpha$-enhancement gives a Fe/$Z$ ratio for $[$Fe/H$]\le-1$, which is a factor of 4.6 lower than at solar $Z$. All other elements were scaled from the solar composition. 

As standard initial rotation rate $40\%$ of critical velocity ($\upsilon_{\rm ini}/\upsilon_{\rm crit}=0.4$) was used. For 15 to $25\,\text{M}_\odot$ stars at solar $Z$, it corresponds to an average equatorial rotation velocity on the main sequence $\langle\upsilon\rangle_{\rm MS}=$200 to $220\,\text{km}\,\text{s}^{-1}$. This is slightly lower than the peak of the velocity distribution, at $\upsilon_{\rm MW,peak}=225\,\text{km}\,\text{s}^{-1}$,  found for O- and B-type stars in the Milky Way \citep{2006A&A...457..265D,2009A&A...496..841H}. Due to their low metal content, low-$Z$ massive stars are more compact and have a higher surface velocity than their equivalents found in the Milky Way. With $\upsilon_{\rm ini}/\upsilon_{\rm crit}=$constant, $\langle\upsilon\rangle_{\rm MS}$ increases with decreasing $Z$ up to about $400\,\text{km}\,\text{s}^{-1}$. 
This view of faster rotating massive stars at low $Z$ is supported by observations of an increasing Be/B-type star ratio with decreasing metallicity \citep{2007A&A...472..577M,1999A&A...346..459M}, by faster rotating massive stars in the SMC compared to the Milky Way \citep{2008A&A...479..541H}, and hydrodynamic models of the first generation of stars \citep{2011MNRAS.tmp..142S}. Thus, $\upsilon_{\rm ini}/\upsilon_{\rm crit}$ being constant is a conservative choice and might turn out to be too slow to reproduce the peak velocity of the velocity distribution at low $Z$, which is unknown. We assess the possible impact of faster rotation at low $Z$ by models C25s5 and D25s6 with $\upsilon_{\rm ini}/\upsilon_{\rm crit}=0.5$ and $0.6$, respectively.

More details about the models, a script to fit reaction rates in the reaclib format and a script to generate initial abundance sets for a given metallicity are available upon request and are described in \citet{2012phdthesis_ursfrischknecht}.

\subsection{Yield calculations}
\label{sec:yields}
In this work, a complete list of pre-supernova (pre-SN) yields is determined. The total pre-SN yields include a wind and a supernova-progenitor contribution. The pre-SN yield of a nucleus $i$ is the net amount produced of it in $\text{M}_\odot$ and can easily be calculated by 
\begin{eqnarray}
m_i&=&\int\limits_{M_{\rm rem}}^{M^*} (X_i(M)-X_{i,0}) \text{d}M \nonumber \\
&&+ \int\limits_0^\tau \dot{M}(t) (X_{i,{\rm s}}(t)-X_{i,0}) \text{d}t,
\label{eq:yields}
\end{eqnarray}
where $M^*$ is the stellar mass before the explosion, $X_i(M)$ the mass fraction of nucleus $i$ at Lagrangian mass coordinate $M$, $X_{i,0}$ the initial mass fraction, $X_{i,{\rm s}}$ the surface mass fraction and $\dot{M}$ the mass loss rate. The first term on the left hand side describes the mass produced or destroyed in the supernova-progenitor and the second term describes what is ejected by the wind. The remnant mass $M_{\rm rem}$ was derived from the relation of $M_{\rm rem}$ to $M_{\rm CO}$, which was originally established in \citet{1992A&A...264..105M}. $M_{\rm CO}$ is the carbon-oxygen core mass determined as the part of the star for which the $^4$He mass fraction is below $10^{-2}$. Both, $M_{\rm rem}$ and $M_{\rm CO}$, are listed in Table~\ref{tab:cores} in units of $\text{M}_\odot$, as well as the final mass, $M_{\rm fin}$, the mass coordinate for which $X(^4{\rm He})>0.75$, $M_\alpha$, the maximal extension of the convective He core $M_{\rm He}^{\rm max}$, and the maximal mass of convective C-burning shell $M_{\rm C}^{\rm max}$. The latter is given because this is the maximal mass coordinate at which the $s$-process produced in the C-shell can be mixed outwards.

The time scales of C-burning and later evolutionary stages are much shorter than those of H and He burning stages. Our models were calculated at least up to the onset of O-burning, hence the wind contribution in Eq.~\ref{eq:yields} is fully determined by our models. The pre-SN term in Eq.~\ref{eq:yields} was calculated from the final profile during O-burning. Changes in the chemical profile during the final phase appear only in the innermost part of the star. We compared our models with \citet{2004A&A...425..649H} and even though our models do not use exactly the same mixing and wind prescription, the lower boundary and the extension of the C-shell as well as the size of convective core during O-burning, are similar. We therefore know that our models would evolve in a similar way as the one of  \citet{2004A&A...425..649H}, up to the onset of core collapse. 
In this case we expect only a weak modification of the yields  for the $15\,\text{M}_\odot$ star. Thus we are confident that running the models only up to O-burning is sufficient for a good approximation of the pre-explosive yields.   

The yields from the SN progenitor are modified by explosive nucleosynthesis activated by SN shock \citep[e.\,g.][]{1996ApJ...460..408T}. 
The total yields of $s$-process nuclei are not strongly modified by the explosion \citep[e.g.,][]{2009ApJ...702.1068T}.
Therefore, the yields calculated here can be taken as a good estimate and are well suited to investigate the galactic chemical enrichment in $s$-process nuclei and light nuclei by massive rotating stars.   
\begin{table}%[ht!]
  \begin{center}
  \caption{Final total mass and different core masses of the models}\label{tab:cores}
  \begin{tabular}{lrrrrrr}
     \hline\hline
Model&$M_{\rm fin}$&$M_\alpha$&$M_{\rm He}^{\rm max}$&$M_{\rm C}^{\rm max}$&$M_{\rm CO}$&$M_{\rm rem}^{\rm a}$\\
     \hline     
     A15s0    &  13.01 & 4.27 & 2.24 & 2.19 & 2.35 & 1.49 \\
     A15s4    &  10.43 & 5.81 & 3.39 & 2.75 & 3.33 & 1.74 \\%2.89 G015z14S413/P015z14S4.y0045851.gz 2.90
     B15s0    &  14.80 & 4.74 & 2.60 & 2.33 & 2.62 & 1.56 \\%2.39 G015z01S013/P015z01S0.y0064881.gz 2.40
     B15s4    &  13.84 & 6.03 & 3.52 & 2.54 & 3.44 & 1.77 \\%2.79 G015z01S413/P015z01S4.y0038941.gz 2.80
     C15s0    &  14.99 & 4.54 & 2.41 & 2.02 & 2.49 & 1.52 \\%2.19 G015zm5S003/P015zm5S0.y0058961.gz 2.20
     C15s4    &  14.84 & 5.70 & 3.41 & 2.06 & 3.34 & 1.74 \\%2.49 G015zm5S413/P015zm5S4.y0035401.gz 2.50
     \hline
     A20s0    &   9.02 & 6.17 & 3.84 & 3.23 & 3.76 & 1.85 \\%3.39 G020z14S013/P020z14S0.y0054241.gz 3.40
     A20s4    &   7.92 & 7.88 & 5.36 & 3.41 & 5.13 & 2.20 \\%3.99 G020z14S413/P020z14S4.y0068061.gz 4.00
     B20s0    &  19.85 & 6.65 & 4.15 & 3.75 & 4.11 & 1.94 \\% 3.79 G020z01S013/P020z01S0.y0065121.gz 3.80
     B20s4    &  10.91 & 8.16 & 5.41 & 4.35 & 5.22 & 2.22 \\%4.59 G020z01S413/P020z01S4.y0044001.gz 4.60
     C20s0    &  20.00 & 6.26 & 3.93 & 3.54 & 3.88 & 1.88 \\%3.59 G020zm5S003/P020zm5S0.y0050341.gz 3.60
     C20s4    &  17.01 & 8.10 & 5.36 & 3.82 & 5.18 & 2.21 \\%4.49 G020zm5S413/P020zm5S4.y0043701.gz 4.50
     \hline
     A25s0    &  10.86 & 8.23 & 5.74 & 4.87 & 5.53 & 2.30 \\% 4.99 G025z14S013/P025z14S0.y0045681.gz 5.00
     A25s4    &  10.04 & 9.99 & 7.40 & 5.97 & 6.97 & 2.66 \\% 6.29 G025z14S413/P025z14S4.y0099721.gz 6.30
     B25s0    &  24.73 & 8.63 & 5.92 & 4.97 & 5.79 & 2.36 \\%5.19 G025z01S013/P025z01S0.y0054341.gz 5.20
     B25s4    &  14.32 &10.96 & 7.93 & 6.62 & 7.56 & 2.81 \\%6.79 G025z01S413/P025z01S4.y0049781.gz 6.80
     C25s0    &  25.00 & 8.03 & 5.61 & 4.47 & 5.57 & 2.31 \\%4.79 G025zm5S003/P025zm5S0.y0066201.gz 4.80
     C25s4    &  24.34 &10.69 & 7.63 & 5.07 & 7.33 & 2.75 \\%5.99 G025zm5S413/P025zm5S4.y0035301.gz 6.00
     C25s4b$^{\rm b}$    &  24.34 &10.69 & 7.65 & 6.33 & 7.25 & 2.73 \\%5.27 -- 6.33 %6.59 G025zm5S404/P025zm5S4.y0049901.gz 6.60
     C25s5    &  24.72 &10.49 & 7.38 & 5.59 & 7.08 & 2.69 \\%6.09 G025zm5S513/P025zm5S5.y0071741.gz 6.10
     C25s5b$^{\rm b}$    &  24.38 &10.49 & 7.37 & 5.10 & 7.12 & 2.70 \\% 5.59 G025zm5S504/P025zm5S5.y0071381.gz 5.60
     D25s0    &  25.00 & 7.39 & 5.72 & 4.09 & 5.56 & 2.31 \\%4.59 G025zm7S003/P025zm7S0.y0051101.gz 4.60
     D25s4    &  25.00 & 8.77 & 5.78 & 4.97 & 5.61 & 2.32 \\%5.09 G025zm7S403/P025zm7S4.y0046881.gz 5.10
     D25s4b$^{\rm b}$    &  25.00 & 8.77 & 5.80 & 4.49 & 5.56 & 2.31 \\%4.79 G025zm7S404/P025zm7S4.y0051381.gz 4.80
     D25s6    &  24.81 & 9.72 & 6.53 & 3.92 & 6.19 & 2.46 \\%4.89 G025zm7S603/P025zm7S6.y0069361.gz 4.90
     D25s6b$^{\rm b}$    &  24.81 & 9.71 & 6.52 & 4.27 & 6.29 & 2.49 \\%% 4.99 G025zm7S604/P025zm7S6.y0062861.gz 5.00
     \hline
     A40s4    &  19.01 &19.01$^{\rm c}$ & 15.23 & 14.10 & 15.04& 4.65 \\%14.09 G040z14S413/P040z14S4.y0067041.gz 14.10
     B40s4    &  25.15 &19.30           & 15.40 & 13.90 & 14.76& 4.57 \\%13.89 G040z01S413/P040z01S4.y0037741.gz 13.90
     C40s4    &  38.49 &19.18           & 14.70 &  6.51 & 14.08& 4.36 \\%11.79 G040zm5S418/P040zm5S4.y0036501.gz 11.80
     \hline
  \end{tabular}
  \end{center}
      Note.\\
      $^{\rm a}$ $M_{\rm rem}$ is estimated following the relation established in \citet{1992A&A...264..105M}.\\$^{\rm b}$ Models calculated with a lower $^{17}$O$(\alpha,\gamma)$, see Sect.~2.1 for details. %As expected, a different value for this rate does not affect significantly the evolution and structure of the model.
\\$^{\rm c}$ This star ends its life as WR star and as a consequence  $M_\alpha=M_{\rm fin}$.

\end{table}

\begin{figure}%[ht!]
  \centering
  \includegraphics[width=0.49\textwidth]{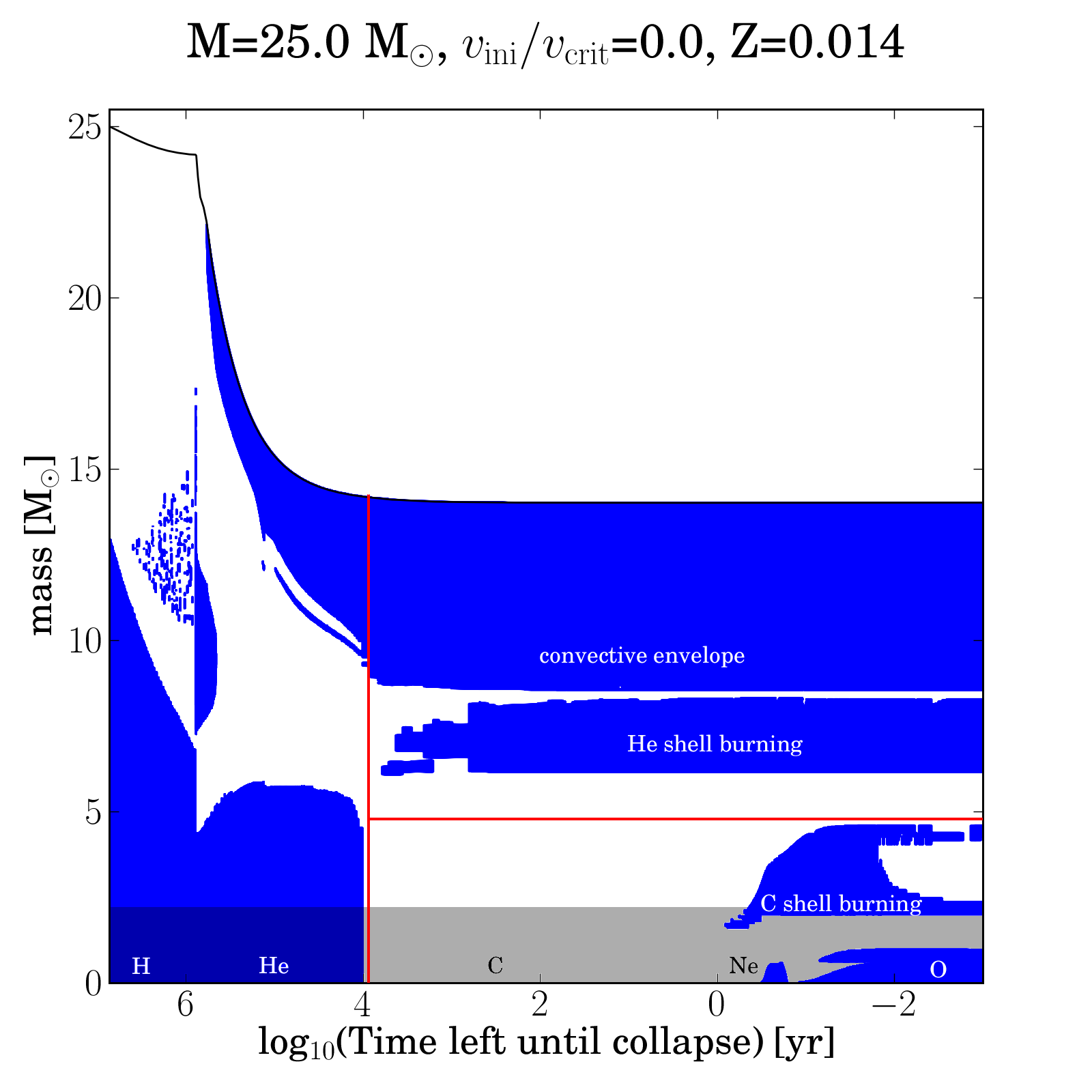}
  \caption[Kippenhahn diagram of 25~$M_\odot$ star with $Z=Z_\odot$ and no rotation]{Kippenhahn diagram of $25\,\text{M}_\odot$ star with $Z=\text{Z}_\odot$ and no rotation (A25s0), to illustrate the $M_{\rm C-He}$ (red horizontal line). The shaded area show the mass ending up inside $M_{\rm rem}$. The red vertical line marks the point in the stellar life where the core He $s$-process yields are calculated. }
  \label{fig:kipm025z14S000}
\end{figure}

We calculated the yields separately for core He, shell He and shell C burning to distinguish between these three contributions to the $s$-process production. For this purpose, we calculated the yields both at the end of core He-burning (He-core contribution) and at the pre-SN stage considering only the material above the final mass cut, $M_{\rm r}>M_{\rm rem}$, as illustrated in Fig.~\ref{fig:kipm025z14S000}.  The separate contributions from shell He and shell C burning are obtained by splitting the pre-SN yields in two parts at mass $M_{\rm C-He}$ (red horizontal line in Fig.~\ref{fig:kipm025z14S000}). $^{20}$Ne is a C burning product and its abundance drop at the outer boundary of the C-burning shell was chosen to determine $M_{\rm C}^{\rm max}$, and finally we set $M_{\rm C-He}=M_{\rm C}^{\rm max}+0.01$. 
      
Besides the yields, the production factors, $f$, will be used in the subsequent discussion. The production factor of an isotope $i$ is defined as
\begin{equation*}
  f_i=\frac{m_{i,{\rm eject}}}{m_{i,{\rm ini}}} = \frac{m_i+m_{i,{\rm ini}}}{m_{i,{\rm ini}}},
\end{equation*}
with $m_i$ the total yield from Eq.~\ref{eq:yields}, $m_{i,{\rm eject}}$ the ejected mass, and $m_{i,{\rm ini}}$ the initial mass of nucleus $i$ in the star. The production factor quantifies if a star is a strong producer of an element or not. 

The yields are available on \url{http://www.astro.keele.ac.uk/shyne/datasets }.

\section{Rotation-induced mixing and production of primary\ {$^{22}$Ne} and {$^{14}$N}}
\label{sec:mixing}
\label{sec:res_ne22_production}
\citet{2002A&A...381L..25M, 2002A&A...390..561M} and \citet{2007A&A...461..571H} find that rotating stars produce important amounts of primary $^{14}$N and $^{22}$Ne via rotation-induced mixing. The production of these nuclei originates from the transport of matter between the He-burning core and the H-burning shell. If the He-burning products $^{12}$C and $^{16}$O reach the proton-rich layers, they are burnt immediately into $^{14}$N via the CNO-cycle. 
A $^{14}$N-rich zone 
is produced in this way at the lower edge of the H-burning shell as shown in Fig.~\ref{fig:xprofile-vs-mr}. Some of this nitrogen is transported back into the He-burning core, where it is further transformed into $^{22}$Ne via two $\alpha$-captures. In this section, we attempt to answer the following questions: Under which conditions is the transport of chemical elements efficient? How much $^{22}$Ne and $^{14}$N is produced in massive stars? 

\subsection{Helium core burning}

\begin{figure}
 \centering
 \includegraphics[width=0.49\textwidth]{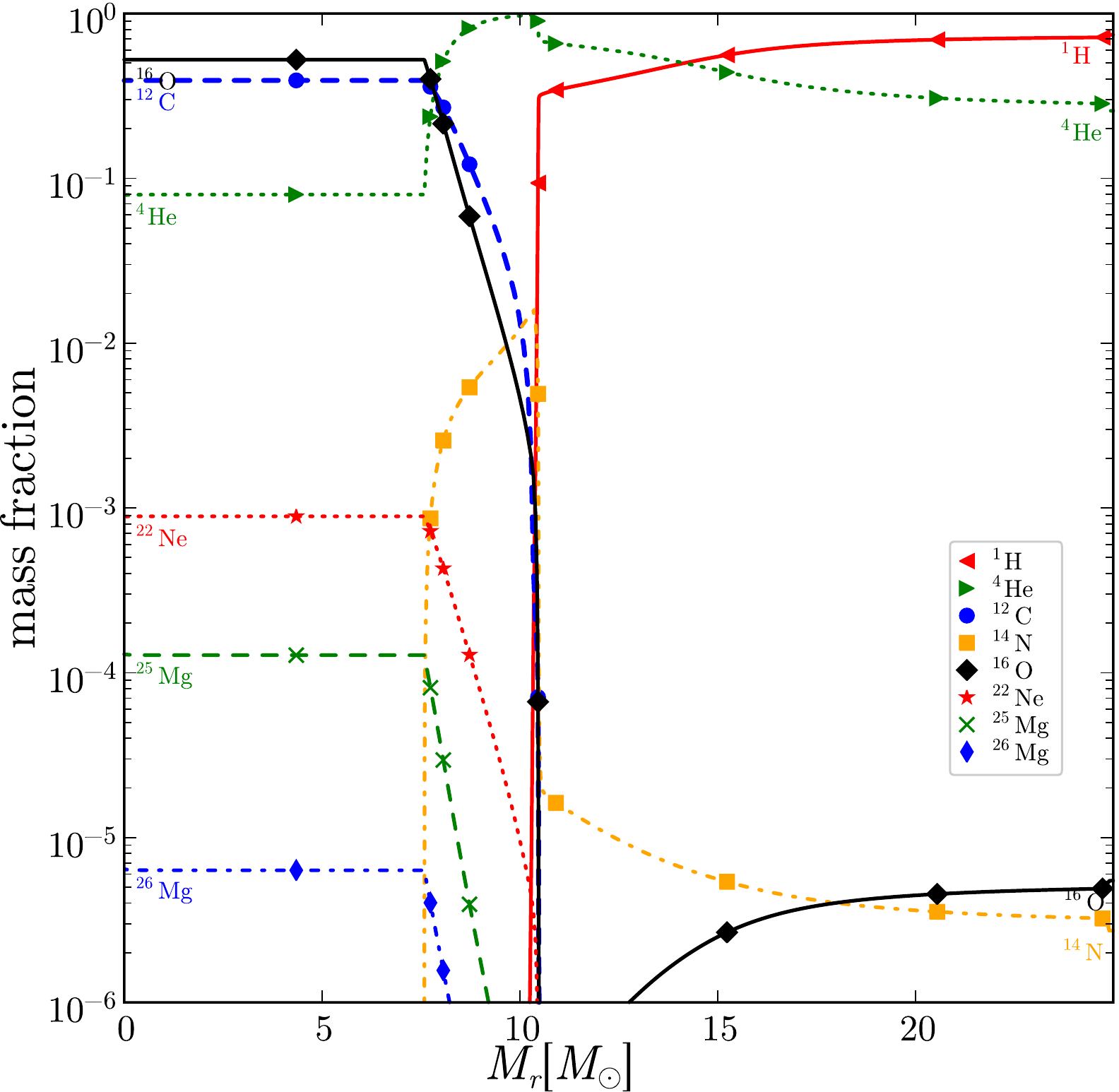}
 \caption[Abundance profile during He-burning of $25\,\text{M}_\odot$ star]{Abundance profiles of the main light isotopes during central He-burning (X$_{\rm c}($He$)\approx0.08$) for the $25\,\text{M}_\odot$ model with rotation 
 and $Z=10^{-5}$ (C25S4). The convective He-burning core extends from the center to about $M_{\rm r}=7.5\,\text{M}_\odot$ (flat abundance profiles). The bottom of hydrogen shell burning is just above $10\,\text{M}_\odot$ (sudden drop of hydrogen abundance). Rotation-induced mixing brings freshly produced $^{12}$C and $^{16}$O from the core into contact with the hydrogen burning shell, where a peak a primary nitrogen ($^{14}$N) develops. Further mixing (both convective and rotation-induced) brings the primary nitrogen down into the He-burning core where it is transformed into $^{22}$Ne, leading to primary production of both $^{14}$N and $^{22}$Ne.}
 \label{fig:xprofile-vs-mr}
\end{figure}

The transport of chemical elements is illustrated for the $25\,\text{M}_\odot$ model  with rotation at $Z=10^{-5}$ in Fig.~\ref{fig:xprofile-vs-mr}, which shows the abundance profiles in this model during core He-burning. The rotation induced mixing, which leads to the production of primary $^{14}$N and $^{22}$Ne, occurs in the region above the convective He core ($M_{\rm r}\approx7.5-10.5\,\text{M}_\odot$). The core itself is identifiable by the flat abundance profile between $M_{\rm r}=0$ and $7.5\,\text{M}_\odot$. Differential rotation develops between the convective He core and H shell mainly because of the core contraction and envelope expansion at the end of the main sequence.
The differential rotation induces secular shear mixing in this radiative zone, in which no mixing would take place in non-rotating models. Shear mixing, a diffusive process, brings primary $^{12}$C and $^{16}$O (blue dashed and black continuous lines) into contact with the H-burning layer and creates a $^{14}$N-pocket ($M_{\rm r}\approx7.5-10.5\,\text{M}_\odot$) via the CNO cycle as explained above. In our models, the transport of $^{14}$N back to the centre is mainly due to the growth of the convective core, incorporating parts of the $^{14}$N-pocket. Indeed, the diffusive transport is not fast enough to produce a $^{22}$Ne mass fraction, $X(^{22}{\rm Ne})$, of $10^{-3}$ to $10^{-2}$ in the core, necessary to boost the $s$~process significantly. 

Secular shear is the main mechanism for the transport between He-core and H-envelope. The diffusion coefficient, $D_{\rm shear}$, used in the models presented here, is the coefficient of \citet{1997A&A...317..749T} and is given by

\begin{eqnarray*}
D_{\rm{shear}} &=& \frac{(K +D_{\rm h})}{\left[\frac{\varphi}{\delta}\nabla_\mu(1+\frac{K}{D_{\rm h}})+(\nabla_{\rm ad}-\nabla_{\rm rad})\right]} \times\\
&&\frac{\alpha H_{\rm p}}{g\delta}\left(\frac{9\pi}{32}\Omega\frac{d\ln\Omega}{d\ln r}\right)^2.
\end{eqnarray*}

Naturally, high $\Omega$-gradient and  $\Omega$ favour shear. The presence of a mean molecular weight gradient, $\nabla_\mu$, on the other hand, has a stabilising effect on shear mixing.
Such a $\nabla_\mu$ is present between H-burning shell and He-rich core and is most prominent at the lower edge of the H-burning zone. Using the formula of \citet{1997A&A...317..749T}, $D_{\rm shear}$ is lowered most efficiently where the thermal diffusivity, $K$, is larger than the horizontal turbulence, $D_\text{h}$ ($K>D_\text{h}$). In our models, just above the convective He-core, where  $K/D_\text{h}$ has typical values between $10$ and $100$, and where $\nabla_\mu$ is highest the term including $\nabla_\mu$ reaches values up to $10^3$, which shows the strong inhibiting effect of $\mu$-gradients on mixing. This can also be seen on the left hand side in Figs.~\ref{fig:diffcoef_case_a},  \ref{fig:diffcoef_case_b} and \ref{fig:diffcoef_case_c} at $M_{\rm r}\approx5-10\,\text{M}_\odot$, where $K$ is the black dotted line and $D_\text{h}$ is the blue dash-dotted line. The $K/D_\text{h}$ ratio does not change significantly in the relevant regions in the course of central He-burning. Regions of strong $\mu$-gradients can be identified by steep slopes in the abundance of hydrogen and carbon on the right-hand side of these figures as discussed below.
There are other formulae for shear mixing, which might lead to different mixing efficiencies. For example, in the formula of \citet{1997A&A...321..134M} for $D_{\rm shear}$, the prefactor $(1+\frac{K}{D_{\rm h}})$ is not present and the inhibiting effect of the $\mu$-gradient is weaker, which means that the shear mixing would be stronger had we used that formula. If the Taylor-Spruit dynamo due to magnetic fields were considered as in for example  \citet{2005ApJ...626..350H}, mixing would also be stronger and often leads to a quasi-homogeneous chemical evolution of rotating low-$Z$ stars \citep{2006A&A...460..199Y}. The mixing considered in this study is thus conservative and mixing could be stronger.

In the grid of models including the effects of rotation that we have calculated, there are three different configurations of the stellar structure that may occur during central He-burning. These cases are illustrated with the help of three evolutionary snap-shots of a rotating $25\,\text{M}_\odot$ $Z=10^{-3}$ star during central He-burning: 
\begin{itemize}
  \item Case (a): In the first configuration, shown in Fig.~\ref{fig:diffcoef_case_a}, the convective H-burning shell ($M_{\rm r}\approx9-13\,\text{M}_\odot$) rotates considerably slower than the regions below (the angular velocity $\Omega$ profile is plotted as an orange dashed line on the left hand side). The steep gradient of $\Omega$ at the lower boundary of the convective shell compensates for the inhibiting effect of $\nabla_\mu$, which is strongest just below the convective shell where the gradient of hydrogen abundance is very steep. In this configuration, $D_{\rm shear}$ has values between $10^4$ and $10^7\,\text{cm}^2\,\text{s}^{-1}$ throughout the radiative region between the convective He-core and the H-shell zones, facilitating a strong production of primary nitrogen. 
  \begin{figure*}%[ht!]
    \includegraphics[width=\textwidth]{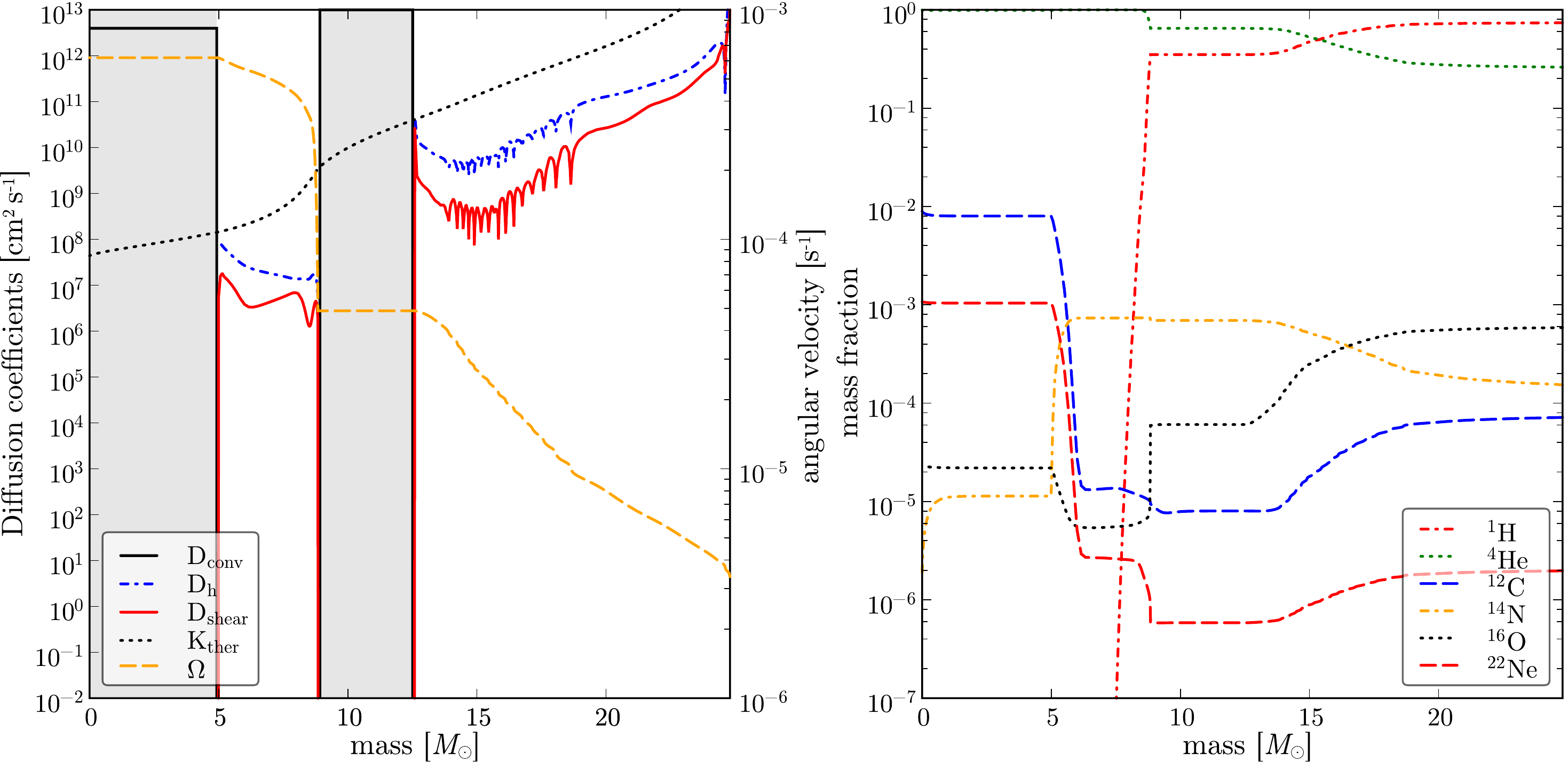}
    \caption[Diffusion coefficient and abundance profiles during central He-burning, when a convective H-shell is present]{Diffusion coefficient profiles on the left hand side and abundance profiles on the right hand side during central He-burning, when a convective H-shell is present, inside the $25\,\text{M}_\odot$ star with rotation at $Z=10^{-3}$ (B25S4). The shear diffusion coefficient (red continuous line) is responsible for the mixing between He-core and H-shell. The convective regions are represented by the grey shaded areas.} 
    \label{fig:diffcoef_case_a}
  \end{figure*}
  \item Case {\bf (b)}: this configuration shown in Fig.~\ref{fig:diffcoef_case_b} is very similar to case (a), i.\,e. there is a convective H-burning shell but with the important difference that the convective H-shell is moving away from its lowest mass coordinate. The upward migration of the lower boundary leaves a shallow $\Omega$-gradient behind, at $M_{\rm r}\approx9.5\,\text{M}_\odot$ on the left hand side in Fig.~\ref{fig:diffcoef_case_b}. In this case, the steep $\Omega$-gradient and the $\mu$-gradient do not coincide, and a region with low values of $D_{\rm shear}$ develops, i.e. $D_{\rm shear}$ between $10$ and $10^4\,\text{cm}^2\,\text{s}^{-1}$. 
The mixing across the bottom of the convective shell is thus less efficient and abundance gradients are steeper below the convective shell (just below $10\,\text{M}_\odot$ in the right panel of Fig.~\ref{fig:diffcoef_case_b})
      
  \begin{figure*}%[ht!]
    \includegraphics[width=\textwidth]{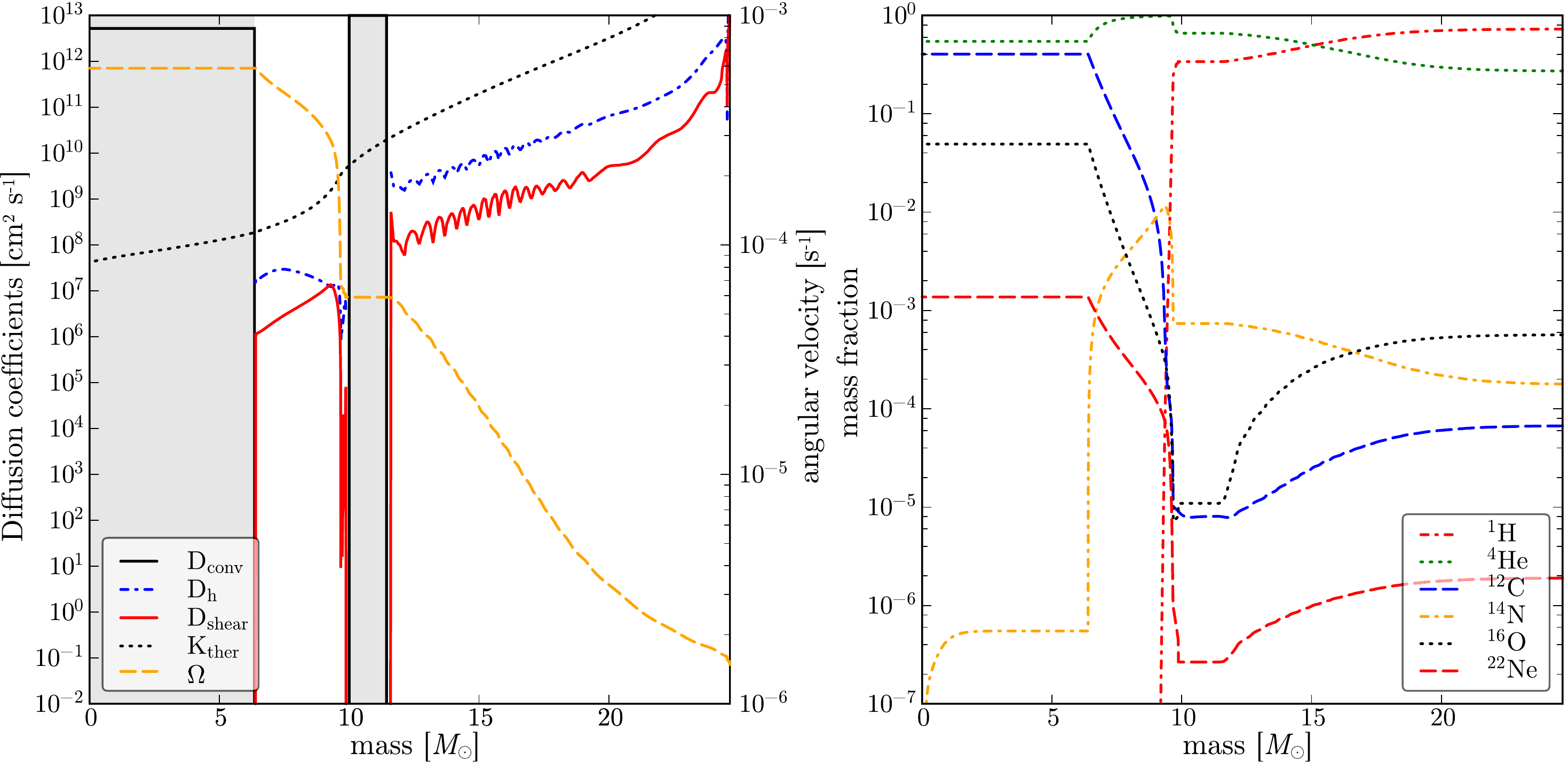}
    \caption[Diffusion coefficient and abundance profiles during central He-burning, when a retracting convective H-shell is present]{Diffusion coefficient profiles on the left hand side and abundance profiles on the right hand side during central He-burning, when a retracting convective H-shell is present, inside the $25\,\text{M}_\odot$ star with rotation at $Z=10^{-3}$ (B25S4). The shear diffusion coefficient (red continuous line) describes the mixing between He-core and H-shell. The convective regions are represented by the grey shaded areas.}
    \label{fig:diffcoef_case_b}
  \end{figure*}
  \item Case (c), shown in Fig.~\ref{fig:diffcoef_case_c}, is the case with no convective zone in the H-rich layers and only a moderate $\Omega$-gradient across the H-burning shell. At the mass coordinate, where abundance gradients are steepest (at 10~$M_\odot$ in the right panel of Fig.~\ref{fig:diffcoef_case_c}), the shear diffusion coefficient is weakest, with $D_{\rm shear}$ between $1$ and $10^3\,\text{cm}^2\,\text{s}^{-1}$. During helium burning, case (c) may follow case (b). In this situation, the $\Omega$-gradient is even lower at the bottom of the H-burning shell 
  and $D_{\rm shear}$ has the lowest values. If there is no convective H-burning shell, then case (c) is the only case the model goes through. 
  \begin{figure*}
    \includegraphics[width=\textwidth]{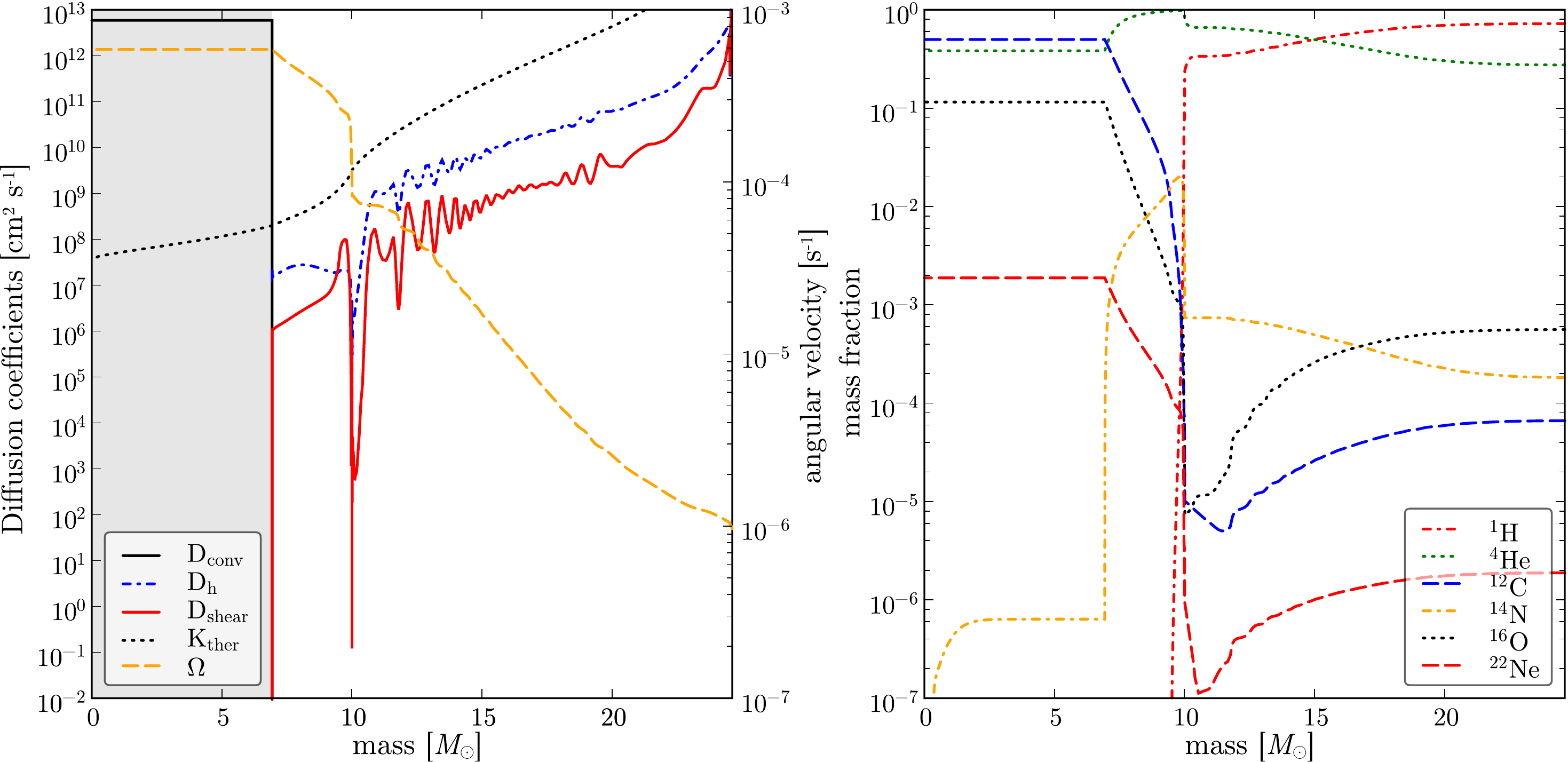}
    \caption[Diffusion coefficient and abundance profiles during central He-burning, when no convective H-shell is present]{Diffusion coefficient profiles on the left hand side and abundance profiles on the right hand side during central He-burning, when no convective H-shell is present, inside the $25\,\text{M}_\odot$ star with rotation at $Z=10^{-3}$ (B25S4). The shear diffusion coefficient (red continuous line) describes the mixing between He-core and H-shell. The convective core is represented by the grey shaded areas.}
    \label{fig:diffcoef_case_c}
  \end{figure*}
\end{itemize}
The rotating solar metallicity 15, 20 and $25\,\text{M}_\odot$ models, as well as the $15\,\text{M}_\odot$ with sub-solar $Z$ do not develop a convective zone at the inner edge of the hydrogen rich layers during central He-burning. Thus, mixing in these models correspond to case (c).
The rotating sub-solar $Z$ models with 20, 25 and $40\,\text{M}_\odot$, as well as the $40\,\text{M}_\odot$ $Z=\text{Z}_\odot$ model develop before the start of central He-burning a convective H-shell where the H-shell burning occurs. It shrinks and retreats when the convective He-core grows. These models follow therefore the sequence: (a)-(b)-(c), but with a basic difference between the models at $Z=10^{-5}$ and those at higher metallicity. While the latter develop case (b) with a very low $D_{\rm shear}$ as soon as the convective shell starts to shrink, the former show strong angular momentum transport at the steep $\Omega$-gradient, which is fast enough to follow the retreating convective zone and therefore develops rather a hybrid case between (a) and (b) when the convective shell shrinks. The mixing is thus strongest in 
$Z=10^{-5}$ models, followed by sub-solar $Z$ models with 20, 25 and $40\,\text{M}_\odot$ and the $40\,\text{M}_\odot$ $Z=\text{Z}_\odot$ model, and finally followed by the $\text{Z}_\odot$ 15, 20 and $25\,\text{M}_\odot$ models and the sub-solar $15\,\text{M}_\odot$ models.
To ensure that the mixing does not depend strongly on our choice of resolution parameters, a $25\,\text{M}_\odot$ $Z=10^{-3}$ rotating model was performed with a much higher resolution, i.e. doubled resolution in the He-core\footnote{The critical value of the luminosity gradient $\Delta L_{\rm crit}$, used to split a mass shell when $\Delta L>\Delta L_{\rm crit}$, was reduced by a factor two.} and 5-times the resolution in the radiative layers between the convective core and H-burning shell\footnote{The critical values of the mass fraction gradients of carbon and helium, $\Delta X_{\rm crit}(^{12}{\rm C})$ and $\Delta X_{\rm crit}(^{4}{\rm He})$, used to split a mass shell when $\Delta X(^{12}{\rm C})>\Delta X_{\rm crit}(^{12}{\rm C})$ or $\Delta X(^{4}{\rm He})>\Delta X_{\rm crit}(^{4}{\rm He})$, were reduced by a factor five. }. The model with higher resolution had a smoother growth of the convective core but it did not affect the  $^{14}$N and $^{22}$Ne production strongly. For example the mass factions $\Delta X(^{22}{\rm Ne})$ of burned $^{22}$Ne during central He-burning decreased only by 2.5\% in the high resolution model compared to the standard resolution. The mass fraction $X_{\rm shell}(^{22}{\rm Ne})$ of $^{22}$Ne in the He-shell at the pre-SN stage differed by 22\% (lower): $X(^{22}{\rm Ne})=0.0246$ and $0.0314$, for the high and the default resolution model, respectively. The slight decrease of transport efficiency when using a higher resolution therefore does not change the $s$~process and only moderately lower the yields of $^{14}$N and $^{22}$Ne. These differences due to resolution are very small compared to the differences between non-rotating and rotating models (see Table \ref{tab:n14ne22}).     

Since $^{22}$Ne is produced and destroyed at the same time in rotating stars, we derived the amount of $^{22}$Ne burned during central He-burning from the sum of the $^{25}$Mg and $^{26}$Mg produced during this stage.

In Table~\ref{tab:n14ne22} the mass factions $\Delta X(^{22}{\rm Ne})$ of burned $^{22}$Ne during central He-burning, $X_{\rm r}(^{22}{\rm Ne})$ of remaining $^{22}$Ne after He-burning, $X_{\rm shell}(^{22}{\rm Ne})$ of $^{22}$Ne in the He-shell at the pre-SN stage, and the yields of $^{22}$Ne and $^{14}$N are tabulated for all models.  
$\Delta X(^{22}{\rm Ne})$ is the $^{22}$Ne destroyed mainly by the (n,$\gamma$) and $\alpha$-capture channels, where the ($\alpha$,n) channel is the $s$-process neutron source in He-burning. 
$X_{\rm r}(^{22}{\rm Ne})$ is the $^{22}$Ne left in the He-core ashes, and it will be destroyed mostly by the (p,$\gamma$) and ($\alpha$,n) channels during C-burning \citep[e.g.,][]{2010ApJ...710.1557P}.
\begin{table*}%[ht!]
  \caption{$^{14}$N and $^{22}$Ne production and destruction. See text for explanations.}
\label{tab:n14ne22}
  \begin{tabular}{lccccc}
     \hline\hline
     Model &  $\Delta X(^{22}{\rm Ne})^{\rm a}$ & $X_{\rm r}(^{22}{\rm Ne})^{\rm a}$ & $X_{\rm shell}(^{22}{\rm Ne})^{\rm a}$ & $m(^{22}$Ne$)^{\rm a}$ &  $m(^{14}$N$)^{\rm a}$ \\
                    &          &          &          & $[$$M_\odot]$ & $[$$M_\odot]$ \\
     \hline     
     A15s0          & 3.06(-3) & 9.70(-3) & 9.23(-3) & 9.11(-3) & 3.19(-2) \\
     A15s4          & 5.59(-3) & 7.42(-3) & 1.38(-2) & 2.78(-2) & 2.63(-2) \\
     B15s0          & 3.54(-4) & 8.02(-4) & 9.24(-4) & 1.28(-3) & 2.91(-3) \\
     B15s4          & 9.37(-4) & 1.02(-3) & 7.34(-3) & 1.49(-2) & 7.17(-3) \\
     C15s0          & 3.75(-6) & 7.70(-6) & 1.02(-5) & 6.42(-5) & 4.77(-5) \\
     C15s4          & 4.84(-4) & 3.92(-4) & 7.55(-3) & 1.39(-2) & 5.25(-3) \\
     \hline
     A20s0          & 5.34(-3) & 7.43(-3) & 1.14(-2) & 2.50(-2) & 3.76(-2) \\
     A20s4          & 7.23(-3) & 5.03(-3) & 1.99(-2) & 4.99(-2) & 3.72(-2) \\
     B20s0          & 6.16(-4) & 5.46(-4) & 1.15(-3) & 2.68(-3) & 4.06(-3) \\
     B20s4          & 3.49(-3) & 1.14(-3) & 3.20(-2) & 7.59(-2) & 9.39(-3) \\
     C20s0          & 5.66(-6) & 5.74(-6) & 1.32(-5) & 1.21(-4) & 5.80(-5) \\
     C20s4          & 1.52(-3) & 4.62(-4) & 1.67(-2) & 4.09(-2) & 4.04(-3) \\
     \hline
     A25s0          & 7.68(-3) & 5.10(-3) & 1.27(-2) & 3.39(-2) & 4.76(-2) \\
     A25s4          & 9.69(-3) & 3.28(-3) & 1.56(-2) & 4.06(-2) & 4.95(-2) \\
     B25s0          & 7.52(-4) & 4.16(-4) & 1.15(-3) & 3.36(-3) & 5.90(-3) \\
     B25s4          & 4.08(-3) & 6.22(-4) & 1.99(-2) & 6.72(-2) & 8.47(-3) \\
     C25s0          & 7.21(-6) & 4.14(-6) & 1.13(-5) & 2.38(-4) & 9.38(-5) \\
     C25s4          & 1.23(-3) & 1.69(-4) & 1.15(-2) & 3.61(-2) & 1.85(-3) \\
    C25s4b$^{\rm b}$& 1.27(-3) & 1.82(-4) & 1.17(-2) & 3.49(-2) & 9.33(-4) \\
     C25s5          & 3.83(-3) & 4.94(-4) & 1.59(-2) & 4.80(-2) & 2.07(-3) \\
    C25s5b$^{\rm b}$& 3.75(-3) & 4.85(-4) & 1.61(-2) & 4.81(-2) & 1.99(-3) \\
     D25s0          & 8.28(-7) & 4.67(-7) & 3.09(-7) & 1.63(-4) & 1.80(-5) \\
     D25s4          & 1.05(-4) & 3.81(-5) & 1.46(-2) & 3.81(-2) & 1.10(-2) \\
    D25s4b$^{\rm b}$& 1.06(-4) & 3.96(-5) & 1.45(-2) & 3.71(-2) & 1.11(-2) \\
     D25s6          & 4.57(-3) & 2.68(-4) & 1.95(-2) & 5.52(-2) & 3.43(-3) \\
    D25s6b$^{\rm b}$& 4.44(-3) & 3.11(-4) & 2.00(-2) & 5.56(-2) & 3.48(-3) \\
     \hline
     A40s4          & 1.23(-2) & 5.29(-4) & 1.21(-2) & 3.34(-2) & 2.23(-2) \\
     B40s4          & 3.31(-3) & 1.06(-4) & 2.08(-2) & 7.99(-2) & 1.84(-2) \\
     C40s4          & 2.70(-3) & 1.93(-5) & 8.75(-3) & 3.21(-2) & 2.07(-3) \\
     \hline
  \end{tabular}
  \begin{flushleft}
     Note.\\
     $^{\rm a}$ Values in brackets are the exponents ($x(y) = x\times10^y$).  \\
     $^{\rm b}$ This model was calculated with the same initial parameters as the model, on the line above, but with the $^{17}$O$(\alpha,\gamma)$ reaction rate of CF88 divided by 10. 
\end{flushleft}
\end{table*}

We can see from Table~\ref{tab:n14ne22}  ($\Delta X(^{22}{\rm Ne, burned})$) that rotating models at all metallicities produce and burn significant amounts of $^{22}$Ne, confirming the results of previous studies \citep{2006A&A...447..623M,2007A&A...461..571H}. At solar metallicity, $^{22}$Ne is predominantly secondary.
At low metallicities, in the models including rotation, mixing is strong enough to produce a pocket of primary $^{14}$N above the convective core, which is then converted to primary $^{22}$Ne. 
The amount of primary $^{22}$Ne in the convective He-core at the end of He-burning, when $s$~process is activated, is between 0.1 and 1\% in mass fractions. Considering a constant value of $\upsilon_{\rm ini}/\upsilon_{\rm crit}=0.4$ at all metallicities, the primary $^{22}$Ne in the He-core decreases slightly with decreasing metallicity.
There is, however, theoretical and observational support to consider a slight increase of $\upsilon_{\rm ini}/\upsilon_{\rm crit}$ with decreasing metallicity as discussed in the previous section. We thus also computed models with $25\,\text{M}_\odot$ and $\upsilon_{\rm ini}/\upsilon_{\rm crit}=0.4$ at $Z =\text{Z}_\odot$ and  $10^{-3}$, $\upsilon_{\rm ini}/\upsilon_{\rm crit}=0.5$ at $Z = 10^{-5}$ and $\upsilon_{\rm ini}/\upsilon_{\rm crit}=0.6$ at $Z = 10^{-7}$, which correspond to a slight increase of $\upsilon_{\rm ini}/\upsilon_{\rm crit}$ with decreasing metallicity. Considering a slightly increasing initial rotation rate with decreasing metallicity, rotating models produce and burn a constant quantity of $^{22}$Ne, around 0.5\% in mass fraction, almost independent of the initial metallicity. 
These results show that significant amounts of $^{22}$Ne are expected to be produced in massive rotating stars over the entire range of masses and all metallicities.

\subsection{Helium shell burning}
The convective He-shell, which follows on the $^{14}$N-rich zone, transforms most of this $^{14}$N into $^{22}$Ne. While the $^{22}$Ne in the He-shell of non-rotating model is purely secondary, in rotating models it is primary at the pre-SN stage and almost independent of metallicity. 
The $^{22}$Ne is only partially destroyed during the He-shell burning and there is a mass fraction of $X(^{22}{\rm Ne})$ between $0.7$ and $3.2$\% in the He layers at the pre-SN stage. 
This is relevant for explosive neutron capture nucleosynthesis in He-shell layers.
This site was investigated by \cite{blake:76}, \citet[][]{1978ApJ...222L..63T} and \citet{1979A&A....74..175T} as a possible $r$~process scenario, but later on found to be unlikely 
\citep{blake:81}. Instead, the explosive shell He-burning in core-collapse supernovae is hosting the $n$~process \citep[e.g.,][]{blake:76}, with typical abundance signatures identified in presolar silicon-carbide grains of type X \citep[e.g.,][]{meyer:00,zinner:14}. It will be worthwhile to explore in the future the impact of these large amounts of primary $^{22}$Ne produced in rotating models at all $Z$, for explosive neutron capture nucleosynthesis.

\subsection{Carbon shell burning}
Carbon shell burning is the second efficient $s$-process production site inside massive stars at solar metallicity \citep[e.g.,][]{1991ApJ...371..665R, 2007ApJ...655.1058T, rauscher:02, 2010ApJ...710.1557P}.
One could think of rotation induced mixing appearing in the same way as in He burning, mixing down some of the primary $^{22}$Ne into the C shell and boosting the $s$-process. However, the time scale of the secular shear mixing, which is still present between convective He and C shells, is of the same order as during central He burning. On the other hand the burning time scale of Ne, O and Si burning are at least 5 to 6 orders of magnitude smaller than the one of He burning. 
This implies that the $^{22}$Ne available to make neutrons via the the $^{22}$Ne($\alpha$,n) reaction in the convective C-burning shell is what is left in the ashes of the previous convective He core, like in non-rotating models.

Rotation, however, affects the CO core sizes and the $^{12}$C/$^{16}$O ratio after He-burning \citep[e.g.,][]{2004A&A...425..649H}. This will indirectly affect all subsequent burning phases and their heavy element production.

\section{Standard weak $s$~process in stars}
\label{sec:standards}

In Table~\ref{tab:he_core_sproc} several characteristic quantities for $s$~process in He burning are presented. 
Note that some of these quantities 
are averaged quantities over the convective core and integrated over the helium-burning phase, encompassing in one number complex processes varying both in space and in time. These quantities are useful in the sense that they allow through a unique number to see the importance of different phases, and also to compare the outputs of different models.

In a one-zone model, a useful quantity is the neutron exposure defined as: 
\begin{equation}
 \tau = \int_{
 t_{\rm ini-He}
 }
 ^{
 t_{\rm end-He}
 } \upsilon_T n_n \text{d}t
 \label{eq:neutronexposure} 
\end{equation}
where $ t_{\rm ini-He}$ and $ t_{\rm end-He}$ are the age of the star at the beginning and the end of the core He-burning phase,  respectively, $n_n$ the neutron density
and $\upsilon_T$ the thermal velocity, $v_T=\sqrt{2kT/m_n}$ with $kT=30$~keV. The value of 30 keV is typical of the conditions at the end of the core He-burning phase.

In multiple-zone simulations, as in stellar models, the neutron number density, $n_n$, varies with time and the mass coordinate in the star.
 For the investigation of $s$~process in convective zones one can define a mean or effective neutron exposure 
\begin{equation}
 \langle \tau \rangle = \int \langle n_n\left(t \right)\rangle v_T \text{d}t.
 \label{eq:neutronexposure_avg}
\end{equation}
In Eq.~\ref{eq:neutronexposure_avg}, $\langle n_n\left(t \right)\rangle $ is an average over the convective core. 
Such a global quantity has to be interpreted with caution since in reality 
the neutrons are captured locally during core He~burning, near the centre of the star and later the $s$-process products are mixed outwards.

Another characteristic $s$-process quantity is the average number of neutron captures per iron ($Z=26$) seed \citep[e.g.][]{1990ApJ...354..630K} 
\begin{equation}
 n_c = \frac{\sum \limits_{A=56}^{209} \left(A-56\right) \left(Y(A)-Y_0(A)\right)}{\sum \limits_{Z=26} Y_0(A)}, \label{eq:neutron_captures_per_seed}
\end{equation}
where $Y(A)$ and $Y_0(A)$ are the final and the initial number abundance respectively of a nucleus with nuclear mass number $A$. 
Additionally, the core averaged (${\bar{n}_{n,{\rm max}}}$) and central (${n_{n,c,{\rm max}}}$) peak neutron density, the amount of $^{22}$Ne burnt during He burning ($\Delta X(^{22}{\rm Ne})$) and the amount of $^{22}$Ne left in the centre at core He exhaustion ($X_{\rm r}(^{22}{\rm Ne})$) are tabulated.

\begin{table*}
  \caption{$s$-process parameters at central He exhaustion}
  \label{tab:he_core_sproc}
  \begin{tabular}{lD{.}{.}{4}D{.}{.}{4}D{.}{.}{2}cccc}
    \hline\hline
    Model$^{\rm a}$ & \multicolumn{1}{c}{${\tau_c}^{\rm b}$} & \multicolumn{1}{c}{${\langle\tau\rangle}^{\rm c}$} & \multicolumn{1}{c}{${n_c}^{\rm d}$} & ${\bar{n}_{n,{\rm max}}}^{\rm e}$ & ${n_{n,c,{\rm max}}}^{\rm f,g}$ & $\Delta X(^{22}{\rm Ne})^{\rm g}$ & $X_{\rm r}(^{22}{\rm Ne})^{\rm g}$ \\ % & ${\log T_c}^d$ & ${\log\rho_c}^d$\\
      &  \multicolumn{1}{c}{$[{\rm mb}^{-1}]$}  & \multicolumn{1}{c}{$[{\rm 10^{-1} mb}^{-1}]$} &   &  $[{\rm ~cm}^{-3}]$ & $[{\rm ~cm}^{-3}]$ &  &  \\
    \hline
     A15s0          & 1.52   & 0.581  &  0.77 & 3.04(5) & 6.58(6) & 3.06(-3) & 9.70(-3) \\
     A15s4          & 2.93   & 1.02   &  1.60 & 4.65(5) & 1.17(7) & 5.59(-3) & 7.42(-3) \\
     B15s0          & 0.883  & 0.427  &  0.53 & 2.32(5) & 4.32(6) & 3.54(-4) & 8.02(-4) \\
     B15s4          & 3.06   & 1.51   &  2.55 & 5.18(5) & 1.07(7) & 9.37(-4) & 1.02(-3) \\
     C15s0          & 0.0157 & 0.0561 &  0.04 & 2.85(3) & 5.59(4) & 3.75(-6) & 7.70(-6) \\
     C15s4          & 2.21   & 1.07   &  2.18 & 3.38(5) & 7.33(6) & 4.84(-4) & 3.92(-4) \\
     \hline
     A20s0          & 2.97   & 0.971  &  1.52 & 5.17(5) & 1.22(7) & 5.34(-3) & 7.43(-3) \\
     A20s4          & 4.66   & 1.43   &  2.57 & 5.89(5) & 1.54(7) & 7.23(-3) & 5.03(-3) \\
     B20s0          & 1.88   & 0.761  &  1.13 & 4.11(5) & 9.10(6) & 6.16(-4) & 5.46(-4) \\
     B20s4          & 9.73   & 4.07   &  9.85 & 8.73(5) & 2.22(7) & 3.49(-3) & 1.14(-3) \\
     C20s0          & 0.0286 & 0.0401 &  0.05 & 6.00(3) & 1.31(5) & 5.66(-6) & 5.74(-6) \\
     C20s4          & 6.55   & 2.80   &  5.87 & 6.84(5) & 1.69(7) & 1.52(-3) & 4.62(-4) \\
     \hline
     A25s0          & 4.42   & 1.33   &  2.42 & 5.85(5) & 1.56(7) & 7.68(-3) & 5.10(-3) \\
     A25s4          & 5.63   & 1.60   &  3.13 & 5.98(5) & 1.72(7) & 9.69(-3) & 3.28(-3) \\
     B25s0          & 2.65   & 0.970  &  1.64 & 4.99(5) & 1.20(7) & 7.52(-4) & 4.16(-4) \\
     B25s4          & 12.1   & 4.80   & 12.7  & 8.03(5) & 2.31(7) & 4.08(-3) & 6.22(-4) \\
     C25s0          & 0.0466 & 0.0829 &  0.08 & 9.36(3) & 2.13(5) & 7.21(-6) & 4.14(-6) \\
     C25s4          & 6.73   & 2.94   &  5.77 & 5.77(5) & 1.53(7) & 1.23(-3) & 1.69(-4) \\
    C25s4b$^{\rm h}$& 16.4   & 7.15   & 23.1  & 8.02(5) & 2.10(7) & 1.27(-3) & 1.82(-4) \\
     C25s5          & 13.5   & 5.73   & 16.5  & 8.27(5) & 2.26(7) & 3.83(-3) & 4.94(-4) \\
    C25s5b$^{\rm h}$& 20.3   & 8.67   & 31.8  & 1.01(6) & 2.74(7) & 3.75(-3) & 4.85(-4) \\
     D25s0          & 0.166  & 0.0866 &  6.31 & 9.61(2) & 2.24(4) & 8.28(-7) & 4.67(-7) \\
     D25s4          & 0.804  & 0.354  & 14.0  & 1.39(5) & 3.38(6) & 1.05(-4) & 3.81(-5) \\
    D25s4b$^{\rm h}$& 2.29   & 1.048  & 16.5  & 3.85(5) & 8.60(6) & 1.06(-4) & 3.96(-5) \\
     D25s6          & 19.2   & 7.78   & 33.5  & 6.77(5) & 2.03(7) & 4.57(-3) & 2.68(-4) \\
    D25s6b$^{\rm h}$& 24.6   & 10.0   & 48.5  & 9.77(5) & 2.76(7) & 4.44(-3) & 3.11(-4) \\
     \hline
     A40s4          &  7.76  & 2.00   & 4.05  & 3.77(5) & 1.42(7) & 1.23(-2) & 5.29(-4) \\
     B40s4          & 12.1   & 4.12   & 10.6  & 6.38(5) & 2.13(7) & 3.31(-3) & 1.06(-4) \\
     C40s4          & 11.6   & 4.67   & 10.4  & 6.12(5) & 1.97(7) & 2.70(-3) & 1.93(-5) \\
     \hline
  \end{tabular}
  \begin{flushleft}
  Notes. \\
  $^{\rm a}$ The A-series models have metallicty of $Z=\text{Z}_\odot$, B-series $Z=10^{-1}$, C-series $Z=10^{-5}$, and D-series $Z=10^{-7}$.\\
  $^{\rm b}$ Central neutron exposure calculated according to Eq.~\ref{eq:neutronexposure}.\\
  $^{\rm c}$ Neutron exposure averaged over He~core (see Eq.~\ref{eq:neutronexposure_avg}). \\
  $^{\rm d}$ Number of neutron capture per seed calculated according to Eq.~\ref{eq:neutron_captures_per_seed}, averaged over the He-core mass.\\
  $^{\rm e}$ Maximum of the mean neutron density.\\
  $^{\rm f}$ Maximum of the central neutron density.\\
  $^{\rm g}$ Values in brackets are the exponents ($x(y) = x\times10^y$).\\
  $^{\rm h}$ This model was calculated with the same initial parameters as the model, on the line above, but with $^{17}$O$(\alpha,\gamma)$ reaction rate of CF88 divided by $10$. 
\end{flushleft}
\end{table*}

\subsection{He-core burning}
Let us begin by discussing the solar metallicity models.
Due to $^{14}$N transformation at the beginning of the core He-burning phase, all models had initially in the He-core\footnote{{\it i.e } before $^{22}$Ne is destroyed by the two reactions $^{22}$Ne$(\alpha,n)$ and $^{22}$Ne$(\alpha,\gamma)$} about $X(^{22}{\rm Ne})=1.3\times10^{-2}$. 
The abundance of $^{22}$Ne will not change during a large fraction of the core He-burning phase.
Only close to the end of central He burning, part of the  $^{22}$Ne will be transformed into $^{25}$Mg and $^{26}$Mg.
When the temperatures for an efficient activation of $^{22}$Ne$(\alpha,n)^{25}$Mg are reached, some $^{22}$Ne has already been destroyed by
the ($\alpha,\gamma)^{26}$Mg reaction. More quantitatively, when $T_8\approx 2.8$ is reached (temperature, at which the $(\alpha,n)$-channel starts to dominate), only $X(^{22}{\rm Ne})=10^{-2}$, $6.8\times 10^{-3}$, $5.7\times10^{-3}$, and $5.0\times10^{-3}$ is left in models A15s0, A25s0, A25s4, and A40s4, respectively. 

Important well-known aspects of the $s$ process during core He-burning are the following: 
\begin{itemize}
  \item Because only a small helium mass fraction, $X(^4{\rm He})$, is left when $^{22}$Ne$+\alpha$ is activated (less than ten percent in mass fraction), the competition with other $\alpha$-captures as the $^{12}$C$(\alpha,\gamma)$ and $3\alpha$ is essential at the end of He burning and will affect the $s$-process efficiency in core He burning. 

\item The low amount of $X(^4{\rm He})$, when the neutron source is activated, means also that not all of $^{22}$Ne is burned and a part of it will be left for subsequent C-burning phase. This depends on the stellar core size. The more massive the core, the more $^{22}$Ne is burned and the more efficient is the $s$~process in core He burning, as can be seen from the increasing number of neutron captures per seed $n_{c}$ from $0.77$, $2.42$, $3.13$ and $4.05$ for the four models mentioned before, which have $M_{\rm CO}$ of $2.35$, $5.53$, $6.97$, and $15.04~M_\odot$, respectively (see Table~\ref{tab:cores}). This is a well-known behaviour already found in previous works \citep{1990A&A...234..211P,1992A&A...258..357B,BT1993,2000A&A...354..740R,2000ApJ...533..998T,2007ApJ...655.1058T,Pumo2010}.

  \item During the late He~burning stages the bulk of the core matter consists of $^{12}$C and $^{16}$O, which are both strong neutron absorbers. 
They capture neutrons to produce $^{13}$C and $^{17}$O, respectively. 
$^{13}$C will immediately recycle neutrons via $^{13}$C$(\alpha,n)$ in He-burning conditions. Instead, we have seen that the relevance of $^{16}$O as a neutron poison depends on the $^{17}$O$(\alpha,\gamma)$ and $^{17}$O$(\alpha,n)$ rates. In particular,
the strength of  primary neutron poisons like $^{16}$O, increases towards lower metallicities, because of the decreasing ratio of seeds to neutron poisons. 
\end{itemize}

\begin{figure*}
  \includegraphics[width=\textwidth]{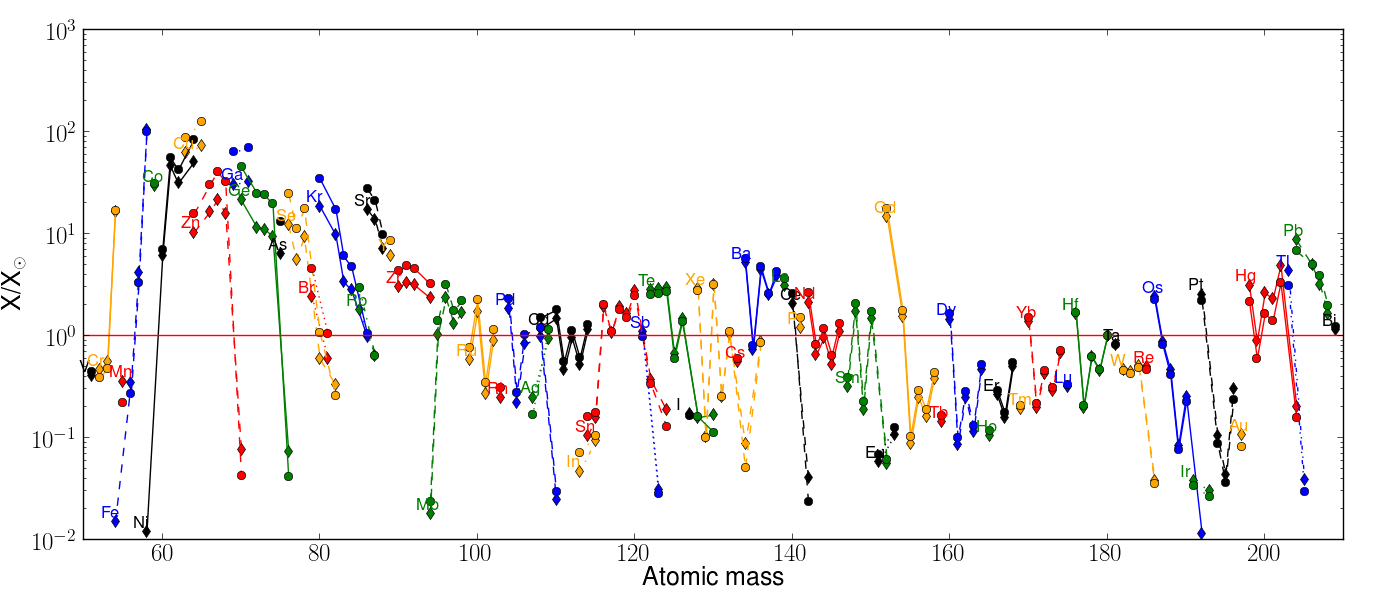}
  \caption[Overproduction factors of $25\,\text{M}_\odot$ models with $Z=\text{Z}_\odot$ after He~burning]{Isotopic overproduction factors (abundances over initial abundances) of $25\,\text{M}_\odot$ models with solar metallicity after He exhaustion. The rotating model (A25s4, circles) has slightly higher factors than the non-rotating model (A25s0, diamonds).}
  \label{fig:overprod_25z14}
\end{figure*}

\begin{figure*}
 \includegraphics[width=\textwidth]{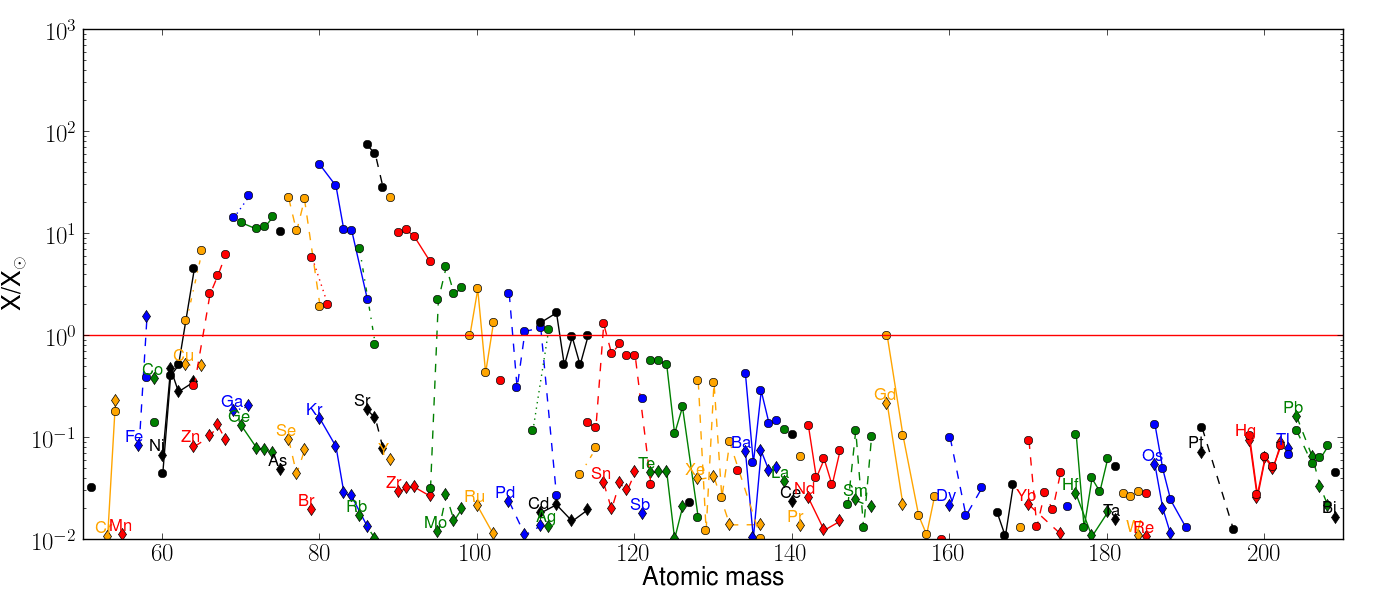}
  \caption[$X/X_\odot$ of $25\,\text{M}_\odot$ models with $Z=10^{-3}$ after He~burning]{Isotopic abundances normalized to solar abundances of $25\,\text{M}_\odot$ models with with $Z=10^{-3}$ after He exhaustion. The rotating model (B25s4, circles) has higher factors than the non-rotating model (B25s0, diamonds).}
  \label{fig:overprod_25z01}
\end{figure*}

\begin{figure*}%[ht!]
 \includegraphics[width=\textwidth]{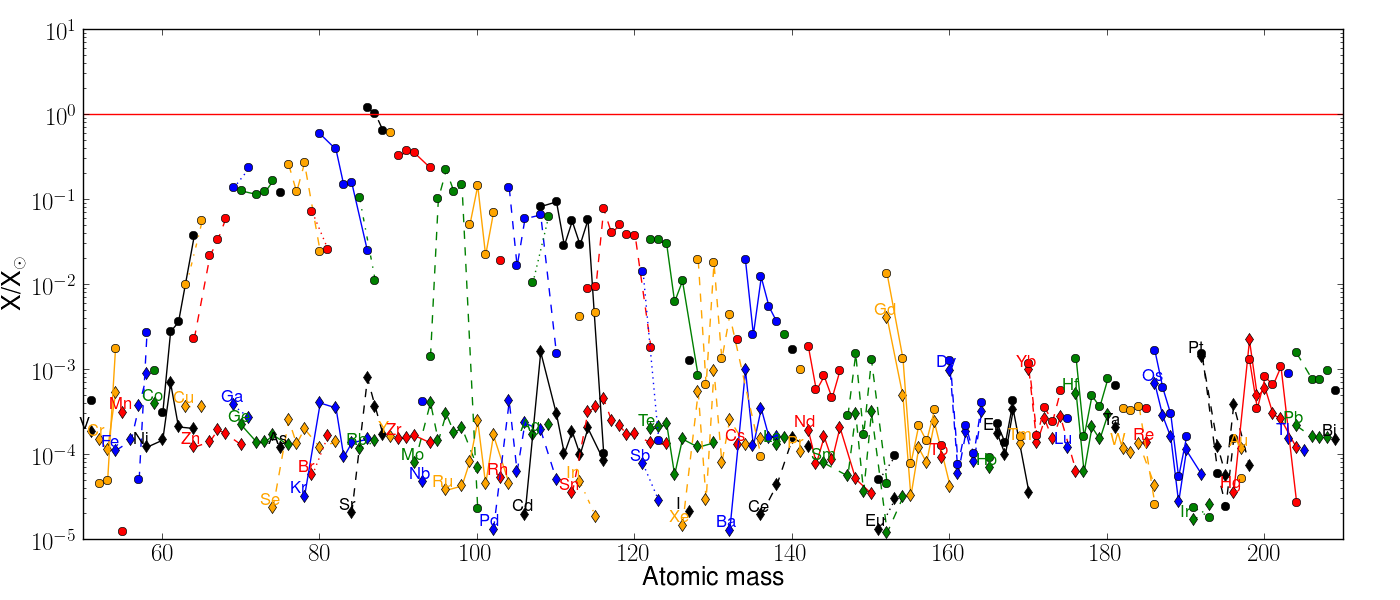}
  \caption[$X/X_\odot$ of $25\,\text{M}_\odot$ models with $Z=10^{-5}$ after He~burning]{Isotopic abundances normalised to solar abundances of $25\,\text{M}_\odot$ models with with $Z=10^{-5}$ after He exhaustion. The rotating model (C25s5, circles) has much higher factors than the non-rotating model (C25s0, diamonds).}
  \label{fig:overprod_25zm5}
\end{figure*}

\begin{figure*}%[ht!]
 \includegraphics[width=\textwidth]{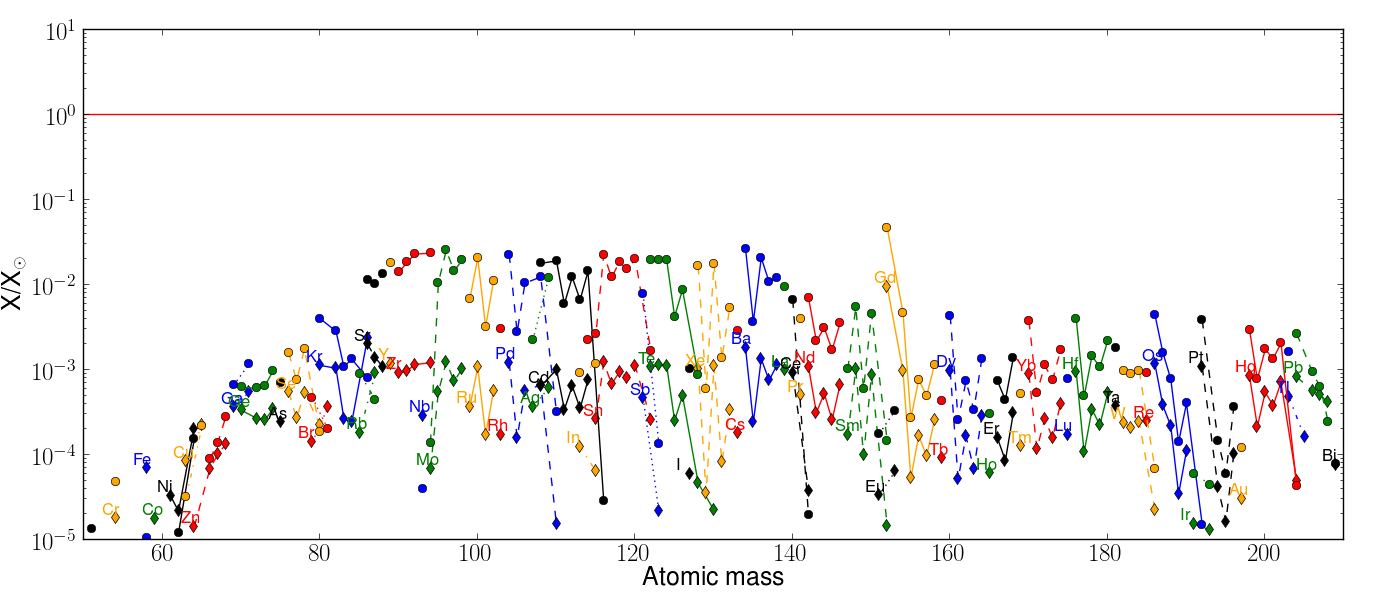}
  \caption[$X/X_\odot$ of $25\,\text{M}_\odot$ models with $Z=10^{-7}$ after He~burning]{Isotopic abundances normalised to solar abundances of $25\,\text{M}_\odot$ models with with $Z=10^{-7}$ after He exhaustion. The rotating model (D25s6, circles) has slightly higher factors than the non-rotating model (D25s0, diamonds).}
  \label{fig:overprod_25zm7}
\end{figure*}

The $s$-process production in the non-rotating models is shown in Figs.~\ref{fig:overprod_25z14} ($Z=\text{Z}_\odot$), \ref{fig:overprod_25z01} ($Z=10^{-3}$), \ref{fig:overprod_25zm5} ($Z=10^{-5}$) and \ref{fig:overprod_25zm7} ($Z=10^{-7}$). In combination with the values given in Table~\ref{tab:he_core_sproc}, we can see that the models confirm the trends expected for 
the $s$~process in non-rotating massive stars, which we will call the {\it standard} $s$~process in the rest of this paper.
The production of nuclei between $A=60$ and $90$ decreases with decreasing metallicity and mass. The decreasing production with decreasing metallicity is due to the secondary nature of both the neutron source ($^{22}$Ne$(\alpha,n)^{25}$Mg) and the seeds (mainly iron) \citep[see e.g.][]{1990A&A...234..211P,1992ApJ...387..263R,pignatari:08}. During helium burning, the neutron poisons are a mixture of secondary (mainly $^{20}$Ne, $^{22}$Ne and $^{25}$Mg) and primary (mainly $^{16}$O) elements. The $s$-process production thus becomes negligible below $Z/\text{Z}_\odot = 10^{-2}$ \citep[][]{1990A&A...234..211P}, which we confirm with our non-rotating models at $Z=10^{-5}$ and $Z=10^{-7}$ (C and D series). The decreasing production with decreasing mass is due to the fact that lower mass stars reach lower temperature at the end of He~burning. Thus less $^{22}$Ne is burnt during He~burning (see Table~\ref{tab:he_core_sproc}).


The only model, which does not follow this trend is the very low metallicity model D25s0. It shows a higher $s$-process efficiency than C25s0. This model has a smooth transition between central H and He burning. When small fractions of hydrogen are still present in the core, temperatures of $T_8=1.4$ are reached and the $3\alpha$-reaction is already activated. It leads to the immediate transformation of the $^{12}$C produced into $^{14}$N by $^{12}$C$(p,\gamma)^{13}$N$(\beta^+)^{13}$C$(p,\gamma)^{14}$N \citep[][]{1992A&A...258..357B} 
and therefore also the consumption of the remaining protons. In this way $X(^{22}{\rm Ne})=1.2\times10^{-6}$ of primary $^{22}$Ne is produced. 
Still as for non rotating $Z=10^{-5}$ models, D25s0 produces negligible amounts of heavy elements. 
This models shows a behaviour a bit similar as pop III (metal-free) stars, which cannot produce enough energy by the pp-chains and therefore go into a state of combined  hydrogen and weak He burning, producing non-negligible amounts of primary $^{14}$N as in previous studies \citep{2008A&A...489..685E,2010ApJ...724..341H}.

\subsection{He-shell burning}
Shell He~burning, similarly to the other burning shells, appear at higher temperatures and lower densities than the equivalent central burning phase. In our models high temperature conditions of $T_8\approx3.5$-$4.5$ and $\rho\approx3$-$5.5\times10^{3}\,\text{g}\,\text{cm}^{-3}$ cause an efficient $^{22}$Ne($\alpha$,n) activation for the $s$~process in shell He burning. However, the highest neutron densities are reached in all our models only in the layers below the convective shell helium burning. Therefore only a narrow mass range, extending  over about $0.2\,\text{M}_\odot$ in non-rotating models, at the bottom of the He shell is strongly affected by neutron capture nucleosynthesis.
The contribution of the s-process in the He-shell amounts to at most $\sim 5\%$ of the total $s$-process yields
for the solar metallicity $25\,\text{M}_\odot$ model. For less massive stars the He~shell gains more weight and produces in $15\,\text{M}_\odot$ models with rotation up to $50\%$ of the total $s$-process rich SN ejecta.
Thus, according to our models for the $15$ to $20\,\text{M}_\odot$ stars the He-shell $s$-process contribution has to be considered \citep[see also][]{2009ApJ...702.1068T}. 

\begin{figure*}
  \includegraphics[width=\textwidth]{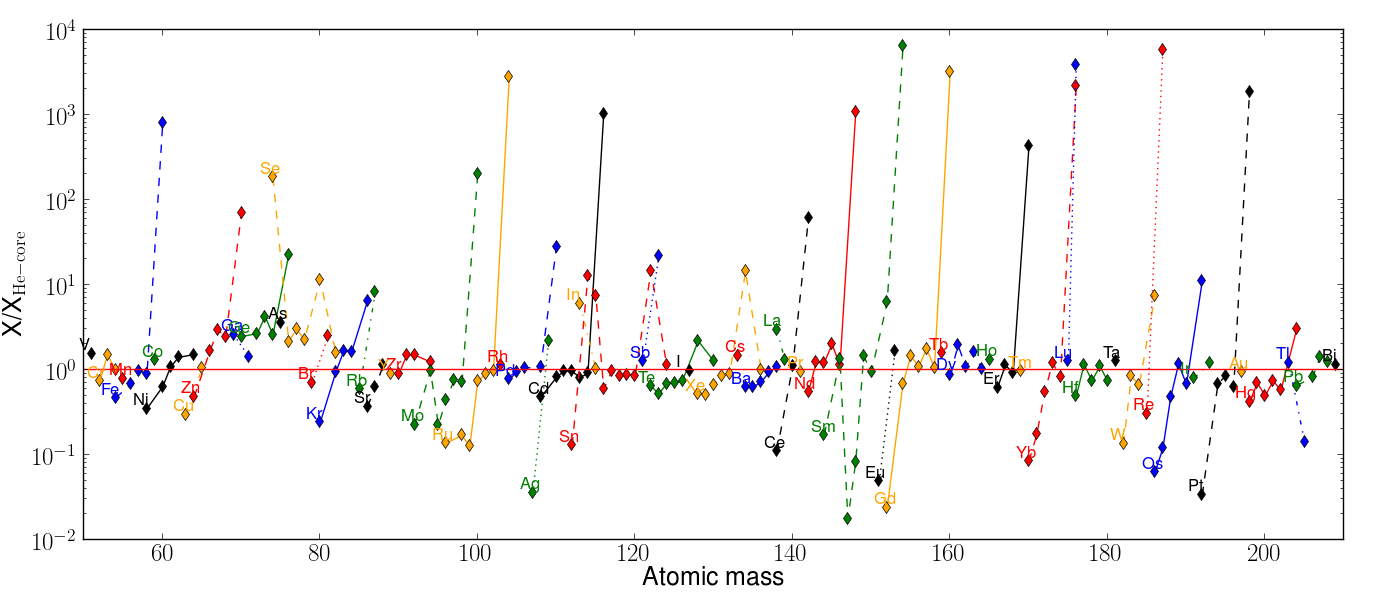}
  \caption[$X_{\rm C}/X_{\rm He}$ in a non-rotating $25~M_\odot$ star at $Z=Z_\odot$]{Ratio of abundances after shell C~burning to the abundances after core He~burning, $X_{\rm C}/X_{\rm He}$, in a non-rotating $25\,\text{M}_\odot$ star at $Z=\text{Z}_\odot$ (A25S0). It illustrates the modification of the abundances by $s$~process in shell C~burning.}
  \label{fig:overprod_25z14S0_cshell}
\end{figure*}

\subsection{C-shell burning}
Shell C-burning occurs in the CO~core (see Table~\ref{tab:cores}) after central C~burning. 
Temperatures and densities at the start of C-shell burning show the same trend with stellar mass as the core burning conditions, i.~e. the temperature increases and the density decreases with stellar mass. 
They vary between $T_9\approx0.8$, $\rho\approx 2\times10^{5}\,\text{g}\,\text{cm}^{-3}$ in $15\,\text{M}_\odot$ models and $T_9\approx1.3$, $\rho\approx8\times10^{4}\,\text{g}\,\text{cm}^{-3}$ in $40\,\text{M}_\odot$ models. These temperatures are higher than in the central C~burning, where $T_9=0.6-0.8$.  

The efficiency of the $s$~process mainly depends on the remaining iron seeds and $^{22}$Ne left after He~burning, $X_{\rm r}(^{22}{\rm Ne})$, in the CO~core. All the remaining $^{22}$Ne is burned quickly with maximal neutron densities between $6\times10^9$ and $10^{12}\,\text{cm}^{-3}$, for the two extremes in models B15s4 and A40s4, respectively. 
The time scale of this $s$~process is in our models of the order of a few tens of years in $15\,\text{M}_\odot$ stars to a few tenth of a year in $40\,\text{M}_\odot$. 

A striking difference between the s-process in the He-shell and in the C-shell is the neutron density, which is much higher in the C-shell than in the He-shell. 
The activation of $^{22}$Ne$(\alpha,n)$ at the start of C-shell burning leads to a short neutron burst with relatively high neutron densities \citep[typically $n_n \sim 10^{10}-10^{12}\,{\rm cm}^{-3}$, see][]{2000ApJ...533..998T,2007ApJ...655.1058T}, compared to He~burning ($n_n \sim 10^{5}-10^7\,{\rm cm}^{-3}$, see Table 4 and references above). 

This leads to a different s-process nucleosynthesis than during the He-shell burning.
The ratio of abundances after shell C~burning to the abundances after core He~burning, $X_{\rm C}/X_{\rm He}$ is plotted for the non-rotating $25\,\text{M}_\odot$ model at $Z=\text{Z}_\odot$ in Fig.~\ref{fig:overprod_25z14S0_cshell}. We can see an overproduction of most isotopes from Zn to Rb. The overproduction during C-burning shell is also found in models of other initial mass, which have both, 
\begin{enumerate}
  \item $X_{\rm r}(^{22}{\rm Ne})\apprge 10^{-3}$ and
  \item $X(^{56}{\rm Fe})\apprge10^{-4}$, at the start of shell C~burning
\end{enumerate}
Therefore, in these calculations only $15$ to $25\,\text{M}_\odot$ stars at solar $Z$  have a 
strong C-shell contribution in term of neutron exposure.

In the mass range $A=60$ to $90$, there are several branching points at $^{63}$Ni, $^{79}$Se, and $^{85}$Kr, respectively. The high neutron densities modify the $s$-process branching ratios, in a way that the neutron capture on the branching nuclei are favoured over the $\beta-$decay channel \citep[see e.g.][and references therein]{2010ApJ...710.1557P}. As a consequence of this, isotopic ratios like $^{63}$Cu/$^{65}$Cu,
$^{64}$Zn/$^{66}$Zn, $^{80}$Kr/$^{82}$Kr, $^{79}$Br/$^{81}$Br, $^{85}$Rb/$^{87}$Rb and $^{86}$Sr/$^{88}$Sr are lowered. Overall, stars with different initial masses show very different final branching ratios. For instance, stars with $15\,\text{M}_\odot$ and with $20\,\text{M}_\odot$ (without rotation) produce $^{64}$Zn, $^{80}$Kr, $^{86}$Sr in the C~shell, while in heavier stars these isotopes are reduced compared to the previous He core.

The impact of the high neutron densities during C-shell can be seen in Fig.~\ref{fig:overprod_25z14S0_cshell}. It causes up to three orders of magnitude overproduction of some $r$-process nuclei, such as $^{70}$Zn, $^{76}$Ge, $^{82}$Se, or $^{96}$Zr, compared to the yields of the ``slower'' $s$~process during He~burning. 
However, the production of $r$-only nuclei in carbon burning compensates only the destruction in the He-core $s$~process when looking at the final yields. Only for the $40\,\text{M}_\odot$ model is $^{96}$Zr weakly produced.

During C-burning, the main neutron poisons are $^{16}$O, $^{20}$Ne, $^{23}$Na, and $^{24}$Mg, which are all primary. Thus the C-shell contribution to the $s$~process will vanish at low metallicities even faster than during He~burning. In our non-rotating stellar models with $Z<\text{Z}_\odot$,
the C-burning shell has a small contribution ($<10\%$).

Many aspects of this phase depend on the rates of a few key nuclear reactions.
First,  how the shells proceed depend on whether central C~burning takes place in a radiative or a convective core. It is thus sensitive to the C/O ratio in the core after He~burning and therefore to the $^{12}$C$(\alpha,\gamma)$ rate. The uncertainty of this rate and its impact on the stellar structure evolution was studied for example in \citet{imbriani:01,2004ApJ...611..452E,2009ApJ...702.1068T}. 
In our models between one and three convective C-burning shells appear in the course of the evolution. 
The last shell has a maximal extension up to $M_r=M_{\rm C}^{\rm max}$ (given in Table \ref{tab:cores}). In most of the models, a large fraction of the He-burning $s$-process material is reprocessed \citep[e.g.,][]{2010ApJ...710.1557P}. Indeed, comparing $M_{\rm C}^{\rm max}$ to $M_{\rm He}^{\rm max}$ in Table~\ref{tab:cores} shows that only $10-20\%$ of the CO~core is not reprocessed and keeps the pure signature of the He-burning $s$~process.   

Second, the $s$-process nucleosynthesis depends on the number of free $\alpha$ particles present in the shell
that can trigger neutron production by $^{22}$Ne$(\alpha,n)$ \citep{1991ApJ...371..665R} or $^{13}$C($\alpha$,n)  \citep{ bennett:12, pignatari:13}. 
In carbon burning $\alpha$ particles are released by the $^{12}$C$+^{12}$C $\alpha$-channel. 
The following studies by \cite{2000ApJS..129..625L,rauscher:02,2007ApJ...655.1058T,2010ApJ...710.1557P}
confirmed that $^{22}$Ne$(\alpha,n)$ is the only important neutron source in C-shell burning, where
the remaining $^{22}$Ne left after central He~burning is consumed in a very short time (time scale $\sim1$~yr).
At shell C~burning temperatures ($T_9\sim1$) the ratio of the $^{22}$Ne$(\alpha,n)$ to $^{22}$Ne$(\alpha,\gamma)$ rates is about $230$. In these conditions, the main competitor is the $^{22}$Ne$(p,\gamma)$, where protons are made by the C-fusion channel $^{12}$C($^{12}$C,p)$^{23}$Na. Alternatively, \cite{bennett:12} and \cite{pignatari:13} showed that for $^{12}$C+$^{12}$C larger than about a factor of 100 compared to the CF88 rate at typical central C-burning temperatures, the $^{13}$C($\alpha$,n)$^{16}$O reaction activated in the C core may strongly affect the final $s$-process yields. 
The $^{12}$C+$^{12}$C rate needs to be better constrained by experiments \citep[e.g.,][]{wiescher:12}. Other neutron sources as $^{17}$O$(\alpha,n)$ and $^{21}$Ne$(\alpha,n)$ recycle most of the neutrons absorbed by $^{16}$O and $^{20}$Ne, respectively \citep[e.g.][]{2000ApJS..129..625L}.

\section{Impact of rotation on the $s$ process}
\label{sec:boosteds}
\subsection{Impact during the various burning stages}
Rotation significantly changes the structure and pre-SN evolution of massive stars \citep{2004A&A...425..649H} and thus also the $s$-process production. Rotating stars have central properties similar to more massive non-rotating stars. In particular they have more massive helium burning and CO~cores (see Table~\ref{tab:cores}), respectively, which is an effect of rotation also found by other studies \citep[e.g.][]{2000ApJ...544.1016H,2013ApJ...764...21C}. Our models with rotation show typically $30\%$ to $50\%$ larger He~cores and CO~cores than the non-rotating models. A $20\,\text{M}_\odot$ star with rotation has thus a core size which is almost as large as the one of a $25\,\text{M}_\odot$ non-rotating star. The higher core size means higher central temperatures at the same evolutionary stage and consequently the $^{22}$Ne$+\alpha$ is activated earlier. 
In these conditions the He-core $s$-process contribution increases at the expense of the C-shell contribution. 
Since in He-burning conditions the amount of neutrons captured by light neutron poisons and not used for the $s$~process is lower compared to C-burning conditions, an overall increase of the $s$-process efficiency is obtained
\citep[see also][]{2010ApJ...710.1557P}.

\begin{figure}%[ht]
  \includegraphics[width=0.49\textwidth]{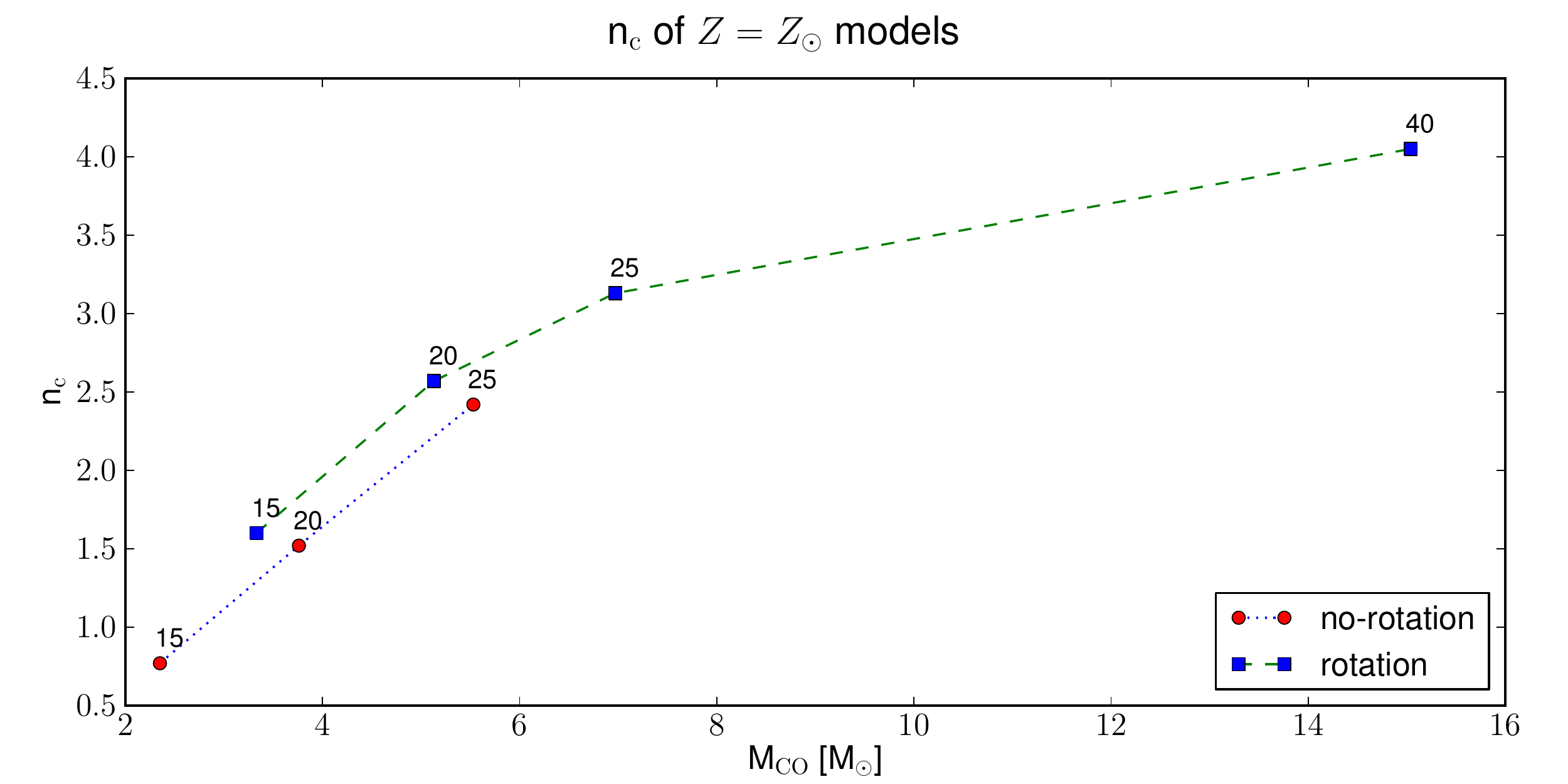}
  \caption[$n_{\rm c}$ vs. $M_{\rm CO}$ for solar $Z$ models]{Average number of neutron captures per seed $n_{c}$ versus $M_{\rm CO}$ for solar metallicity models after central He~burning. Blue squares show rotating stars and red circles non-rotating stars. The initial mass of each star is written above the symbol.}
  \label{fig:nc_vs_mco}
\end{figure}

At solar metallicity the difference between rotating and non-rotating stars is mainly found in the core size, but not in the amount of available $^{22}$Ne. This becomes clear if one compares $X(^{22}{\rm Ne})=\Delta X(^{22}{\rm Ne})+X_{\rm r}(^{22}{\rm Ne})$ of the A-series models in Table~\ref{tab:he_core_sproc}. In mass fraction, $X(^{22}{\rm Ne})\approx1.3\times10^{-2}$ is available for $\alpha$-captures, which is therefore mainly secondary. Similar values are obtained in both rotating and non-rotating models. The difference in $s$-process efficiency is therefore mainly due to the rotation-induced larger core size and the related impact on temperature (higher) and density (lower). The difference in the neutron exposure is due to

higher fraction of burned $^{22}$Ne. The difference in $s$~processing between rotating and non-rotating stars is the smallest at $25\,\text{M}_\odot$ (A25s0 vs A25s4), when comparing $15$ to $25\,\text{M}_\odot$ models. 
It is related to the saturation of the $s$~process towards higher core/initial masses, which was already found by \citet{1989A&A...210..187L} and can be seen in Fig.~\ref{fig:nc_vs_mco}. 
This figure shows $n_{c}$ after He~burning versus CO-core mass of rotating (blue squares) and non-rotating stars (red circles).  We see that $n_{c}$ saturates for $M_{\rm CO}>7\,\text{M}_\odot$ (initial mass $>25\,\text{M}_\odot$). The saturation is caused by the exhaustion of $^{22}$Ne. Typically, the model 
A40s4 has burned $96\%$ of available $^{22}$Ne after He burning.

In Fig.~\ref{fig:overprod_25z14} the overproduction factors of $25\,\text{M}_\odot$ models (A25s0 and A25s4) with solar metallicity after the end of He burning are shown. Model A25s4 (circles) shows only a moderate increase of the $s$-process production with respect to A25s0 (diamonds). Both models produce heavy isotopes from iron seeds up to the Sr-peak ($A\approx90$). In A25s0 model, $66\%$ of Fe is destroyed, and in A25s4 $73\%$. The varying overproduction factors ($\ne 1$) beyond $A=90$ are 
the signature of a local redistribution of pre-existing heavy nuclei. This figure therefore illustrates that not only the $s$-process quantities given in Table~\ref{tab:he_core_sproc} are similar, but also the abundances pattern of rotating and non-rotating models at solar $Z$ are almost identical. The difference in the efficiency is mostly caused by the larger core size in the rotating models. 

\begin{figure*}
  \includegraphics[width=\textwidth]{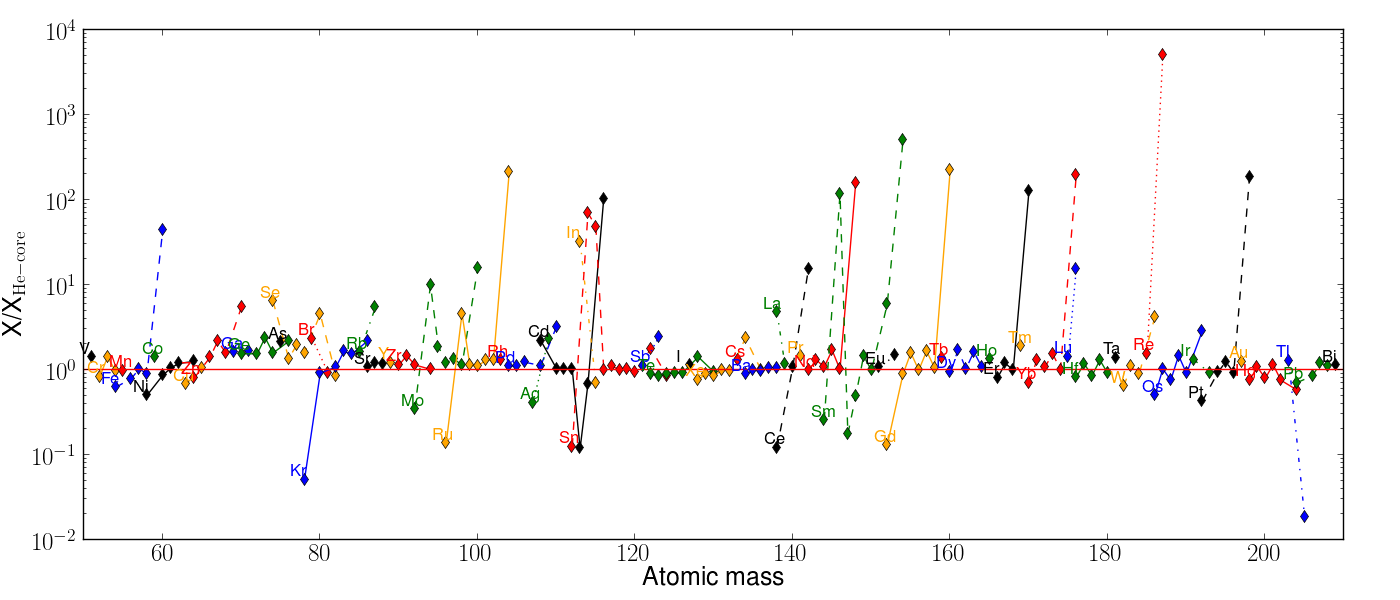}
  \caption[$X_{\rm C}/X_{\rm He}$ in a rotating $25~M_\odot$ star at $Z=Z_\odot$]{Ratio of abundances after shell C~burning to the abundances after core He~burning, $X_{\rm C}/X_{\rm He}$, in a rotating $25~M_\odot$ star at $Z=Z_\odot$ (A25S4). It illustrates the modification of the abundances by $s$~process in shell C~burning.}
  \label{fig:overprod_25z14S4_cshell}
\end{figure*}
%./overprodfact.py 3.3 ../data/G025z14S013/P025z14S0.y0022071.gz  ../data/G025z14S013/P025z14S0.y0045681.gz
%./overprodfact.py 3.3 ../data/G025z14S413/P025z14S4.y0079991.gz ../data/G025z14S413/P025z14S4.y0099721.gz

At sub-solar metallicities the differences between rotating and non-rotating models are much more striking. Rotating models have much higher neutron exposures compared to non-rotating stars, which is due to the primary $^{22}$Ne produced and burned during central He burning (see Section~\ref{sec:res_ne22_production}). This is also illustrated by the $3$ to $270$ times higher amount of $^{22}$Ne burned in rotating stars up to central He exhaustion, depending on the initial mass (or $M_{\rm CO}$) and metallicity. 
The large production of neutrons by $^{22}$Ne is partially compensated by the larger concentration of $^{25}$Mg and $^{22}$Ne itself, which become primary neutron poisons in rotating massive stars \citep{2008ApJ...687L..95P}.
Figures~\ref{fig:overprod_25z01}, \ref{fig:overprod_25zm5} and \ref{fig:overprod_25zm7} show the abundance normalised to solar in the CO~core of $25\,\text{M}_\odot$ stars with $Z=10^{-3}$, $Z=10^{-5}$ and $Z=10^{-7}$ just after central He exhaustion, each for a rotating (circles) and a non-rotating model (diamonds). Going from $Z=\text{Z}_\odot$ (Fig.~\ref{fig:overprod_25z14}) to $Z=10^{-3}$ and $10^{-5}$ (Fig.~\ref{fig:overprod_25z01} and \ref{fig:overprod_25zm5}) the production of nuclei between $A=60$ and $90$ vanishes in the non-rotating models, which is what is expected from the combination of secondary neutron source, secondary seeds and primary neutron poisons. The non-rotating model at $Z=10^{-7}$ (D25s0, diamonds in Fig.~\ref{fig:overprod_25zm7}) is special with its small amount of primary $^{22}$Ne. The rotating models at sub-solar $Z$ produce efficiently up to Sr ($Z=10^{-3}$), Ba ($Z=10^{-5}$) and finally up to Pb ($Z=10^{-7}$). At the same time the consumption of iron seeds increases from $74\%$ at $Z=\text{Z}_\odot$ (A25s4) to $96\%$ (B25s4), $97\%$ (C25s4) and $99\%$ (D25s6) at $Z=10^{-3}$, $Z=10^{-5}$ and $Z=10^{-7}$, respectively. Also with the standard rotation rate $\upsilon_{\rm ini}/\upsilon_{\rm crit}=0.4$ around $90\%$ of initial Fe is destroyed in models with $25\,\text{M}_\odot$ and $Z<\text{Z}_\odot$. Hence already from the $s$~process in He~burning one can conclude, that the primary neutron source in the rotating models is sufficient to deplete all the seeds and the production is limited by the seeds (not the neutron source any more). The other stellar masses show similar trends with $Z$. 
It is interesting to look at the rotation dependence of the non-standard $s$-process production. At $Z=10^{-5}$ the faster rotating model (C25s5) does not produce more heavy isotopes beyond iron compared to the one with standard rotation (C25s4). Instead, what happens is that not only iron is depleted but elements up to Sr are partially destroyed (after being produced) and heavier elements like Ba are produced. Even at the lowest metallicities in a very fast rotating model (D25s6 and D25s6b, $\upsilon_{\rm ini}/\upsilon_{\rm crit}= 0.6$ instead of the standard $0.4$), and thus with a larger primary neutron source, there is no additional production of $s$-process elements starting from light element seeds like $^{22}$Ne. Indeed, going from $[$Fe/H$]= -3.8$ (C25s4) to $[$Fe/H$]= -5.8$ (D25s4), the Sr yield decreases by a factor of $\sim 9$, while the Ba yield increases by a factor of $5$. Hence, the production is limited mainly by the iron seeds. 

Models with a reduced $^{17}$O$(\alpha,\gamma)$ (C25s4b, C25s5b, D25s4b and D25s6b) produce more neutrons. Actually, reducing this rate has similar consequences to increasing the amount of $^{22}$Ne.
Already a reduction of $^{17}$O$(\alpha,\gamma)$ by a factor of $10$ boosts the $s$~process up to Ba more (model C25s4b) than going from standard (C25s4) to faster rotation (C25s5). Models C25s4b, C25s5b, D25s4b and D25s6b show [Sr/Ba] of about $+1$, $+0.3$, $0$, and $-0.6$.
These models therefore emphasize the importance of $^{16}$O as a neutron poison, 
as discussed in \citet{2012A&A...538L...2F}.   
Note that the models with a reduced $^{17}$O$(\alpha,\gamma)$ are still limited by seeds.

The normalisation to solar composition allows to compare the low $Z$ models in Figs.~\ref{fig:overprod_25z01}, \ref{fig:overprod_25zm5}, and \ref{fig:overprod_25zm7} to the solar $Z$  models in Fig.~\ref{fig:overprod_25z14} with respect to their total production. Model B25s4 produces overall similar amounts of heavy nuclei in the range $A=60$-$90$ as models A25s0 and A25s4. A closer look reveals that the solar metallicity models produce higher amounts beyond Fe up to Ge. For isotopes of As, Se, Br and Kr, A25s0, A25s4 and B25s4 produce similar amounts, while for Sr, Y and  Zr B25s4 produces more. However, here one has to keep in mind, that for the final picture also the shell C burning contribution has to be taken into account. 
The impact on GCE of these results have been discussed elsewhere \citep[e.g.,][]{2013A&A...553A..51C}. However, according to models A25s0, A25s4 and B25s4 compared to C25s5 (Fig.~\ref{fig:overprod_25zm5}), 
rotating stars at  $Z=10^{-5}$ (initial $[$Fe/H$]=-3.8$)  
probably does not contribute significantly to the $s$-process chemical enrichment at solar $Z$, because the $X/X_\odot$ values are only around 1 or lower for C25s5. This is 
confirmed for the model D25s6 in  Fig.~\ref{fig:overprod_25zm7}. For the Sr, Y, and Zr, a small contribution from rotating stars with $Z$ between $10^{-3}$ (initial $[$Fe/H$]=-1.8$) and $10^{-5}$ can nevertheless be expected. Instead, for the non-rotating stars the $s$-process contribution is already negligible at $10^{-3}$. 

Rotation only has a mild impact on the He-shell contribution. Rotation-induced mixing widens the radiative zone where $^{22}$Ne$(\alpha,n)$ is activated to about $0.4\,\text{M}_\odot$ in rotating stars (compared to  $0.2\,\text{M}_\odot$ in non-rotating models).
As explained in the previous section, the contribution to the total $s$-process yields is therefore low in our models, and only in the region of $5\%$ for solar metallicity $25\,\text{M}_\odot$ stars with and without rotation. For less massive stars the He~shell gains more weight and produces in $15\,\text{M}_\odot$ models with rotation up to $50\%$ of the total yields.

In Fig.~\ref{fig:overprod_25z14S4_cshell}, the ratio of abundances after shell C~burning to the abundances after core He~burning, $X_{\rm C}/X_{\rm He}$ is plotted for the rotating $25\,\text{M}_\odot$ model at $Z=\text{Z}_\odot$ (A25s4). As in the non-rotating $Z=\text{Z}_\odot$ model, 
the high neutron densities lower the $s$-process branching ratios. Rotating models with $15\,\text{M}_\odot$ still produce $^{64}$Zn, $^{80}$Kr, $^{86}$Sr in the C~shell, while in $20\,\text{M}_\odot$ and heavier stars these isotopes are 
depleted due to the large neutron densities favouring the neutron capture channel at the $s$-process branching points $^{63}$Ni, $^{79}$Se and $^{85}$Kr \citep[e.g.,][]{2010ApJ...710.1557P}. 
This effect mainly occurs 
at solar $Z$ (or higher), 
but it is still relevant also at lower metallicities to calculate the complete $s$-process pattern. 

\subsection{Relative contributions and total yields}
\begin{figure*}%[ht!]
  \includegraphics[width=0.49\textwidth]{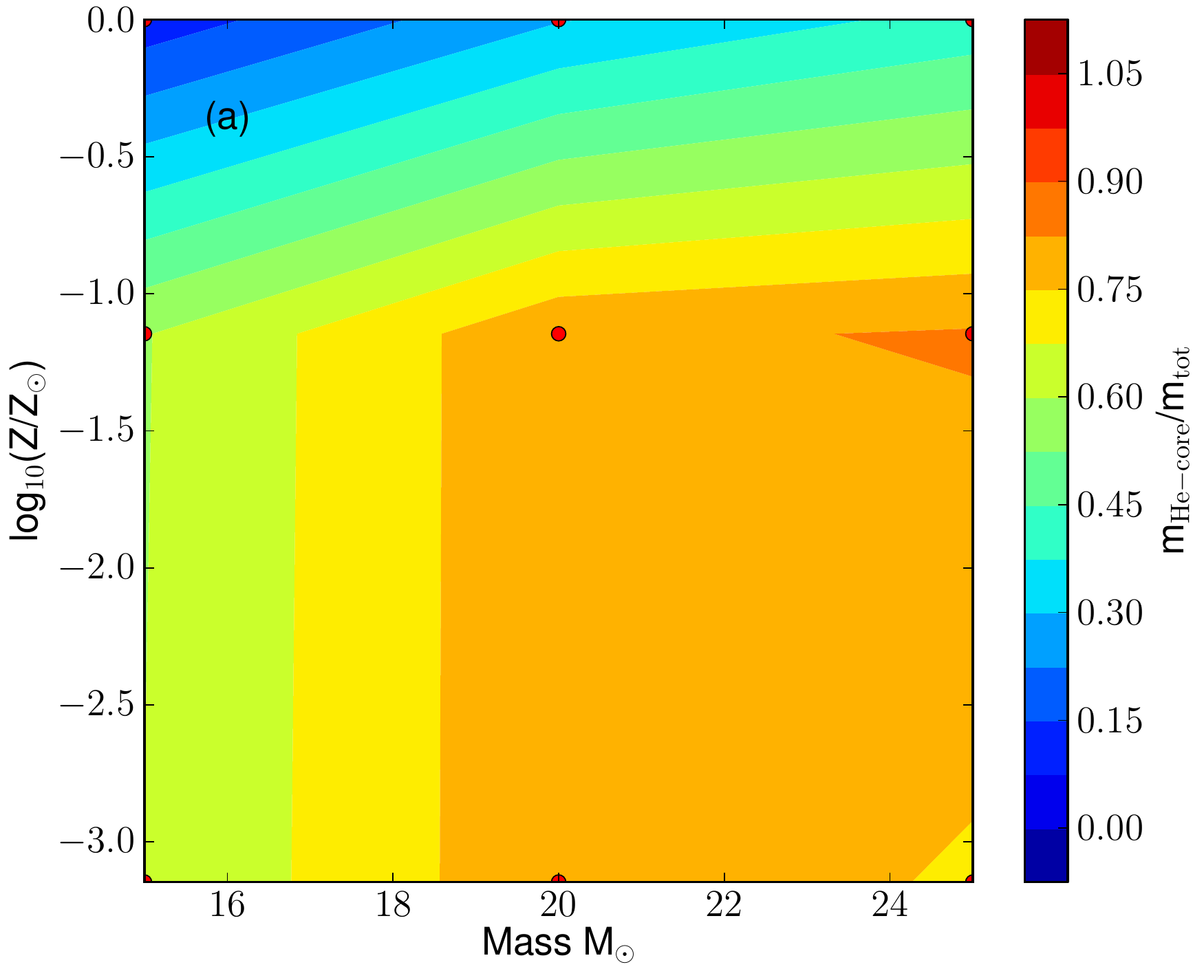}
  \includegraphics[width=0.49\textwidth]{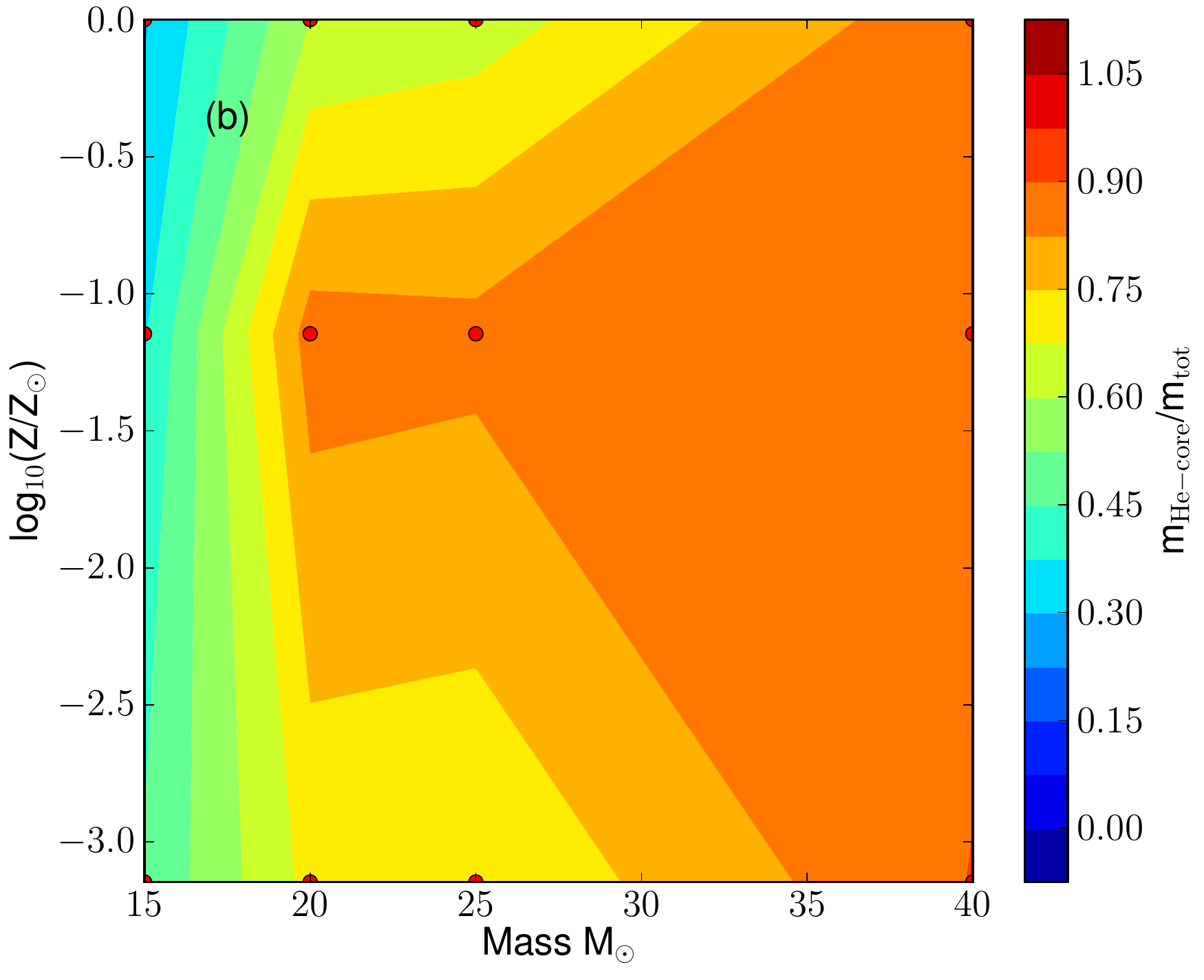} \\
  \includegraphics[width=0.49\textwidth]{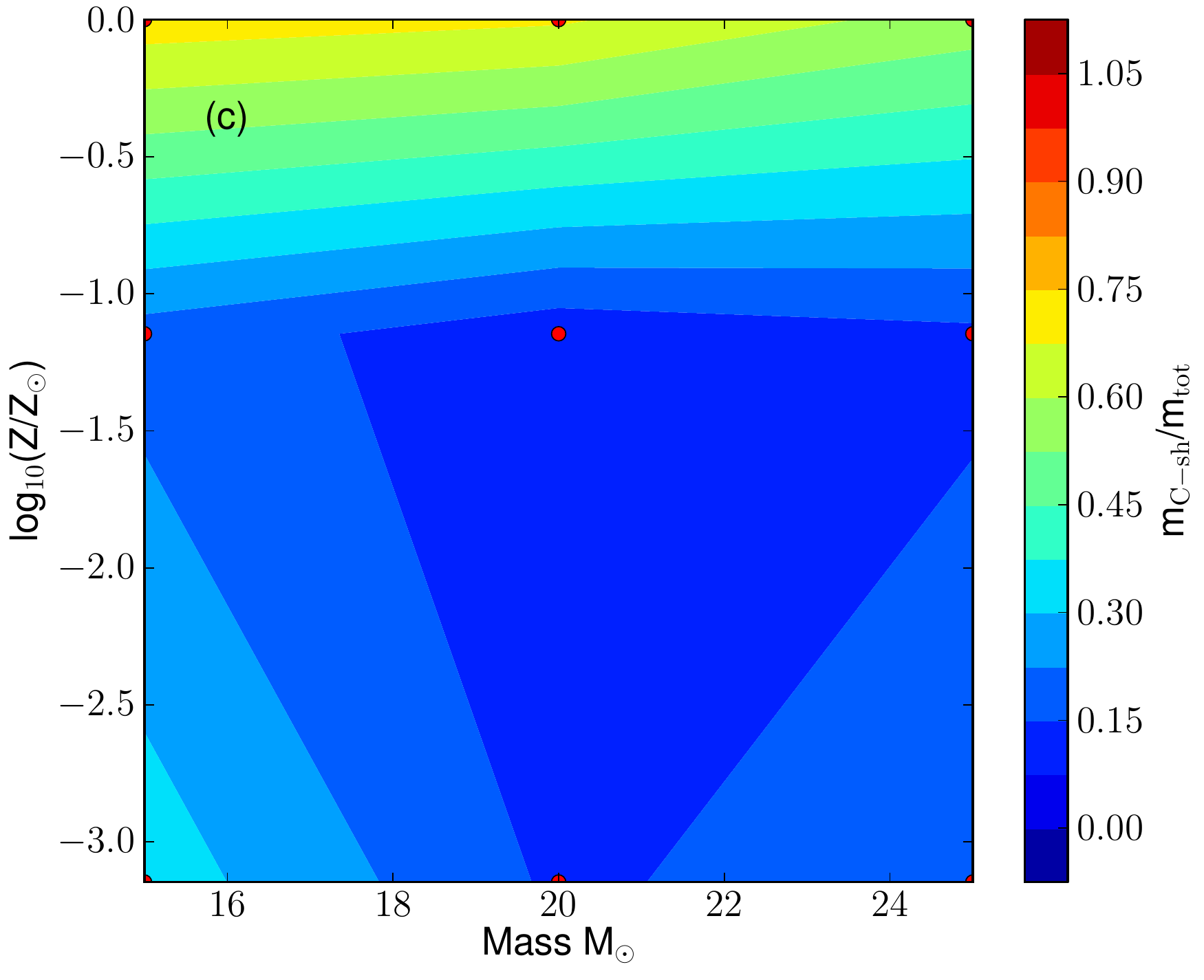}
  \includegraphics[width=0.49\textwidth]{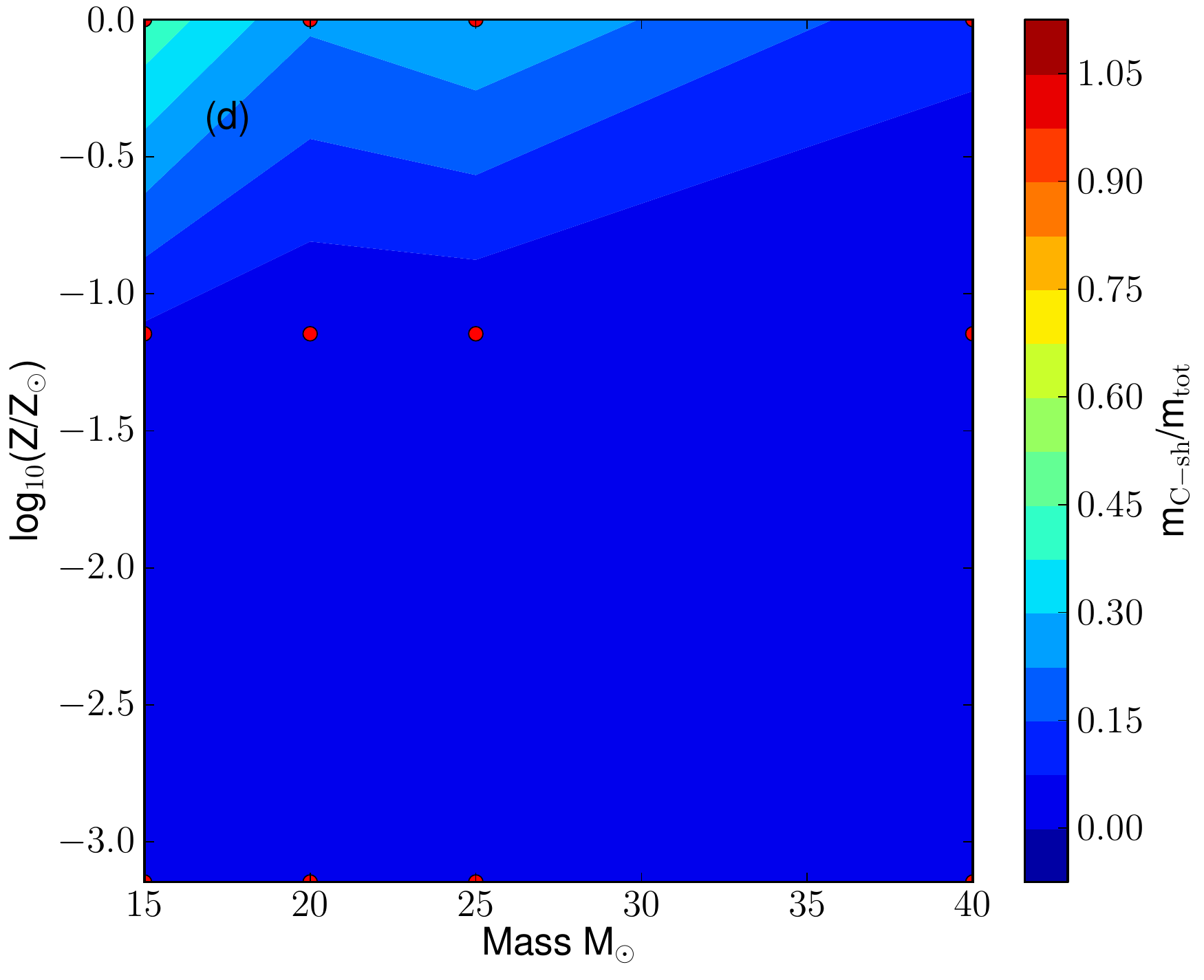} \\
  \includegraphics[width=0.49\textwidth]{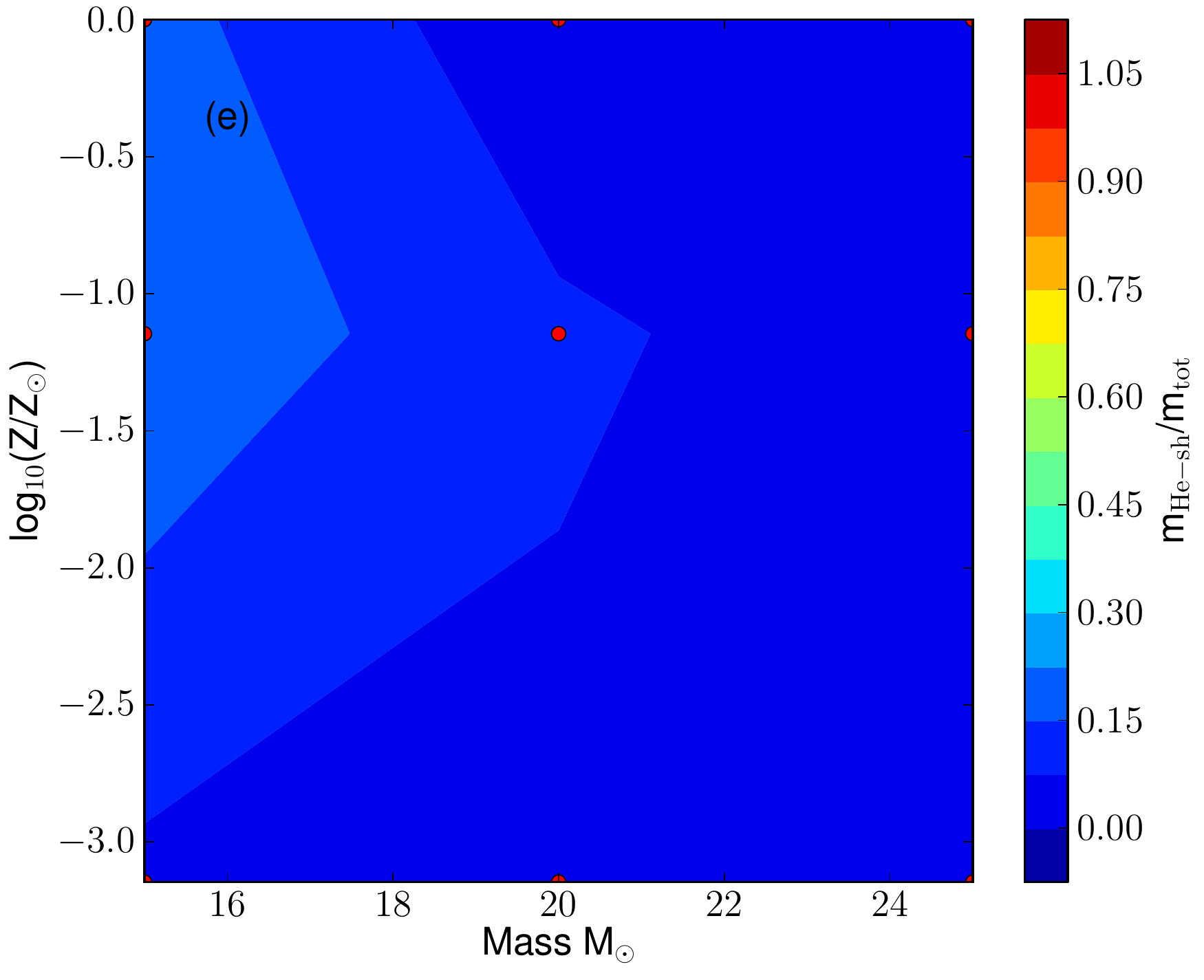}
  \includegraphics[width=0.49\textwidth]{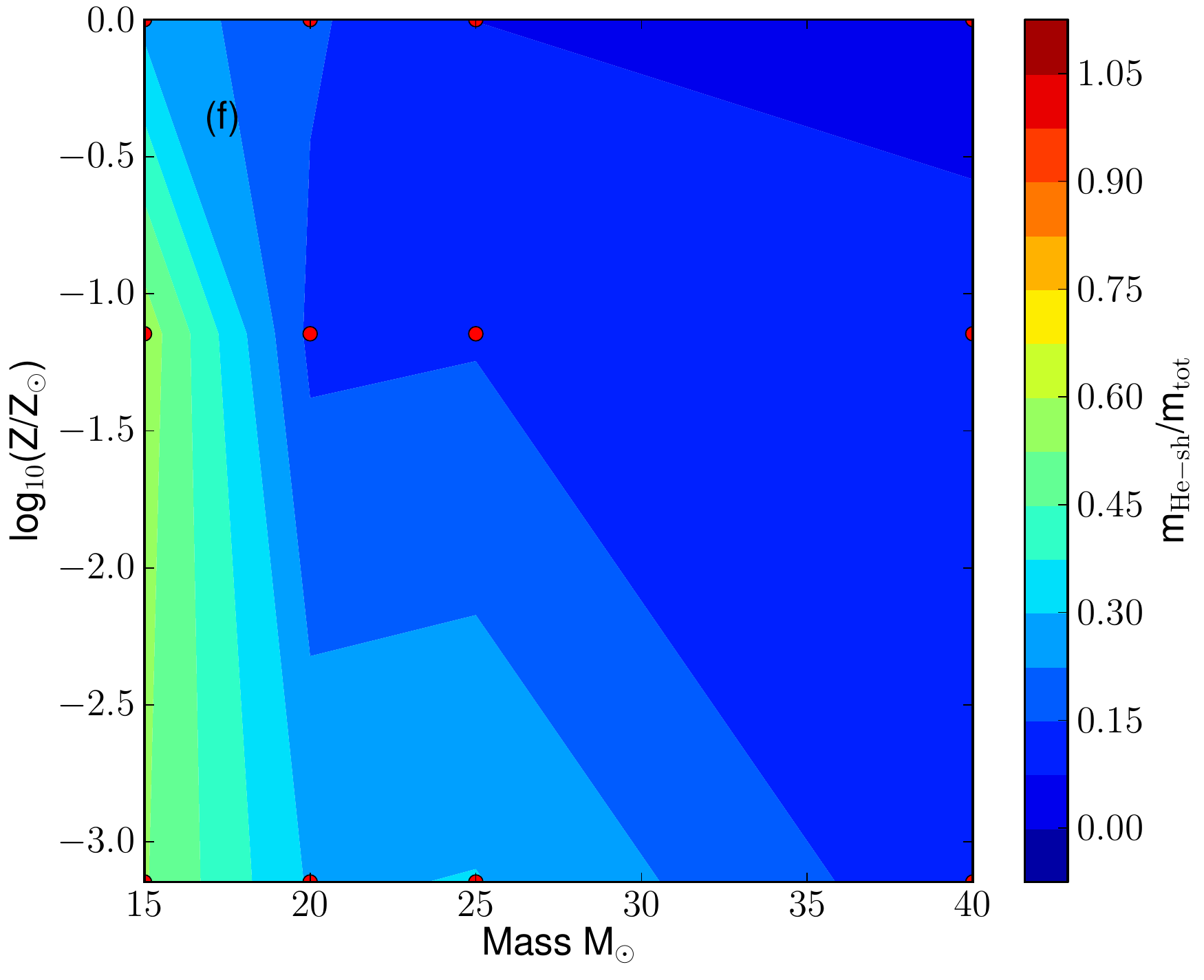} \\
  \caption[$s$-process site yields normalised to the total yields of $^{68}$Zn]{$s$-process site yields of $^{68}$Zn normalised to the total yields to illustrate the different relative contributions as a function of mass and metallicity $Z$, for He~core without (a) and with rotation (b), for C~shell without rotation (c) and with rotation (d), and the He~shell without (e) and with rotation (f). The red circles display the location of our models in the mass-metallicity space. Note that decayed yields are plotted in this figure. 
  The values in between the data points are interpolated linearly in $\log(m)$.}
  \label{fig:sproc_site_contributions}
\end{figure*}

In Fig.~\ref{fig:sproc_site_contributions} the yields of $^{68}$Zn of the three $s$-process sites normalised to the total yields are displayed, for non-rotating stars on the left hand side and rotating stars on the right hand side, and from top to the bottom for He-core, C-shell and He-shell burning yields. We plotted $^{68}$Zn as a representative for the isotopes in range $A=60$-$80$, because it is produced by the $s$~process in all three phases. 
This figure allows to compare the contributions of the three different sites to the  total yields. The following points can be derived:
\begin{enumerate}
  \item In general, the contribution from He-core burning (colours yellow to red in Fig.~\ref{fig:sproc_site_contributions}a and \ref{fig:sproc_site_contributions}b) dominates over the other two phases overall. 
  
  \item Shell carbon burning is, compared to the other two sites, only efficient at solar metallicity (see Fig.~\ref{fig:sproc_site_contributions}c and Fig.~\ref{fig:sproc_site_contributions}d). The weak contribution at low-$Z$ is due to the low amount of $^{22}$Ne left, the smaller amount of seeds and the primary neutron poisons, which have an increased strength towards lower $Z$ in C-shell conditions.
  The only mass-metallicity range for which the C-shell dominates is at solar $Z$ with $M \lesssim 25\,\text{M}_\odot$ for non-rotating models and with $M \lesssim 20\,\text{M}_\odot$ for rotating models. Such a dominant contribution from C-shell was not seen in previous literature \citep[e.g.][]{2007ApJ...655.1058T}. 
This may be due to the high $^{22}$Ne$(\alpha,\gamma)$ rate of NACRE, which is in strong competition to the neutron source during central He~burning and dominates for stars with $M \lesssim 20\,\text{M}_\odot$. This inhibition during He-core burning is weaker for rotating stars since they have higher central temperatures.
  \item Shell He~burning contributes only a small fraction but typically $5\%$ to the final yields (see Fig.~\ref{fig:sproc_site_contributions}e and Fig.~\ref{fig:sproc_site_contributions}f). The exceptions are the rotating $15$ to $25\,\text{M}_\odot$ stars at low $Z$ and rotating $15$ to $20\,\text{M}_\odot$ stars at solar $Z$. It is the effect of decreasing contribution from the He~core towards lower masses and the higher burning temperatures in the shell compared to the He~core, which allows an efficient activation of $^{22}$Ne$(\alpha,n)$ in the $15\,\text{M}_\odot$ models. Additionally the He~shell is not limited by the diminished iron seeds consumed by $s$~process in He~core but occurs in a region still containing its initial iron content. Note that decayed yields are plotted in this figure. 
\end{enumerate}

\begin{figure*}%[ht!]
  \includegraphics[width=0.49\textwidth]{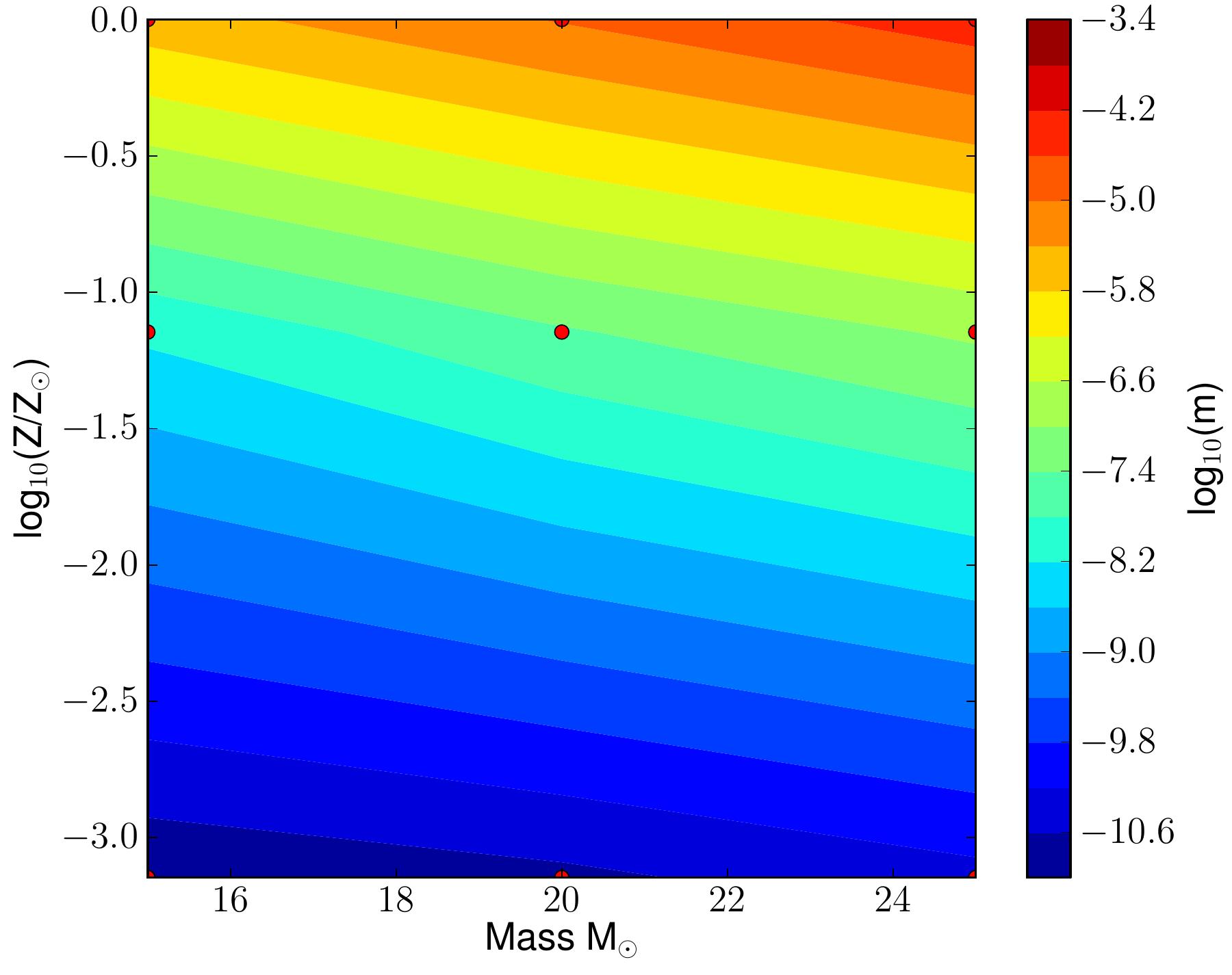}
  \includegraphics[width=0.49\textwidth]{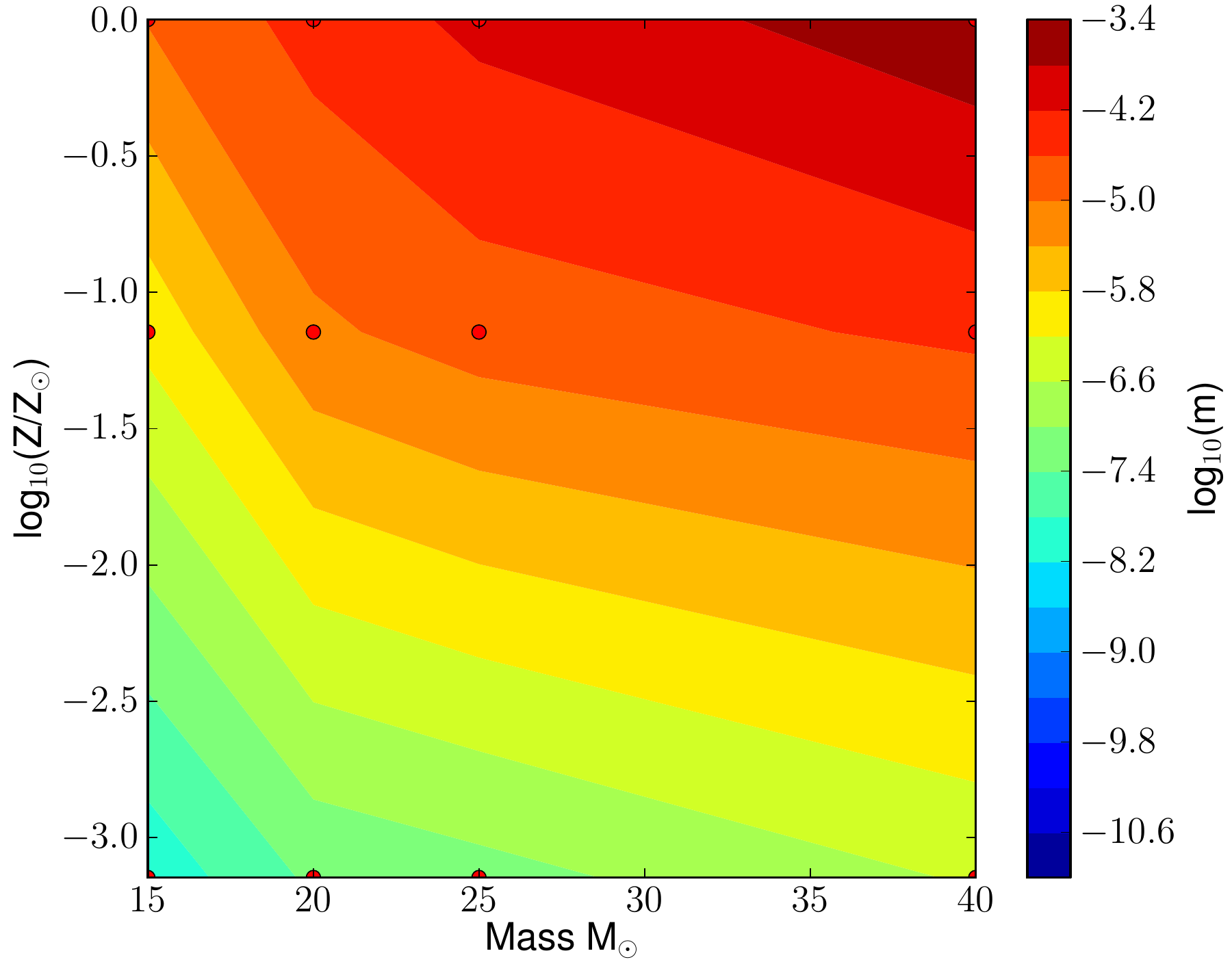} \\
  \caption[$s$-process yields of $^{68}$Zn]{$s$-process yields, $m$, of $^{68}$Zn in $\text{M}_\odot$ to illustrate the mass and metallicity dependence of the $s$~process, without rotation on the left hand side and with rotation on the right hand side. The red circles display the location of our models in the mass-metallicity space. The values in between the data points are interpolated linearly in $\log(m)$. }
  \label{fig:sproc_zn68_yield}
\end{figure*}

\begin{figure*}%[ht!]
  \includegraphics[width=0.49\textwidth]{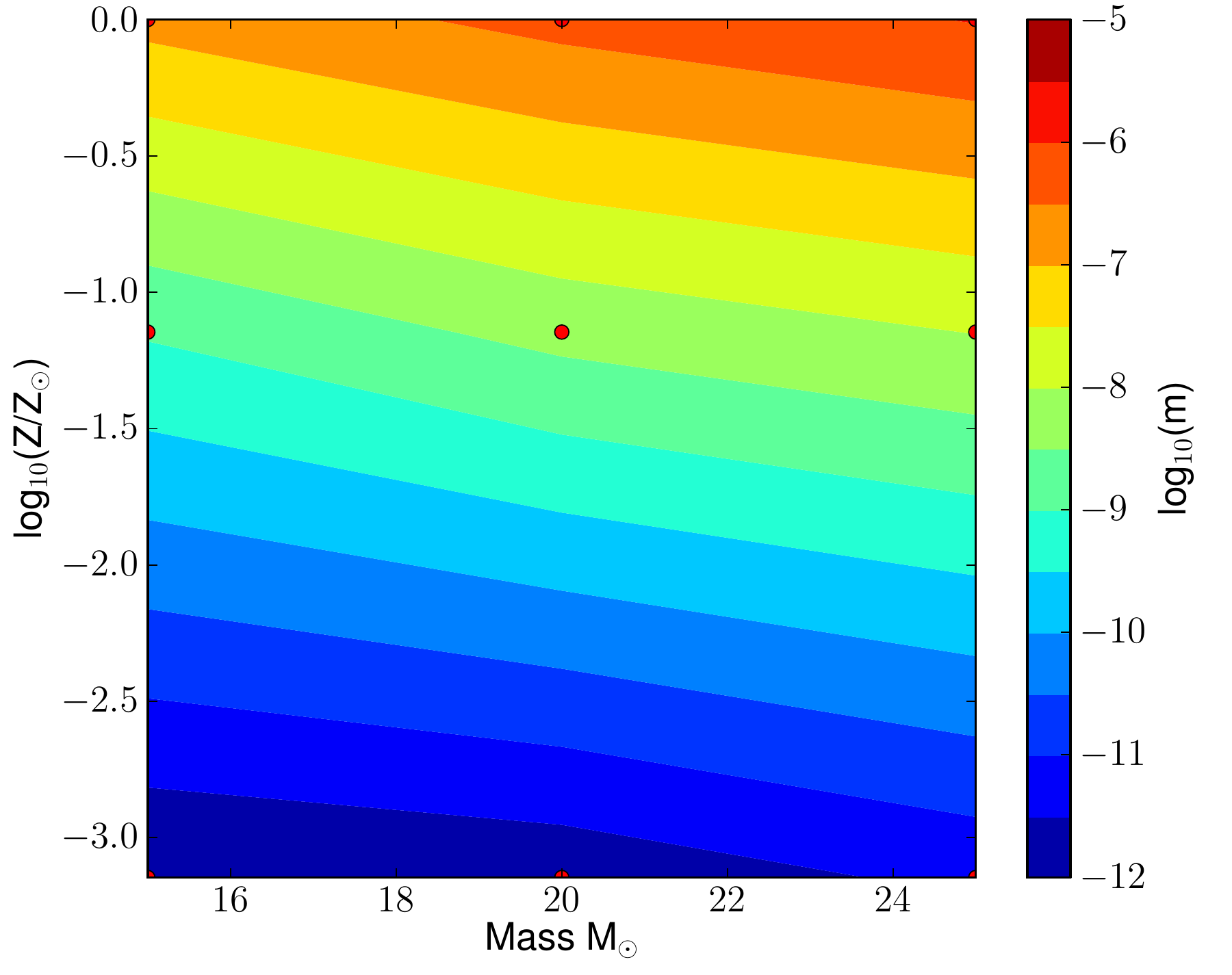}
  \includegraphics[width=0.49\textwidth]{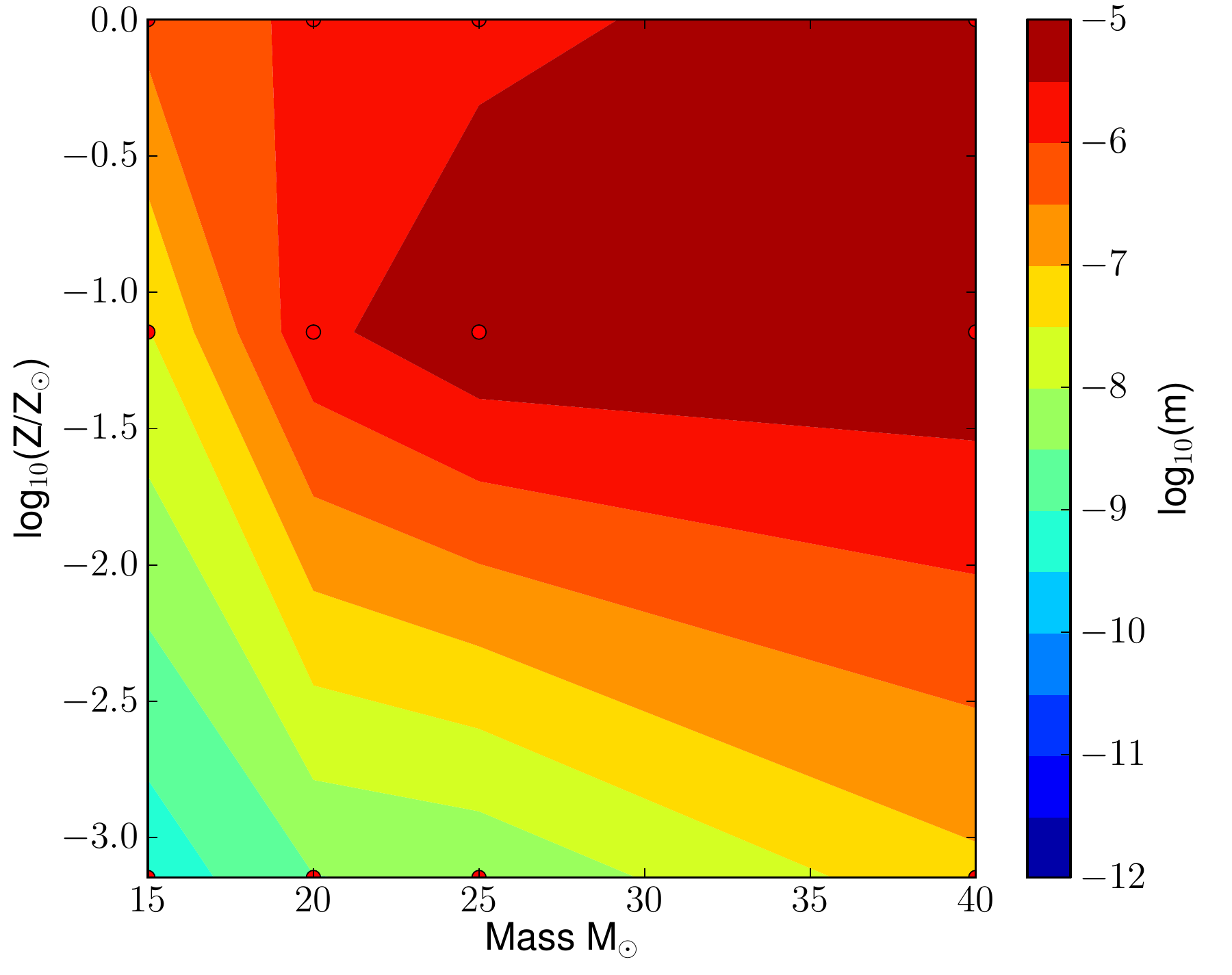} \\
  \caption[$s$-process yields of $^{88}$Sr]{$s$-process yields, $m$, of $^{88}$Sr in $\text{M}_\odot$ to illustrate the mass and metallicity dependence of the $s$~process, without rotation on the left hand side and with rotation on the right hand side. The red circles display the location of our models in the mass-metallicity space. The values in between the data points are interpolated linearly in $\log(m)$.}
  \label{fig:sproc_sr88_yield}
\end{figure*}

In Fig.~\ref{fig:sproc_zn68_yield}, the dependence of total $^{68}$Zn yields on the mass and metallicity are displayed for rotating stars with standard rotation rate ($\upsilon_{\rm ini}/\upsilon_{\rm crit}=0.4$) on the right-hand side and for non-rotating stars on the left-hand side. The red circles display the location of our models in the mass-metallicity space. The values in between the data points are interpolated linearly in $\log(m)$. As mentioned above, $^{68}$Zn is representative for the isotopes in range $A=60$-$80$. A similar plot for the neutron-magic isotope $^{88}$Sr is presented in Fig.~\ref{fig:sproc_sr88_yield} to show the dependence of the Sr-peak production on rotation ($^{86}$Sr, $^{87}$Sr, $^{89}$Y, and $^{90}$Zr show the same trends as $^{88}$Sr). Several differences between the standard and rotation boosted $s$~process can be seen: 
\begin{enumerate}
  \item Rotating models clearly produce more $s$-process elements at all metallicities.
 
  \item Whereas the $s$-process production in non-rotating model decreases steeply with metallicity 
\citep[dependence steeper than linear, e.g.,][]{pignatari:08}, the $^{68}$Zn yields of rotating stars show a secondary-like behaviour, going from reddish to blueish colours towards lower $Z$. 
 While the $^{68}$Zn yields of non-rotating stars drop by five orders of magnitude when the metallicity goes down by a factor $10^3$, the yields from rotating stars drop only by a factor $10^3$. The scaling with metallicty is less steep for rotating models. 

  \item Furthermore, the Sr-peak isotopes do not show a secondary behaviour for stars with rotation and $M>15\,\text{M}_\odot$ in the metallicity range between solar ($\log(Z/\text{Z}_\odot)=0$) and about one hundredth ($Z=1.4\times10^{-4}$, $\log(Z/\text{Z}_\odot)=-2$) of solar metallicity, but they eject maximal absolute yields around one tenth of solar metallicity (dark red around $\log(Z/\text{Z}_\odot)=-1$) for $20$ to $30\,\text{M}_\odot$ stars.
  
\end{enumerate}

\section{Comparison to the literature and observations}
\label{sec:comparison}
\subsection{Comparison to the literature}
\label{sec:comparison_literature}
In Table~\ref{tab:overproduction_heburning} the overproduction factors $X_i/X_{ i,{\rm ini}}$ in the centre of solar metallicity $25\,\text{M}_\odot$ models after the end of central He~burning are presented. It shows $X_i/X_{ i,{\rm ini}}$ for isotopes between Cu and Zr for the models with (A25s4) and without rotation (A25s0), as well as for models $1$ and $2$ from \citet{2010ApJ...710.1557P}, models 25K and 25C from \citet{2007ApJ...655.1058T} which are based on stellar models of \citet{2004ApJ...611..452E}, and the model from \citet{1991ApJ...367..228R}.  
\begin{table*}%[ht!]
  \caption{Production factors$^{\rm a}$ of $25\,\text{M}_\odot$ models after central He exhaustion$^{\rm b}$}
  \label{tab:overproduction_heburning}
  \begin{tabular}{lD{.}{.}{1}D{.}{.}{1}D{.}{.}{1}D{.}{.}{1}D{.}{.}{1}D{.}{.}{1}D{.}{.}{1}}
  \hline\hline
  Model & \mbox{A25s0$^{\rm c}$} & \mbox{A25s4$^{\rm c}$} & \mbox{Pi10-1} & \mbox{Pi10-2} & \mbox{T07-25K} & \mbox{T07-25C} & \mbox{Ra91a}\\
  \hline
  Isotope       & \multicolumn{7}{c}{Overproduction factors} \\
  \hline
  $^{63}$Cu     &  62.0 &  88.0 &  127   & 134   &  60.8 &  78.2 &  91.8 \\
  $^{65}$Cu     &  73.7 & 125   &  280   & 317   & 128   & 205   & 226.3 \\
  $^{64}$Zn     &  10.4 &  15.7 &   34.1 &  36.8 &  30.7 &  43.6 &  41.0 \\
  $^{66}$Zn     &  16.5 &  30.0 &   76.3 &  88.7 &  59.6 & 107   & 118.9 \\
  $^{67}$Zn     &  21.8 &  40.7 &  109   & 127   &  82.9 & 153   & 171.7 \\
  $^{68}$Zn     &  16.0 &  32.1 &   99.1 & 121   &  73.1 & 158   & 164.7 \\
  $^{70}$Zn     &   0.1 &   0.0 &    0.3 &   0.3 &   0.4 &   0.6 & \dots \\
  $^{69}$Ga     &  30.8 &  63.1 &  126   & 156   & \dots & \dots & 208.6 \\
  $^{71}$Ga     &  32.6 &  69.4 &  147   & 187   & \dots & \dots & 263.9 \\
  $^{70}$Ge     &  21.5 &  45.2 &  154   & 193   & 112   & 270   & 253.7 \\
  $^{72}$Ge     &  11.7 &  24.9 &  88.0  & 114   &  75.2 & 201   & 190.7 \\
  $^{73}$Ge     &  11.2 &  24.0 &  82.4  & 107   &  46.9 & 128   & 128.8 \\
  $^{74}$Ge     &   9.5 &  19.5 &  71.0  &  94.2 &  37.5 & 110   &  99.3 \\
  $^{76}$Ge     &   0.1 &   0.0 &   0.0  &   0.0 & \dots & \dots & \dots \\
  $^{75}$As     &   6.4 &  13.1 &  45.3  &  60.2 &  27.4 &  81.9 &  59.6 \\
  $^{76}$Se     &  12.3 &  24.6 &  99.4  & 133   &  78.2 & 241   & 212.2 \\
  $^{77}$Se     &   5.6 &  11.1 &  44.0  &  59.1 & \dots & \dots &  88.6 \\
  $^{78}$Se     &   9.3 &  17.6 &  67.4  &  91.7 & \dots & \dots & 108.9 \\
  $^{80}$Se     &   0.6 &   1.1 &   1.5  &   1.9 &   1.3 &   4.0 & \dots \\
  $^{82}$Se     &   0.3 &   0.3 &   0.1  &   0.1 & \dots & \dots & \dots \\
  $^{79}$Br     &   2.4 &   4.5 &  15.6  &  21.3 & \dots & \dots &  36.6 \\
  $^{81}$Br     &   0.6 &   1.0 &  15.4  &  21.1 & \dots & \dots & \dots \\
  $^{80}$Kr     &  18.7 &  34.6 &  169   & 232   & 183   & 618   & 480.7 \\
  $^{82}$Kr     &   9.8 &  17.4 &   79.1 & 108   &  77.9 & 277   & 210.3 \\
  $^{83}$Kr     &   3.4 &   6.0 &   25.9 &  35.5 & \dots & \dots &  63.0 \\
  $^{84}$Kr     &   2.8 &   4.7 &   22.0 &  29.9 & \dots & \dots &  52.6 \\
  $^{86}$Kr     &   1.0 &   1.1 &    0.7 &   0.8 &   2.6 &   5.7 & \dots \\
  $^{85}$Rb     &   1.8 &   2.9 &   14.8 &  20.0 & \dots & \dots &  28.6 \\
  $^{87}$Rb     &   0.6 &   0.6 &    0.8 &   0.8 &   1.3 &   3.0 & \dots \\
  $^{86}$Sr     &  17.5 &  27.8 &   79.9 & 107   &  60.7 & 232   & 147.3 \\
  $^{87}$Sr     &  13.8 &  21.1 &   68.8 &  91.4 &  50.4 & 190   & 129.2 \\
  $^{88}$Sr     &   7.2 &   9.9 &   21.5 &  26.8 &  14.9 &  45.3 &  34.8 \\
  $^{89}$Y      &   6.2 &   8.6 &   15.6 &  18.9 & \dots & \dots &  22.3 \\
  $^{90}$Zr     &   3.0 &   4.3 &    6.9 &   8.2 & \dots & \dots & \dots \\
  $^{91}$Zr     &   3.3 &   4.8 &    8.6 &  10.1 & \dots & \dots & \dots \\
  $^{92}$Zr     &   3.2 &   4.6 &    7.3 &   8.5 & \dots & \dots & \dots \\
  $^{94}$Zr     &   2.4 &   3.2 &    5.4 &   6.3 & \dots & \dots & \dots \\
  \hline
  \end{tabular}
\begin{flushleft}
  References.  Pi10-$x$ - model $x$ from \citet{2010ApJ...710.1557P}, T07-25K/C - model 25K/C from \citet{2007ApJ...655.1058T}, Ra91a - \citet{1991ApJ...367..228R}.\\
  Notes.\\
  $^{\rm a}$ Production factors are defined as the mass fractions/abundances $X$ normalised to the initial ones $X_{\rm ini}$. Since we have here $Z=\text{Z}_\odot$ models,  the production factors are $X/X_\odot$. \\
  $^{\rm b}$ In our models, the overproduction factors are constant throughout the convective core (due to the very fast convective mixing) so our central values are directly comparable with the literature where the ``core-averaged'' overproduction factors are reported. \\
  $^{\rm c}$ The other authors used the solar abundances of \citet{1989GeCoA..53..197A}, but we used the one of \citet{2005ASPC..336...25A}.
\end{flushleft}
\end{table*}

First of all, the overproduction factors in Table~\ref{tab:overproduction_heburning} show a wide spread between the models. For 
Cu and Zn isotopes, the most efficient models (Pi10-2, T07-25C, Ra91a) produce four to seven times more than the least efficient model (A25s0). This difference becomes even more pronounced for heavier isotopes, e.g. $^{86}$Sr, where the difference from the least efficient (A25s0) to the most efficient models (T07-25C, Ra91a) can exceed a factor of twenty. Model Pi10-2 produces large amounts of Cu isotopes, while for heavier elements the production factors are lower than T07-25C and Ra91a results.     

In Table~\ref{tab:sproc_param_heburning} we show the characteristic $s$-process parameters of the same models. The central neutron exposure $\tau_c$ and the convective core averaged neutron exposure $\langle\tau\rangle$ together with the average number of neutron captures per seed $n_c$ describe the $s$-process efficiency. These $s$-process quantities show a similar picture as the overproduction factors in Table~\ref{tab:overproduction_heburning}. The most efficient models are again Pi10-2 and T07-25C, Ra91a, and the least efficient model is A25s0.
\begin{table*}%[ht!]
  \caption{$s$-process parameters in the centre of $25\,\text{M}_\odot$ stars after central He exhaustion.}
  \label{tab:sproc_param_heburning}
  \begin{tabular}{lD{.}{.}{2}D{.}{.}{3}D{.}{.}{2}D{.}{.}{2}D{.}{.}{3}D{.}{.}{2}D{.}{.}{3}}
    \hline\hline
    Model     & \multicolumn{1}{c}{${\tau_c}^{\rm a}$} & \multicolumn{1}{c}{${\langle\tau\rangle}^{\rm b}$} & \multicolumn{1}{c}{${n_c}^{\rm c}$} & \multicolumn{1}{c}{${\bar{n}_{n,{\rm max}}}^{\rm d}$} & \multicolumn{1}{c}{${n_{n,c,{\rm max}}}^{\rm e}$} & \multicolumn{1}{c}{$\Delta X(^{22}{\rm Ne})$} & \multicolumn{1}{c}{$X_{\rm r}(^{22}{\rm Ne})$} \\ 
      &  \multicolumn{1}{c}{$[{\rm mb}^{-1}]$}  & \multicolumn{1}{c}{$[{\rm mb}^{-1}]$} &   &  \multicolumn{1}{c}{$[10^5{\rm ~cm}^{-3}]$} & \multicolumn{1}{c}{$[10^7{\rm ~cm}^{-3}]$} & \multicolumn{1}{c}{$(\times10^{-2})$} & \multicolumn{1}{c}{$(\times10^{-2})$}\\
    \hline
    A25s0     & 3.80  & 0.133 & 2.34 & 5.85 & 1.56  & 0.77  &  0.51  \\ 
    A25s4     & 4.86  & 0.160 & 3.06 & 5.98 & 1.72  & 0.97  &  0.33  \\ 
    Pi10-1    & \dots & 0.197 & 4.95 & 11.4 & 3.22  & 1.03  &  1.14  \\  
    Pi10-2    & \dots & 0.209 & 5.35 & 10.1 & 2.88  & 1.47  &  0.70  \\ 
    T07-25K   & 5.00  & 0.15  & 3.63 & 2.53 & \dots & 1.39^{\rm f}  &  0.78  \\ 
    T07-25C   & 5.43  & 0.30  & 5.14 & 1.95 & \dots & 1.19^{\rm f}  &  0.98  \\ 
    Ra91a     & \dots & 0.206 & 5.67 & 6.79 & 1.80  & 1.06  &  0.96  \\ 
    \hline   
  \end{tabular}
\begin{flushleft}
  Notes.\\
  $^{\rm a}$ Central neutron exposure calculated according to Eq.~\ref{eq:neutronexposure}.\\
  $^{\rm b}$ Neutron exposure averaged over He~core (see Eq.~\ref{eq:neutronexposure_avg}).\\
  $^{\rm c}$ Number of neutron capture per seed calculated according to Eq.~\ref{eq:neutron_captures_per_seed}, averaged over the He-core mass.\\
  $^{\rm d}$ Maximum of the mean neutron density.\\
  $^{\rm e}$ Maximum of the central neutron density.\\
  $^{\rm f}$ Assuming for the secondary $^{22}$Ne a mass fraction of $X(^{22}{\rm Ne})=2.17\times10^{-2}$ at the start of He burning as in \citet{2010ApJ...710.1557P}.
\end{flushleft}
\end{table*}

There are several important differences between our models (A25s0, A25s4) and the others, namely in the initial composition and the nuclear reaction input, which explain the big differences. Here these differences are listed. 
\begin{itemize}
  \item We used for our models with solar-like composition the initial chemical composition from \citet{2005ASPC..336...25A} with a metallicity $Z=0.014$. The other authors used the solar composition from \citet{1989GeCoA..53..197A} with $Z\approx0.019$. It means that in our models the secondary $^{22}$Ne and the iron seeds are reduced by about $35\%$. From a reduction of the $^{22}$Ne neutron source and the seeds a reduction of the $s$-process production is expected. However, if one uses a solar-like composition with lower $Z$, this is partially compensated in the overproduction factors by the normalisation to the smaller initial abundances. It is only partially compensated, because the source and the seeds are reduced while the primary poisons not, and the standard $s$~process scales therefore less than secondary.
  The impact of a similar change, from \citet{1989GeCoA..53..197A} composition to the one of \citet{2003ApJ...591.1220L} with $Z=0.0149$, was investigated by \citet{2009ApJ...702.1068T}. They found that the change of initial composition can modify the final production factors by $0.2$ to $0.5$~dex for $25\,\text{M}_\odot$ stars. Since we used $Z=0.014$ in our solar $Z$ models, the reduction in the overproduction factors is even higher.

  \item In Table~\ref{tab:literature_rates_comparison} the sources of the reaction rates used in the works, compared here, are listed. The neutron source and the $^{22}$Ne$(\alpha,n)$/$^{22}$Ne$(\alpha,\gamma)$ ratio, respectively, of our models is only similar in Pi10-1, but they use the lower rate for $^{22}$Ne$(\alpha,\gamma)$ of \citet{2006ApJ...643..471K}, which is lower than the NACRE rate we used. The rates for the neutron source of CF88 and NACRE are both considerably higher \citep[see discussion in NACRE and][]{2001PhRvL..87t2501J}. Therefore all other models used more favourable combinations of $^{22}$Ne$+\alpha$ rates for the $s$~process. There is an indication that our choice of rates leads to a too weak $s$~process at solar metallicity,  
because most isotopes (except for copper) are less overproduced compared to $^{16}$O \citep[see][for more details]{2010ApJ...710.1557P}. 
  
  \item In the mass region $A=50-90$ many $(n,\gamma)$ rates, relevant for the $s$~process, were found to be lower by new measurements in the past $15$ years. 
  Thus the neutron capture rates also changed over the time frame of the different studies. \citet{2010ApJ...710.1557P} used the same rates of KADoNiS v0.3, as we did in our models. The rate reduction of several $s$-process path bottlenecks, in particular at $^{63}$Cu hinder the $s$~process and reduce the overproduction factors above the copper isotopes, when using the newer rate compilation.
  
  \item The $^{12}$C$(\alpha,\gamma)^{16}$O rate sources are listed in table~\ref{tab:literature_rates_comparison}. The rate of \citet{2002ApJ...567..643K} is the lowest and about $10\%$ to $20\%$ smaller than the NACRE rate in the relevant temperature region for core He~burning. A higher rate means that the star can obtain the same amount of energy at lower temperatures. In this way a lower rate supports the $s$~process. \citet{2009ApJ...702.1068T} studied the impact of the uncertainty in the $^{12}$C$(\alpha,\gamma)^{16}$O rate.  And a reduction of this rate by $10$ to $20\%$ increases the overproduction factors on average by $0.1$ to $0.2$~dex.
 
  \item Neglecting mass loss 
  means that the core is larger during the core He-burning phase, and consequently has higher temperatures. \citet{2007ApJ...655.1058T} point out the possible impact of such a change with their models 25N and 25NM. \citet{2010ApJ...710.1557P} and \citet{1991ApJ...367..228R} used stellar models calculated with the Frascati Raphson Newton Evolutionary Code (FRANEC), which did not include mass loss \citep{1994ApJ...437..396K}. The maximal core size of their model during He~burning is $M_{\rm He}^{\rm max}=6.17\,\text{M}_\odot$ (priv. comm. M. Pignatari). It lies thus between the core sizes of our models A25s0 and A25s4 (see Table~\ref{tab:cores}). The mass loss introduces therefore a rather moderate uncertainty, but still reduces the overproduction factors, $n_c$ and $\langle\tau\rangle$ by about $10\%$.

\end{itemize} 
These various differences in the nuclear reaction input as well as the stellar models make it 
difficult to disentangle the impact of the different parameters quantitatively. On the qualitative side, our models are consistent with the previous publications considering the differences discussed above. 
\begin{table*}%[ht!]
  \caption{Reaction rates used in $25\,\text{M}_\odot$ $Z=\text{Z}_\odot$ models.}
  \label{tab:literature_rates_comparison}
  \begin{tabular}{lccccccc}
  \hline\hline
  Model & \mbox{A25s0} & \mbox{A25s4} & \mbox{Pi10-1} & \mbox{Pi10-2} & \mbox{T07-25K} & \mbox{T07-25C} & \mbox{Ra91a}\\
  \hline
  $^{22}$Ne$(\alpha,n)$        & Ja01  & Ja01 & Ja01   & NACRE  & NACRE & CF88  & CF88   \\
  $^{22}$Ne$(\alpha,\gamma)$   & NACRE & NACRE& Ka06   & NACRE  & NACRE & K94   & CF88   \\
  $^{12}$C$(\alpha,\gamma)$    & Ku02  & Ku02 & CFHZ85 & CFHZ85 & Ku02  & CF88  & CFHZ85 \\
  n-captures                   & K0.3  & K0.3 & K0.3   & K0.3   & Be92  & Be92  &        \\
  \hline
  \end{tabular}
  \begin{flushleft}
  \footnotesize
  References.  Pi10-$x$ - model $x$ of \citet{2010ApJ...710.1557P}, T07-25K/C - model 25K/C of \citet{2007ApJ...655.1058T}, Ra91a - \citet{1991ApJ...367..228R}, Ja01 - \citet{2001PhRvL..87t2501J}, NACRE - \citet{1999NuPhA.656....3A}, CF88 - \citet{1988ADNDT..40..283C}, Ka06 - \citet{2006ApJ...643..471K}, K94 - \citet{1994ApJ...437..396K}, Ku02 - \citet{2002ApJ...567..643K}, CFHZ85 - \citet{1985ADNDT..32..197C}, K0.3  - KADoNiS v0.3, Be92 - \citet{1992ApJS...80..403B} \\
  \end{flushleft}
\end{table*}

If we compare the difference between our two models (A25s0, A25s4) and the other model we can also conclude, that the effect of rotation at solar metallicity is rather moderate and well within the nuclear reaction rate uncertainties. This is the case because $^{22}$Ne production by rotation induced mixing does not play a role at $Z=\text{Z}_\odot$. As discussed above, the rotation still leads to a stronger production at solar metallicity. The impact of rotation becomes stronger and stronger as the initial metallicity decreases.

Recently, \cite{chieffi:15} presented preliminary results where their models for fast-rotating massive stars at low metallicity can efficiently produce elements also up to Pb. In their models, the $s$-process production is due to the mixing of $^{13}$C into the helium core, which provide additional neutrons. A comparison is not possible at this stage since the models are not described in details in that study.

\subsection{Comparison to observations}
\subsubsection{Production of elements at the Sr and Ba peaks}
Spectroscopic observations have shown a secondary trend of [Cu/Fe] \citep[e.g.,][and references therein]{2005NuPhA.758..284B, sobeck:08}, in agreement with $s$-process calculations which predicts that a major part of Cu come from the $s$~process in massive stars \citep[e.g.,][]{2010ApJ...710.1557P}. The same trend is expexcted for Ga, for which only few observations and upper limits are available from low-metallicy stars and not a real comparison can be made, and for Ge \citep[see discussion in][]{2010ApJ...710.1557P}. More data is available for Ge compared to Ga \citep[][]{2005ApJ...627..238C}, but the metallicity range of interest is still not fully covered by observations. As mentioned before, we show that rotation would not change the secondary nature of the $s$-process production of these elements. 

\citet{2004ApJ...601..864T} compared the spectroscopic observations of the Sr- peak elements Sr, Y, and Zr at different metallicities with the $s$-process distribution in the solar system obtained from GCE calculations. They proposed that a Lighter Element Primary Process (or LEPP) was responsible for both the observations and the missing $s$-process abundances in the solar distribution. Later, \citep[][]{2007ApJ...671.1685M} compared the  ``stellar LEPP'' signature at low metallicity with the ``solar LEPP'' in the solar system, concluding that while they are compatible, also explosive nucleosynthesis processes can be responsible for the same elemental signature in the early galaxy. While the existence of the solar LEPP have been recently questioned \citep[][]{maiorca:12,cristallo:15}, we cannot exclude that an additional $s$-process component is needed to contribute to its total amount. We have seen in this work that it is quite unlikely that the $s$-process in fast rotating massive stars is the responsible, due to its secondary nature and its significance only at much lower metallicities for elements in the Sr mass region and heavier.  
On the other hand, \cite{2013A&A...553A..51C,barbuy:14} showed that $s$-process in fast rotating massive stars is compatible with observations at low metallicity \citep[e.g.,][]{hansen:13}.
Alternative or complementaty theoretical scenarios proposed to explain the stellar LEPP are explosive nucleosynthesis components, mainly associated to neutrino-driven winds on top of the forming neutron star \citep[e.g.,][]{2006PhRvL..96n2502F,2008ApJ...687..272Q,2009ApJ...694L..49F,2011ApJ...731....5A}.

We have seen a 
scatter in the production up to Ba, which is 
strongly affected by nuclear uncertainties.
Additionally, a scatter in Sr production is intrinsic to the rotation boosted $s$~process, since a varying rotation rate would lead to a varying amount of primary $^{22}$Ne and thus to a varying neutron exposure and $s$-process production, respectively. Typically the $s$~process in massive stars produces only minor amounts of Ba and [Sr/Ba] is around $+2$, with an upper limit of $\approx+2.3$. However, due to the seed limitation and the larger neutron capture per iron seed, the enhanced $s$~process in fast-rotating massive stars can produce more significantly also elements at the Ba neutron-magic peak.
On the other hand, as shown by \cite{pignatari:13} the intrinsic nature of $^{22}$Ne as a neutron source and neutron poison does not allow to efficiently feed also heavier elements along the $s$-process path, up to Pb. 

\subsubsection{The very low-Z stars: the case of CEMP-no stars}

At metallicities [Fe/H]$\lesssim$-2  
it is possible to observe a large number of
``carbon--enhanced--metal--poor''  (CEMP; [C/Fe]$> 0.7$ \citep{Aoki2007} stars, which exhibit large excesses of carbon with values of  [C/Fe] reaching more than 4.0 dex. At the same time, the  abundances of nitrogen, oxygen and other elements
are also largely overabundant. These stars are very old low--mass stars
(about $0.8\,\text{M}_{\odot}$) still surviving and exhibiting the particular  nucleosynthetic 
products of the first stellar generations. 

CEMP stars were classified in CEMP--s, CEMP--r/s and CEMP--r \citep[e.g.,][]{Beers1992,2005ARA&A..43..531B},
depending on the observed abundances of s--elements (mainly Ba), r--elements (mainly Eu). 
Another group was identified the CEMP--no stars, with  
much weaker overabundances of n--capture elements (typically [Ba/Fe]$<1$).  Nevertheless, a fraction of them
contains measurable amounts of heavy s-elements. 
Recent catalogs may be found by \citet{Masseron2010}, \citet{Allen2012}, \citet{bisterzo:12}, \citet{lugaro:12}, \citet{Norris4}, \citet{bonifacio:15} and \citet{Hansen2015}. The CEMP--no stars clearly dominate for low metallicity stars with
[Fe/H] $< -3.0$. Several of these stars are still Main Sequence or subgiant objects, thus their particular abundances are 
likely not resulting from self--enrichment, but  from 
the nucleosynthetic contributions of previous massive stars,
called the source stars, possibly belonging to the first stellar generations.

Many different kinds of models have been suggested to explain the properties  of the CEMP--no stars, see a review of these models by \citet{Nomotoaraa2013}. Two kinds of models are presently emerging \citep{Norris4}: the mixing and fallback models of faint supernovae \citep{Nomotoaraa2013,Tominaga2014} and the models of spinstars, \emph{i.e.} of massive stars with fast rotation and mass loss \citep{2006A&A...447..623M},  a combination of both sets of models being  also possible.
Recently \citet{Maeder2015} have provided many tests showing that the
particular CNO abundances of CEMP--no stars result from products of He--burning (mainly C and O) having undergone partial mixing and processing in the H--burning shell before being ejected into the interstellar medium.
This result is based on the analysis of the $^{12}$C/$^{13}$C, [C/N] and  [O/N] ratios as well as on the study of the elements involved in the Ne--Na and Mg--Al cycles of H--burning, which all show large excesses and a behavior completely different from that of the $\alpha$--elements.  At the same time, 
some of these CEMP--no stars show the presence of s--elements.  As shown by the models presented in previous sections, the mixing processes, by successive back and forth motions between the He-- and H--burning regions,  may also   lead to the $^{22}$Ne($\alpha$,n)$^{25}$Mg reaction
which produces s-elements by neutron captures on seed heavy elements.

The present models show a great sensitivity to both metallicity and rotation  of the ratio of s--elements of the first peak (like Sr) to s-elements  of the second peak (like Ba). 
Specifically, 
the models of $25\,\text{M}_{\odot}$ with $Z=10^{-3}$ , corresponding to [Fe/H]= -1.8,  without rotation predict a ratio [Sr/Ba]=0.13, with rotation [Sr/Ba]=2.12 (see Fig.  \ref{fig:overprod_25z01}). For the models with   $Z=10^{-5}$ ([Fe/H]= -3.8),  the corresponding values are [Sr/Ba]= 
0.03 and 1.17 respectively  (see Fig.  \ref{fig:overprod_25zm5}). For the models with $Z=10^{-7}$ ([Fe/H] = -5.8),  the ratios become [Sr/Ba]= 
0.05 and -0.08   (see Fig.  \ref{fig:overprod_25zm7}).
Thus, we notice that for non rotating models the ratio [Sr/Ba] decreases slightly for lower $Z$, nevertheless  still remaining positive. For rotating models, at [Fe/H]= -1.8, [Sr/Ba] is very high, decreasing first slightly for lower $Z$ and then very steeply, become 
negative at  [Fe/H] = -5.8. As shown by the above models, the physical reason of these changes  is that at  lower $Z$ the many free neutrons produced by $(\alpha$,n) captures
can more easily saturate the less abundant seeds and thus the succession of n--captures may proceed to nuclei of higher atomic masses.
According to Sect. 5.1, the trend with rotation mainly results from the larger cores and thus higher temperatures, which produce  higher fractions of burned $^{22}$Ne.

\begin{table} 
\vspace*{2mm}
 \caption{Strontium and barium abundances for CEMP--no stars with  [Sr/Fe] $>$0. } \label{alphalm}
\hspace{-3mm} \begin{tabular}{lcccrrrrrrrrrr}

\hspace{3mm} Star$^{\rm Ref}$  &  $ T_{\mathrm{eff}}$ & $ \log g $ & $\left[\frac{\rm Fe}{\rm H}\right]$ & $\left[\frac{\rm Sr}{\rm Fe}\right]$  &   $\left[\frac{\rm Ba}{\rm Fe}\right]$   &  $\left[\frac{\rm Sr}{\rm Ba}\right]$ \\
&   &  & &   \\
\hline
&   &  & & \\
\scriptsize{BS 16929-005$^1$   }   & 5229   & 2.61  & -3.34           & 0.54         &   -0.41  &  0.95       \\
\scriptsize{CS 22949-037$^1$   }   & 4958 & 1.84  & -3.97         & 0.55         &   -0.52   & 1.07      \\
\scriptsize{HE 0100-1622$^3$   }   & 5400  & 3.0  & -2.93      & 0.25        & $<$-1.80    &  $>$2.05     \\
\scriptsize{HE 0233-0343$^3$   }   & 6100  & 3.4  & -4.68      &   0.32        &   $<$0.80 &  $>$-0.48        \\
\scriptsize{HE 1300-2201$^2$   }     &   6332 & 4.64 &  -2.61      & 0.28             &   -0.04&   0.32   \\
\scriptsize{HE 1327-2326$^{1,2,4}$ }       & 6180 & 3.70  & -5.76         & 1.04            &   $<$1.46  & $>$-0.42     \\
\scriptsize{HE 1330-0354$^2$   }     &  6257& 4.13  &  -2.29           & 0.01            &  -0.47 &   0.48   \\
\scriptsize{53327-2044-515$^1$ } & 5703 & 4.68 & -4.05       & 1.09           &   $<$0.34  &   $>$0.75  \\
   &   &  & & &  \\

\hline
 \label{SRBA}
\end{tabular}
\vspace*{-1mm}
\footnotesize{Ref: 1. \citet{Norris4}; 2. \citet{Allen2012};  3.   \citet{Hansen2015}; 4. \citet{Frebel2005}.}
\vspace*{2mm}
\end{table}

In the sample of 46 CEMP--no stars we may collect from the mentioned catalogs \citep{Maeder2015}, 39 stars have [Sr/Fe] data measured. Their mean value is 
[Sr/Fe]= -0.36.   There are 8 CEMP--no stars with a significant excess of the ratio [Sr/Fe], say with [Sr/Fe] $>0$ . Table \ref{SRBA} shows 
their  [Sr/Fe], [Ba/Fe] and [Sr/Ba] ratios. We see that all stars in the range of [Fe/H]= -2.2 to  -4.0 have clearly {\emph{positive}}
values of the [Sr/Ba] ratios, up to more than 2.05. The two stars with the lowest [Fe/H] values , HE 0233-0343 ([Fe/H]=-4.68) and HE 1327-2326  ([Fe/H]=-5.76),
both have {\emph{negative}} values of their lower limits for [Sr/Ba] of -0.48 and -0.42. 

There is, for now, only one  star known with a [Fe/H] ratio lower than  those quoted above, this is SMSS 0313-6708 \citep{Keller2014}. Its chemical abundances are mainly given in the form of upper limits: [Fe/H]=$< -7.3$, [Sr/H]$< -6.7$, [Ba/H]$< -6.1$. We notice that these limits may also support a positive [Sr/Fe] together with a negative [Sr/Ba] ratio for this object with an extremely low metallicity, but, since these are only upper limits, it is not possible at the moment to interpret the heavy elements abundances in this star.

Despite the fact that the sample of these most extreme objects is limited, we may note an impressive agreement between the model
predictions and the observations with the following conclusions. 
\begin{itemize}
 \item If we consider both models without and with rotation, the ranges of theoretical and observed  [Sr/Ba] ratios correspond very well lying between [Sr/Ba] $\sim$ - 0.5 and + 2.0.
 \item Without the effects of rotation, the predicted range of [Sr/Ba] ratios lies between 0.0 and 0.2, being much shorter than the observed range. Thus, non--rotating models are unable to account for the observed range of [Sr/Ba]  \citep{2013A&A...553A..51C}.
 \item The range of [Sr/Ba] ratios predicted by rotating models is much broader extending from -0.5 to 2.1, in agreement with observations.
Thus rotating models are needed for accounting the abundances of s--elements in very low metallicity stars, as shown in the last reference.
 \item Not only the observed range are correctly predicted by the models, but also the observed trend of lower [Sr/Ba]
for stars with the lower [Fe/H] ratios.  
\end{itemize}

We note that this last effect is quite consistent with the so--called  ``Ba--floor''
recently found by \citet{Hansen2015}. This is  a plateau in the absolute Ba abundances of CEMP stars  for stars with [Fe/H] $<-3.0$. Indeed, the existence of this Ba--floor
implies that for the lower [Fe/H] ratios the observed [Ba/Fe] ratios become
larger, and thus [Sr/Ba] lower as shown by the present models.

This confirms the many evidences \citep{Maeder2015} 
consistent with a significant role of rotation in stars of low metallicities, an effect with a high impact on the early chemical and spectral evolution of galaxies.

\section{Conclusions}
\label{sec:conclusions}

We calculated a large grid of rotating massive star models to determine the impact of rotation on slow neutron captures from solar down to very low metallicities following our previous exploratory studies. The main results of this study are the following:

\begin{itemize}
 \item Our models show that rotation not only enables the production of primary nitrogen, but also of important quantities of primary $^{22}$Ne at all metallicities. Whereas the neutron source 
for the $s$~process in 
non-rotating models is secondary, the neutron source is primary in rotating models.
 \item At solar metallicity, rotation-induced mixing increases the weak $s$-process production but its impact is modest (within a factor of 2)
and the production in rotating models stops at the strontium peak as in standard models.
 \item As the metallicity decreases, the amount of iron seeds decreases and the iron seeds are the main limitation to the production of heavier elements in rotating models, in which the neutron source is primary. The decreasing amount of seeds does not prevent the production of heavier elements though. On the other hand, the lack of seeds means that not only the seeds get depleted but elements in the mass range $A=60-80$ also get depleted as the production peak shifts to the strontium peak by $Z=10^{-3}$ and elements up to the barium peak are efficiently produced at that metallicity and very low metallicities. The final [Sr/Ba] ratio that we obtain is covering the range between roughly -0.5 and 2.1. 
 \item The strong dependence of production of the barium peak on metallicity and initial rotation rate means that our models provide a natural explanation for the observed scatter for the [Sr/Ba] ratio at the low metallicities.
 \item The general decrease with metallicity of the [Sr/Ba] ratio in our models also matches the decreasing ratio observed in the small current sample of CEMP-no stars at extremely low [Fe/H].
 \item Although they are challenging to measure, isotopic ratios, for example for magnesium isotopes, have a great potential for constraining stellar models. 
\end{itemize}

There are important uncertainties that affect the results presented in this paper. On the nuclear side, the dominant uncertainties are the exit channel ratios between $n$ and $\gamma$ for alpha captures on $^{17}$O and $^{22}$Ne. The first ratio determines whether $^{16}$O is a strong neutron poison or only a strong absorber, while the second determine the strength of the neutron source $^{22}$Ne$(\alpha,n)$. 
On the stellar side, the interplay of mean molecular weight and magnetic fields with rotation-induced instabilities and mixing is the main uncertainty. Concerning the stabilising effect of mean molecular weight on shear mixing, we have used a conservative prescription for shear mixing. It is not fully clear yet whether magnetic fields would increase or decrease rotation-induced mixing. 
If we compare models computed with the Tayler-Spruit dynamo and models without, we observed that
starting from the same initial conditions (mass metallicity, rotation), models with the Tayler-Spruit dynamo are more mixed \citep[see for instance][]{2005A&A...440.1041M}. On the other hand this does not imply that the models with the Tayler-Spruit dynamo would produce more primary $^{14}$N and $^{22}$Ne.  Primary nitrogen production needs
strong enough mixing in a very specific region of the star, i.e. between the helium core and the hydrogen burning shell. Whether this mixing will be strong enough depends on the gradients of the angular velocity and of the mean molecular weight in this region, as explained 
above in this paper \citep[see also][]{2013LNP...865....3M}.

These uncertainties affect quantitatively the results obtained in this study and new models will be required, e.\,g. when updated reaction rates become available. Nevertheless, the results will remain true qualitatively and their ability to explain many observed abundance features provides a strong support for the impact of rotation-induced mixing at low metallicities.

\section*{Acknowledgments}
The research leading to these results has received funding from the European Research Council under the European Union's Seventh Framework Programme (FP/2007-2013) / ERC Grant Agreement n. 306901. 
R. Hirschi acknowledges support from the World Premier International Research
Center Initiative (WPI Initiative), MEXT, Japan and from the Eurogenesis EUROCORE programme.
M. Pignatari thanks the support from the "Lend\"ulet-2014" Programme of the Hungarian Academy of Sciences (Hungary) and from SNF (Switzerland). This work was partially supported by the European Research Council GA 321263-FISH and the UK Science and Technology Facilities Council (grant ST/M000958/1).

  \bibliographystyle{mn2e}
%  {\footnotesize
%  \bibliography{ref_sproc}
%  }
%\appendix

\label{lastpage}

\end{document}